\documentclass[preprint,prd,aps,showpacs,showkeys,nofootinbib]{revtex4}
\usepackage{graphicx}
\usepackage{dcolumn}
\usepackage{bm}
\begin{document}

%\preprint{hep-ph/0412147}

\title{The two-loop supersymmetric corrections to lepton
anomalous magnetic and electric dipole moments}

\author{Tai-Fu Feng$^{a,b,c}$, Xue-Qian Li$^{d}$, Lin Lin$^a$, Jukka Maalampi$^b$,
He-Shan Song$^a$}

\affiliation{$^a$ Department of Physics, Dalian University of Technology, Dalian 116024, China}

\affiliation{$^b$ Department of Physics, 40014 University of Jyv\"askyl\"a,Finland}

\affiliation{$^c$ Center for Theoretical Physics, Seoul National University,
Seoul, 151-742, Korea}

\affiliation{$^d$ Department of Physics, Nankai University, Tianjin 300071, China}

\date{\today}

\begin{abstract}

Using the effective Lagrangian method, we analyze the electroweak
corrections to the anomalous dipole moments of  lepton from some
special two-loop topological diagrams which are composed of
neutralino (chargino) - slepton (sneutrino) in the minimal
supersymmetric extension of the standard model (MSSM). Considering
the translational invariance of the inner loop momenta and the
electromagnetic gauge invariance, we get all dimension 6 operators
and derive their coefficients. After applying equations of motion
to the external leptons, the anomalous dipole moments of lepton
are obtained. The numerical results imply that there is a
parameter space where the two-loop supersymmetric corrections to
the muon anomalous dipole moments may be significant.

\end{abstract}

\pacs{11.30.Er, 12.60.Jv,14.80.Cp}

\maketitle

\section{Introduction}
\indent\indent At both aspects of experiment and theory, the
magnetic dipole moment of lepton as well as the electric dipole
moment draw great attention of physicists because of their obvious
importance. The anomalous dipole moments of muon not only can be
used for testing loop effect in the standard model (SM), but also
provide a potential window to detect new physics
beyond the SM. The current experimental world average of the muon
magnetic dipole moment is \cite{exp}
\begin{eqnarray}
&&a_{_\mu}^{exp}=11\;659\;203\;\pm\;8\;\times 10^{-10}\;.
\label{data}
\end{eqnarray}

Contributions to the muon magnetic dipole moment are generally
divided into three sectors: QED loops, hadronic contributions as
well as electroweak  corrections. With the hadronic contributions
which are driven from the most recent $e^+e^-$ data, we can get
the following SM predictions\cite{sm1,sm2,sm3}
\begin{eqnarray}
&&a_{_\mu}^{SM}=11\;659\;180.9\;\pm\;8.0\;\times 10^{-10}
\nonumber\\
&&a_{_\mu}^{SM}=11\;659\;175.6\;\pm\;7.5\;\times 10^{-10}
\nonumber\\
&&a_{_\mu}^{SM}=11\;659\;179.4\;\pm\;9.3\;\times 10^{-10}\;.
\label{sm}
\end{eqnarray}
The deviations between the above theoretical predictions and the
experimental data are all approximately within error range of
$\sim2\sigma$. Although this $\sim2\sigma$ deviation cannot be
regarded as a strong evidence for new physics, along with the
experimental measurement precision and theoretical prediction
accuracy being constantly improved, this deviation may turn more
significant in near future.

In fact, the current experimental precision ($8\times 10^{-10}$)
already puts very restrictive bounds on new physics scenarios.
In the SM, the electroweak one- and two-loop contributions amount to $19.5\times
10^{-10}$ and $-4.4\times10^{-10}$ \cite{sm-2l} respectively.
Comparing with the standard electroweak corrections, the supersymmetric
corrections are generally suppressed by $\Lambda_{_{\rm EW}}^2/\Lambda_{_{\rm NP}}^2$,
where $\Lambda_{_{\rm EW}}$ denotes the electroweak energy scale and
$\Lambda_{_{\rm NP}}$ denotes the supersymmetric energy scale.
However, there is a parameter space where the one-loop supersymmetric
corrections are comparable to that from the SM \cite{susy1}.
Since the one-loop contribution can be large, the two-loop supersymmetric
corrections are possibly quite important \cite{susy2}.

Utilizing the heavy mass expansion approximation (HME) together
with the corresponding projection operator method, the two-loop
standard electroweak corrections to muon anomalous magnetic dipole
moment (MDM) have been evaluated \cite{czarnecki}. Within the
framework of CP conservation, the authors of Ref. \cite{heinemeyer} present the
supersymmetric corrections from some special two-loop diagrams
where a close chargino (neutralino) or scalar fermion loop is
inserted into those two-Higgs-doublet one-loop diagrams. Ref. \cite{geng}
discusses the contributions to muon MDM from the effective
vertices $H^\pm W^\mp\gamma, h_0(H_0)\gamma\gamma$ which induced
by the scalar quarks of the third generation.

In this work, we calculate the corrections from some special
two-loop diagrams which are composed of internal neutralino
(chargino) and (scalar) lepton lines. Since the electric dipole
moment (EDM) of muon is also of special interest in both
theoretical and experimental aspects \cite{nexp}, we as well
present the lepton EDM  here by keeping all possible CP violation
phases. All the diagrams which we are going to calculate,
were not discussed in literature. Besides, we first express our
results in the form which explicitly satisfies the Ward identity
requested by the QED gauge theory. In order to rationally
predict the muon EDM, we certainly
need to take the current upper experimental bounds on electron and
neutron EDMs as rigorous constraints into account. Nevertheless,
if we invoke a cancellation mechanism among different
supersymmetric contributions \cite{edm1}, or assume those
sfermions of the first generation to be heavy enough \cite{edm2},
the loop inducing lepton and neutron EDMs restrict the argument of
the $\mu$ parameter to be $\le \pi/(5\tan\beta)$, but no
constraints on other explicit $CP$ violation phases are enforced.

Here, we apply the effective Lagrangian method to get the
anomalous dipole momentums of lepton in this work. In concrete
calculation, we assume that all external leptons as well as photon
are off-shell, then expand the amplitude of corresponding triangle
diagrams according to the external momenta of leptons and photon.
Using loop momentum translational invariance, we write the sum
of the triangle diagrams which correspond to the corresponding self-energy in
the form which explicitly satisfies the Ward identity required by
the QED gauge symmetry. Then we can get all dimension six operators
together with their coefficients. After applying the equations of
motion for external leptons,  higher dimensional operators, such
as dimension eight operators, also contribute to the muon MDM and
EDM in principle. However, the contributions of dimension eight
operators contain an additional suppression factor
$m_{_\mu}^2/\Lambda_{_{\rm NP}}^2$ compared to that of dimension
six operators, where $m_{_\mu}$ is the mass of muon. Setting
$\Lambda_{_{\rm NP}}\sim100{\rm GeV}$, this suppression factor is
about $10^{-6}$. Under current experimental precision, it implies
that the contributions of all higher dimension operators ($D\ge8$)
can be neglected safely.

We adopt the naive dimensional regularization with the
anti-commuting $\gamma_{_5}$ scheme, where there is no distinction
between the first 4 dimensions and the remaining $D-4$ dimensions.
Since the bare effective Lagrangian contains the ultra-violet
divergence which is induced by divergent sub-diagrams, we give the
renormalized results in the $\overline{MS}$ scheme and
on-mass-shell scheme \cite{onshell} respectively. The two-loop
theoretical prediction certainly relies on our concrete choice of
regularization scheme and renormalization scheme, however, our
numerical results show that there is only tiny difference between
the theoretical predictions by different regularization and
renormalization schemes. We will discuss this problem in our other
work.

Through repeating the supersymmetric one-loop results, we
introduce the effective Lagrangian method and our notations in
next section. We will demonstrate how to obtain the supersymmetric
two-loop corrections to the lepton MDMs and EDMs in Section
\ref{sec3}. In the Section \ref{sec4} we study the dependence of
the lepton  MDMs and EDMs on the supersymmetry parameters
numerically. Conclusions are presented in the last Section.

\section{Our notations and the supersymmetric one-loop results \label{sec2}}
\indent\indent
The lepton MDMs and EDMs that we will calculate can actually be expressed
as the operators
\begin{eqnarray}
&&{\cal L}_{_{MDM}}={e\over4m_{_l}}\;a_{_l}\;\bar{l}\sigma^{\mu\nu}
l\;F_{_{\mu\nu}}
\;,\nonumber\\
&&{\cal L}_{_{MDM}}=-{i\over2}\;d_{_l}\;\bar{l}\sigma^{\mu\nu}\gamma_5
l\;F_{_{\mu\nu}}\;.
\label{adm}
\end{eqnarray}
Here, $l$ denotes the lepton fermion, $F_{_{\mu\nu}}$ is the
electromagnetic field strength, $m_{_l}$ is the lepton mass and
$e$ represents the electric charge respectively. Note that the
lepton here is on-shell.

%%%%%%%%%%%%%%%%%%%%%%%%%%%%%%%%%%%%%%%%%%%%%%%%%%%%%%%%%%%%%%%%%%%
\begin{figure}[t]
\setlength{\unitlength}{1mm}
\begin{center}
\begin{picture}(0,50)(0,0)
\put(-82,-150){\includegraphics{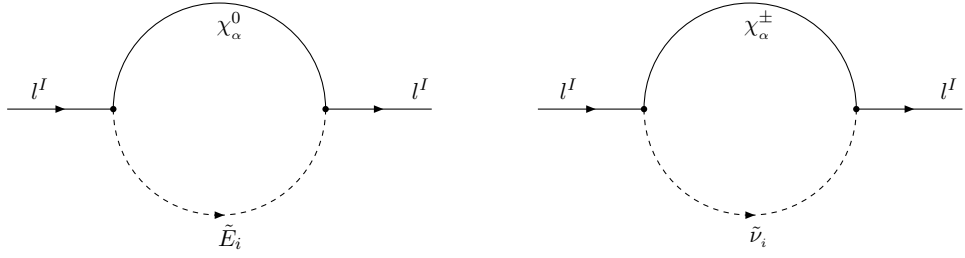}}
\end{picture}
\caption[]{The one-loop self energy diagrams which lead to the
lepton MDMs and EDMs in  MSSM, the corresponding triangle diagrams
are obtained by attaching a photon in all possible ways
to the internal particles.} \label{fig1}
\end{center}
\end{figure}
%%%%%%%%%%%%%%%%%%%%%%%%%%%%%%%%%%%%%%%%%%%%%%%%%%%%%%%%%%%%%%%%%%%

In fact, it is convenient  to get the corrections of the
loop-diagrams to lepton MDMs and EDMs in terms of the effective
Lagrangian method, if the masses of all the internal lines  are
much heavier than the external lepton mass. Assuming external
leptons as well as photon all are off-shell, we expand the
amplitude of corresponding triangle diagrams according to the
external momenta of leptons and photon. Then we can get all higher
dimension operators together with their coefficients. As discussed
in the introduction, it is enough to retain only those dimension 6
operators  in later calculations :
\begin{eqnarray}
&&{\cal O}_{_1}^\mp={1\over(4\pi)^2}\;\bar{l}\;(i/\!\!\!\!{\cal D})^3
\omega_\mp\;l\;,\nonumber\\
&&{\cal O}_{_2}^\mp={e\over(4\pi)^2}\;\overline{(i{\cal D}_{_\mu}l)}
\gamma^\mu F\cdot\sigma\omega_\mp l\;,\nonumber\\
&&{\cal O}_{_3}^\mp={e\over(4\pi)^2}\;\bar{l}F\cdot\sigma\gamma^\mu
\omega_\mp(i{\cal D}_{_\mu}l)\;,\nonumber\\
&&{\cal O}_{_4}^\mp={e\over(4\pi)^2}\;\bar{l}(\partial^\mu F_{_{\mu\nu}})
\gamma^\nu\omega_\mp l\;,\nonumber\\
&&{\cal O}_{_5}^\mp={m_{_l}\over(4\pi)^2}\;\bar{l}\;(i/\!\!\!\!{\cal D})^2
\omega_\mp\;l\;,\nonumber\\
&&{\cal O}_{_6}^\mp={em_{_l}\over(4\pi)^2}\;\bar{l}\;F\cdot\sigma
\omega_\mp\;l\;,\nonumber\\
\label{ops}
\end{eqnarray}
with ${\cal D}_{_\mu}=\partial_{_\mu}+ieA_{_\mu}$ and
$\omega_\mp=(1\mp\gamma_5)/2$. At one-loop level, there are two
triangle diagrams which contribute to lepton MDMs and EDMs
(Fig.\ref{fig1}). After expanding the amplitude according to
external momenta, the triangle diagrams determine following
dimension 6 operators together with their coefficients as
\begin{eqnarray}
&&{\cal L}_{_{\chi_\alpha^0\tilde{E}_i}}^{1L}=
-{(4\pi)^2e^2\over2s_{_{\rm w}}^2c_{_{\rm w}}^2}\int{d^4q\over i(2\pi)^4}
{1\over(q^2-m_{_{\chi_\alpha^0}}^2)(q^2-m_{_{\tilde{E}_i}}^2)}
\nonumber\\
&&\hspace{1.6cm}\times
\Bigg\{|(\xi_{_N}^I)_{_{i\alpha}}|^2\Bigg[-m_{_{\tilde{E}_i}}^2
{q^2\over(q^2-m_{_{\tilde{E}_i}}^2)^3}{\cal O}_{_1}^-
+{m_{_{\tilde{E}_i}}^2\over4}{q^2\over(q^2-m_{_{\tilde{E}_i}}^2)^3}
\Bigg({\cal O}_{_2}^-+{\cal O}_{_3}^-\Bigg)
\nonumber\\
&&\hspace{1.6cm}
+{1\over6}{(q^2)^2\over(q^2-m_{_{\tilde{E}_i}}^2)^3}{\cal O}_{_4}^-\Bigg]
\nonumber\\
&&\hspace{1.6cm}
+|(\eta_{_N}^I)_{_{i\alpha}}|^2\Bigg[-m_{_{\tilde{E}_i}}^2
{q^2\over(q^2-m_{_{\tilde{E}_i}}^2)^3}{\cal O}_{_1}^+
+{m_{_{\tilde{E}_i}}^2\over4}{q^2\over(q^2-m_{_{\tilde{E}_i}}^2)^3}
\Bigg({\cal O}_{_2}^++{\cal O}_{_3}^+\Bigg)
\nonumber\\
&&\hspace{1.6cm}
+{1\over6}{(q^2)^2\over(q^2-m_{_{\tilde{E}_i}}^2)^3}{\cal O}_{_4}^+\Bigg]
\nonumber\\
&&\hspace{1.6cm}
-{m_{_{\chi_\alpha^0}}m_{_{\tilde{E}_i}}^2\over2m_{_{l^I}}}(\eta_{_N}^I)_{_{i\alpha}}
(\xi_{_N}^I)_{_{\alpha i}}^\dagger{1\over(q^2-m_{_{\tilde{E}_i}}^2)^2}
\Bigg(-2{\cal O}_{_5}^-+{\cal O}_{_6}^-\Bigg)
\nonumber\\
&&\hspace{1.6cm}
-{m_{_{\chi_\alpha^0}}m_{_{\tilde{E}_i}}^2\over2m_{_{l^I}}}(\xi_{_N}^I)_{_{i\alpha}}
(\eta_{_N}^I)_{_{\alpha i}}^\dagger{1\over(q^2-m_{_{\tilde{E}_i}}^2)^2}
\Bigg(-2{\cal O}_{_5}^++{\cal O}_{_6}^+\Bigg)\Bigg\}
\;,\nonumber\\
%%%%%%%%%%%%%%%%%%%%%%%%%%%%%%%%%%%%%%%%%%%%%%%%%%%%%%%%%%%%%%%%%%%%%%%%%%%%%%%%%%%
&&{\cal L}_{_{\chi_{_\alpha}^\pm\tilde{\nu}_i}}^{1L}=
-{(4\pi)^2e^2\over s_{_{\rm w}}^2}\int{d^4q\over i(2\pi)^4}
{1\over(q^2-m_{_{\chi_\alpha^\pm}}^2)(q^2-m_{_{\tilde{\nu}_i}}^2)}
\nonumber\\
&&\hspace{1.6cm}\times
\Bigg\{|(\xi_{_C}^I)_{_{i\alpha}}|^2
\Bigg[m_{_{\chi_\alpha^\pm}}^4{1\over(q^2-m_{_{\chi_\alpha^\pm}}^2)^3}
{\cal O}_{_1}^--{m_{_{\chi_\alpha^\pm}}^2\over4}{q^2\over(q^2-m_{_{\chi_\alpha^\pm}}^2)^3}
({\cal O}_{_2}^-+{\cal O}_{_3}^-)
\nonumber\\
&&\hspace{1.6cm}
-{1\over6}\Bigg({(q^2)^2\over(q^2-m_{_{\chi_\alpha^\pm}}^2)^3}
-{3q^2\over(q^2-m_{_{\chi_\alpha^\pm}}^2)^2}\Bigg){\cal O}_{_4}^-\Bigg]
\nonumber\\
&&\hspace{1.6cm}
+{m_{_{l^I}}^2\over2m_{_{\rm w}}^2c_{_\beta}^2}|(\eta_{_C}^I)_{_{i\alpha}}|^2
\Bigg[m_{_{\chi_\alpha^\pm}}^4{1\over(q^2-m_{_{\chi_\alpha^\pm}}^2)^3}
{\cal O}_{_1}^+-{m_{_{\chi_\alpha^\pm}}^2\over4}{q^2\over(q^2-m_{_{\chi_\alpha^\pm}}^2)^3}
({\cal O}_{_2}^++{\cal O}_{_3}^+)
\nonumber\\
&&\hspace{1.6cm}
-{1\over6}\Bigg({(q^2)^2\over(q^2-m_{_{\chi_\alpha^\pm}}^2)^3}
-{3q^2\over(q^2-m_{_{\chi_\alpha^\pm}}^2)^2}\Bigg){\cal O}_{_4}^+\Bigg]
\nonumber\\
&&\hspace{1.6cm}
-{m_{_{\chi_\alpha^\pm}}\over\sqrt{2}m_{_{\rm w}}c_{_\beta}}
(\eta_{_C}^I)_{_{i\alpha}}(\xi_{_C}^I)_{_{\alpha i}}^\dagger
\Bigg[-m_{_{\chi_\alpha^\pm}}^2
{1\over(q^2-m_{_{\chi_\alpha^\pm}}^2)^2}{\cal O}_{_5}^-
+{1\over2}{q^2\over(q^2-m_{_{\chi_\alpha^\pm}}^2)^2}{\cal O}_{_6}^-
\Bigg]
\nonumber\\
&&\hspace{1.6cm}
-{m_{_{\chi_\alpha^\pm}}\over\sqrt{2}m_{_{\rm w}}c_{_\beta}}
(\xi_{_C}^I)_{_{i\alpha}}(\eta_{_C}^I)_{_{\alpha i}}^\dagger
\Bigg[-m_{_{\chi_\alpha^\pm}}^2
{1\over(q^2-m_{_{\chi_\alpha^\pm}}^2)^2}{\cal O}_{_5}^+
+{1\over2}{q^2\over(q^2-m_{_{\chi_\alpha^\pm}}^2)^2}{\cal O}_{_6}^+
\Bigg]\Bigg\}\;,
%%%%%%%%%%%%%%%%%%%%%%%%%%%%%%%%%%%%%%%%%%%%%%%%%%%%%%%%%%%%%%%%%%%%%%%%%%%%%%%%%%%
\label{1loop}
\end{eqnarray}
with
\begin{eqnarray}
&&(\xi_{_N}^I)_{_{i\alpha}}=(R_{_{\tilde E}})_{Ii}\Bigg(({\cal N}^\dagger)
_{_{\alpha1}}s_{_{\rm w}}+({\cal N}^\dagger)_{_{\alpha2}}c_{_{\rm w}}\Bigg)
-{m_{_l}c_{_{\rm w}}\over m_{_{\rm w}}c_{_\beta}}
(R_{_{\tilde E}})_{(3+I)i}({\cal N}^\dagger)_{_{\alpha3}}\;,
\nonumber\\
&&(\eta_{_N}^I)_{_{i\alpha}}=
2s_{_{\rm w}}(R_{_{\tilde E}})_{(3+I)i}({\cal N})
_{_{1\alpha}}+{m_{_l}c_{_{\rm w}}\over m_{_{\rm w}}c_{_\beta}}
(R_{_{\tilde E}})_{Ii}({\cal N})_{_{3\alpha}}
\;,\nonumber\\
&&(\xi_{_C}^I)_{_{i\alpha}}=(R_{_{\tilde\nu}})_{Ii}({\cal V})_{_{\alpha1}}
\;,\nonumber\\
&&(\eta_{_C}^I)_{_{i\alpha}}=(R_{_{\tilde\nu}})_{Ii}({\cal U}^\dagger)_{_{2\alpha}}\;.
\label{coup1}
\end{eqnarray}
Here, ${\cal N},\;{\cal V},\;{\cal U}$ denote the mixing matrices
of neutralinos and charginos respectively, $I,\;J=1,\;2,\;3$ are
the indices of generations. We also adopt the shortcut
notations: $c_{_{\rm w}}=\cos\theta_{_{\rm w}},\;s_{_{\rm w}}
=\sin\theta_{_{\rm w}},\;c_{_\beta}=\cos\beta,\;$ where
$\theta_{_{\rm w}}$ is the Weinberg angle, and
$\tan\beta=\upsilon_{_2}/\upsilon_{_1}$ is the ratio between the
vacuum expectation values of two Higgs doublets. As for the mixing
matrices of sleptons and sneutrinos, $R_{_{\tilde
E}},\;R_{_{\tilde\nu}}$ are:
\begin{eqnarray}
&&R_{_{\tilde E}}^\dagger\left(\begin{array}{cc}
(M_{_{LL}}^2)&(M_{_{LR}}^2)\\(M_{_{LR}}^2)^\dagger&
(M_{_{RR}}^2)\end{array}\right)R_{_{\tilde E}}
=m_{_{\tilde{E}_i}}^2\delta_{_{ij}}\;,\;(i,\;j=1,\;\cdots,\;6)
\nonumber\\
&&R_{_{\tilde\nu}}^\dagger(M_{_{\tilde \nu}}^2)R_{_{\tilde\nu}}
=m_{_{\tilde{\nu}_i}}^2\delta_{_{ij}}\;,\;(i,\;j=1,\;2,\;3).
\label{mix1}
\end{eqnarray}
Those $3\times3$ matrices are defined as
\begin{eqnarray}
&&(M_{_{LL}}^2)_{_{IJ}}=(M_{_{\tilde L}}^2)_{_{IJ}}+m_{_{l^I}}^2\delta_{_{IJ}}
+m_{_{\rm Z}}^2(s_{_{\rm w}}^2-{1\over2})\cos2\beta\delta_{_{IJ}}\;,
\nonumber\\
&&(M_{_{RR}}^2)_{_{IJ}}=(M_{_{\tilde R}}^2)_{_{IJ}}+m_{_{l^I}}^2\delta_{_{IJ}}
-m_{_{\rm Z}}^2s_{_{\rm w}}^2\cos2\beta \delta_{_{IJ}}\;,
\nonumber\\
&&(M_{_{LR}}^2)_{_{IJ}}=-(\mu_{_{\rm H}}m_{_{l^I}}\tan\beta)\delta_{_{IJ}}
+m_{_{l^I}}(A_{_e})_{_{IJ}}\;,
\nonumber\\
&&(M_{_{\tilde \nu}}^2)_{_{IJ}}=(M_{_{\tilde L}}^2)_{_{IJ}}
+{1\over2}m_{_{\rm Z}}^2\cos2\beta\;\;(I,\;J=1,\;2,\;3)\;,
\label{matrix}
\end{eqnarray}
where $M_{_{\tilde L}}^2,\;M_{_{\tilde R}}^2,\;A_{_e}$ are the
bilinear and trilinear soft breaking parameters in the lepton
sector separately, and $\mu_{_{\rm H}}$ denotes the
$\mu$-parameter in the soft supersymmetry breaking terms. Applying
the equations of motion for leptons in Eq.(\ref{1loop}), we can
get the lepton MDMs as
\begin{eqnarray}
&&\Delta a_{_{l^I}}^{\chi_\alpha^0\tilde{E}_i}=
-{e^2\over2(s_{_{\rm w}}c_{_{\rm w}})^2}\int{d^4q\over i(2\pi)^4}
{1\over(q^2-m_{_{\chi_\alpha^0}}^2)(q^2-m_{_{\tilde{E}_i}}^2)}
\nonumber\\
&&\hspace{1.9cm}\times
\Bigg\{m_{_{l^I}}^2m_{_{\tilde{E}_i}}^2\Big(|(\xi_{_N}^I)_{_{i\alpha}}|^2
+|(\eta_{_N}^I)_{_{i\alpha}}|^2\Big){q^2\over(q^2-m_{_{\tilde{E}_i}}^2)^3}
\nonumber\\
&&\hspace{1.9cm}
-2m_{_{l^I}}m_{_{\chi_\alpha^0}}m_{_{\tilde{E}_i}}^2{\bf Re}
\Big((\xi_{_N}^I)_{_{i\alpha}}(\eta_{_N}^I)_{_{\alpha i}}^\dagger\Big)
{1\over(q^2-m_{_{\tilde{E}_i}}^2)^2}\Bigg\}
\nonumber\\
%%%%%%%%%%%%%%%%%%%%%%%%%%%%%%%%%%%%%%%%%%%%%%%%%%%%%%%%%%%%%%%%%%%%%%%%%%
&&\hspace{1.5cm}=
-{e^2\over12(4\pi)^2(s_{_{\rm w}}c_{_{\rm w}})^2}\Bigg\{
\Big(|(\xi_{_N}^I)_{_{i\alpha}}|^2+|(\eta_{_N}^I)_{_{i\alpha}}|^2\Big)x_{_{l^I}}
\rho_1(x_{_{\chi_\alpha^0}},x_{_{\tilde{E}_i}})
\nonumber\\
&&\hspace{1.9cm}
+6(x_{_{l^I}}x_{_{\chi_\alpha^0}})^{1/2}{\bf Re}
\Big((\xi_{_N}^I)_{_{i\alpha}}(\eta_{_N}^I)_{_{\alpha i}}^\dagger\Big)
\rho_2(x_{_{\tilde{E}_i}},x_{_{\chi_\alpha^0}})\Bigg\}
\;,\nonumber\\
%%%%%%%%%%%%%%%%%%%%%%%%%%%%%%%%%%%%%%%%%%%%%%%%%%%%%%%%%%%%%%%%%%%%%%%%%%
%%%%%%%%%%%%%%%%%%%%%%%%%%%%%%%%%%%%%%%%%%%%%%%%%%%%%%%%%%%%%%%%%%%%%%%%%%
&&\Delta a_{_{l^I}}^{\chi_{_\alpha}^\pm\tilde{\nu}_i}=
{e^2\over s_{_{\rm w}}^2}\int{d^4q\over i(2\pi)^4}
{1\over(q^2-m_{_{\chi_\alpha^\pm}}^2)(q^2-m_{_{\tilde{\nu}_i}}^2)}
\nonumber\\
&&\hspace{1.9cm}\times
\Bigg\{m_{_{l^I}}^2m_{_{\chi_\alpha^\pm}}^2\Big(|(\xi_{_C}^I)_{_{i\alpha}}|^2
+{m_{_{l^I}}^2\over2m_{_{\rm w}}^2c_{_\beta}^2}|(\eta_{_C}^I)_{_{i\alpha}}|^2\Big)
{q^2\over(q^2-m_{_{\chi_\alpha^\pm}}^2)^3}
\nonumber\\
&&\hspace{1.9cm}
+{\sqrt{2}m_{_{l^I}}^2m_{_{\chi_\alpha^\pm}}\over m_{_{\rm w}}c_{_\beta}}
{\bf Re}\Big((\eta_{_C}^I)_{_{i\alpha}}(\xi_{_C}^I)_{_{\alpha i}}^\dagger\Big)
{q^2\over(q^2-m_{_{\chi_\alpha^\pm}}^2)^2}\Bigg\}\;,
\nonumber\\
%%%%%%%%%%%%%%%%%%%%%%%%%%%%%%%%%%%%%%%%%%%%%%%%%%%%%%%%%%%%%%%%%%%%%%%%%%
&&\hspace{1.5cm}=
{e^2\over6(4\pi)^2s_{_{\rm w}}^2}x_{_{l^I}}\Bigg\{\Big(|(\xi_{_C}^I)_{_{i\alpha}}|^2
+{m_{_{l^I}}^2\over2m_{_{\rm w}}^2c_{_\beta}^2}|(\eta_{_C}^I)_{_{i\alpha}}|^2\Big)
\rho_1(x_{_{\tilde{\nu}_i}},x_{_{\chi_\alpha^\pm}})
\nonumber\\
&&\hspace{1.9cm}
-6\sqrt{2}{m_{_{\chi_\alpha^\pm}}\over m_{_{\rm w}}c_{_\beta}}
{\bf Re}\Big((\eta_{_C}^I)_{_{i\alpha}}(\xi_{_C}^I)_{_{\alpha i}}^\dagger\Big)
\varphi_3(x_{_{\chi_\alpha^\pm}},x_{_{\tilde{\nu}_i}})\Bigg\}\;,
\label{MDM1}
\end{eqnarray}
with $x_{_i}=m_{_i}^2/\Lambda_{_{\rm NP}}^2$, and $\Lambda_{_{\rm
NP}}$ denotes the new physics scale. The definitions of the
functions $\rho_{1,2}(x,y),\;\varphi_{1,2,3}(x,y)$ can be found in
appendix \ref{ap3}. If we change our notations to that of Ref.
\cite{susy1}, one can notice that those expressions are completely
the same as the corresponding equations given in Ref.
\cite{susy1}. In order to obtain the above expressions, we have
used the following identities which originate from the loop
momentum translational invariance:
\begin{eqnarray}
%%%%%%%%%%%%%%%%%%%%%%%%%%%%%%%%%%%%%%%%%%%%%%%%%%%%%%%%%%%%%%%%%%%%%%%%%%%%%%%%%%%
&&\int{d^Dq\over(2\pi)^D}{q^2\over(q^2-m^2)^2}-{D\over 2}\int{d^Dq\over(2\pi)^D}
{1\over q^2-m^2}\equiv0\;,\nonumber\\
&&\int{d^Dq\over(2\pi)^D}{(q^2)^2\over(q^2-m^2)^2}-{D+2\over 2}\int{d^Dq\over(2\pi)^D}
{q^2\over q^2-m^2}\equiv0\;.
%%%%%%%%%%%%%%%%%%%%%%%%%%%%%%%%%%%%%%%%%%%%%%%%%%%%%%%%%%%%%%%%%%%%%%%%%%%%%%%%%%%
\label{iden1}
\end{eqnarray}
In the CP conservation framework, the supersymmetric one-loop contribution
is approximately given by
\begin{eqnarray}
\Delta a_{_\mu}^{1L}\simeq13\times10^{-10} \bigg({100\;{\rm
GeV}\over\Lambda_{_{\rm NP}}}\bigg)^2\tan\beta, \label{1L-approxi}
\end{eqnarray}
when all supersymmetric masses are assumed to be equal to
$\Lambda_{_{\rm NP}}$, and $\tan\beta\gg1$.

Correspondingly, the one loop supersymmetric contributions to the lepton EDMs
can also be written as
\begin{eqnarray}
&&\Delta d_{_{l^I}}^{\chi_\alpha^0\tilde{E}_i}=
{e^3\over2(s_{_{\rm w}}c_{_{\rm w}})^2}m_{_{\chi_\alpha^0}}
m_{_{\tilde{E}_i}}^2{\bf Im}\Big((\eta_{_N}^I)_{_{i\alpha}}
(\xi_{_N}^I)_{_{\alpha i}}^\dagger\Big)\int{d^4q\over i(2\pi)^4}
{1\over(q^2-m_{_{\chi_\alpha^0}}^2)(q^2-m_{_{\tilde{E}_i}}^2)^3}
\nonumber\\
%%%%%%%%%%%%%%%%%%%%%%%%%%%%%%%%%%%%%%%%%%%%%%%%%%%%%%%%%%%%%%%%%%%%%%%%%%
&&\hspace{1.5cm}=
-{e^3\over4(4\pi)^2(s_{_{\rm w}}c_{_{\rm w}})^2\Lambda_{_{\rm NP}}}
{\bf Im}\Big((\eta_{_N}^I)_{_{i\alpha}}(\xi_{_N}^I)_{_{\alpha i}}^\dagger\Big)
x_{_{\chi_\alpha^0}}^{1/2}\rho_2(x_{_{\tilde{E}_i}},x_{_{\chi_\alpha^0}})
\;,\nonumber\\
%%%%%%%%%%%%%%%%%%%%%%%%%%%%%%%%%%%%%%%%%%%%%%%%%%%%%%%%%%%%%%%%%%%%%%%%%%
%%%%%%%%%%%%%%%%%%%%%%%%%%%%%%%%%%%%%%%%%%%%%%%%%%%%%%%%%%%%%%%%%%%%%%%%%%
&&\Delta d_{_{l^I}}^{\chi_{_\alpha}^\pm\tilde{\nu}_i}=
{e^3\over\sqrt{2}s_{_{\rm w}}^2}{m_{_{l^I}}m_{_{\chi_\alpha^\pm}}\over m_{_{\rm w}}
c_{_\beta}}{\bf Im}\Big((\eta_{_C}^I)_{_{i\alpha}}(\xi_{_C}^I)_{_{\alpha i}}^\dagger\Big)
\int{d^4q\over i(2\pi)^4}{q^2\over(q^2-m_{_{\chi_\alpha^\pm}}^2)^3
(q^2-m_{_{\tilde{\nu}_i}}^2)}
\nonumber\\
%%%%%%%%%%%%%%%%%%%%%%%%%%%%%%%%%%%%%%%%%%%%%%%%%%%%%%%%%%%%%%%%%%%%%%%%%%
&&\hspace{1.5cm}=
-{e^3\over\sqrt{2}(4\pi)^2s_{_{\rm w}}^2\Lambda_{_{\rm NP}}}
{m_{_{l^I}}\over m_{_{\rm w}}c_{_\beta}}
{\bf Im}\Big((\eta_{_C}^I)_{_{i\alpha}}(\xi_{_C}^I)_{_{\alpha i}}^\dagger\Big)
x_{_{\chi_\alpha^\pm}}^{1/2}\varphi_3(x_{_{\chi_\alpha^\pm}}
,x_{_{\tilde{\nu}_i}})\;.
\label{EDM1}
\end{eqnarray}
Certainly, supersymmetry inducing operators ${\cal
O}_{_{2,\;3,\;6}}^\mp$ also contribute to the lepton MDMs. As we
have seen above, only the operators ${\cal O}_{_{6}}^\mp$
contribute to the EDMs of lepton at one loop level. However, we
will find that operators ${\cal O}_{_{2,\;3}}^\mp$ also contribute
to both lepton MDMs and EDMs at the two loop order.

\section{The two-loop supersymmetric corrections \label{sec3}}
\begin{figure}[t]
\setlength{\unitlength}{1mm}
\begin{center}
\begin{picture}(0,100)(0,0)
\put(-62,-100){\includegraphics{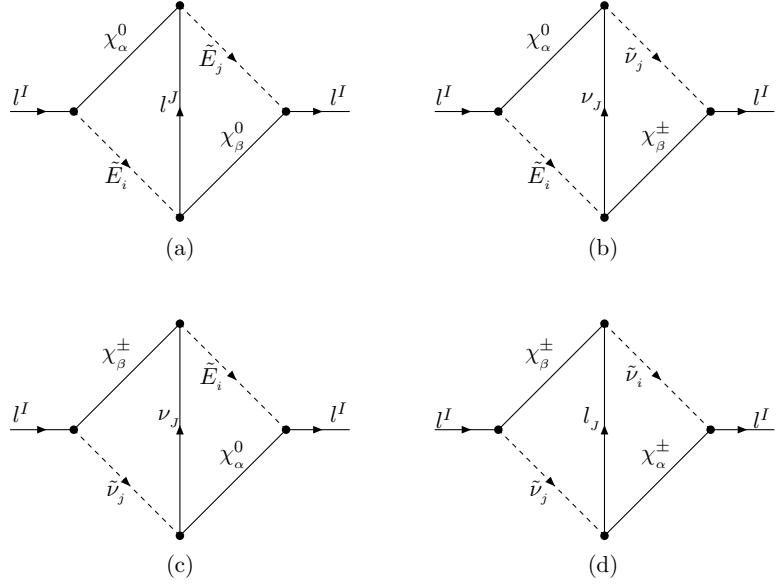}}
\end{picture}
\caption[]{The two-loop self energy diagrams which lead to the
lepton MDMs and EDMs in  MSSM, the corresponding "triangle" diagrams
are obtained by attaching a photon in all possible ways
to the internal particles.} \label{fig2}
\end{center}
\end{figure}

\indent\indent In this sector, we analyze the two-loop
supersymmetric corrections to lepton anomalous dipole moments. The
two-loop supersymmetric corrections to the coefficients  of those
operators in Eq. (\ref{ops}) originate from the two-loop
self-energy diagrams of leptons, which are depicted in Fig.
\ref{fig2}. The corresponding dipole moment diagrams are obtained
by attaching a photon to these diagrams in all possible ways. In
these diagrams there is no new suppression factor, except a factor
arising from loop integration, and the divergence caused by the
sub-diagrams can be subtracted in the $\overline{MS}$ or on-shell
schemes safely. It turns out that for some regions of the
parameter space the two-loop results are comparable with the
one-loop contributions \cite{heinemeyer,geng}. The reason for this
is that the dependence of the two-loop results on the relevant
parameters differs from that of the one-loop results. Among those
two-loop contributions which have been analyzed in the literature,
the corrections to muon MDM from the effective vertices
$\gamma\gamma H_0,\;\gamma ZH_0$  induced by the scalar quarks of
the third generation can be very well approximated by the
formulaes
\begin{eqnarray}
&&\Delta a_{_\mu}^{\tilde{t},\;2L}\simeq-0.013\times10^{-10}
{m_{_t}\mu_{_{\rm H}}\tan\beta\over m_{_{\tilde t}}M_{_H}}
\;{\rm sign}(A_{_t})\;,\nonumber\\
&&\Delta a_{_\mu}^{\tilde{b},\;2L}\simeq-0.0032\times10^{-10}
{m_{_b}A_{_b}\tan^2\beta\over m_{_{\tilde b}}M_{_H}}
\;{\rm sign}(\mu_{_{\rm H}})\;,
\label{2L-approxi1}
\end{eqnarray}
where $m_{_{\tilde t}}$ and $m_{_{\tilde b}}$ are the masses of
the lighter ${\tilde t}$ and ${\tilde b}$, $A_{_{t,b}}$ denote the
trilinear soft breaking parameters of the $t$ and $b$ quarks,
respectively, and $M_{_H}$ is the mass of the heavy CP-even Higgs
bosons. As for the two-loop diagrams where a close chargino
(neutralino) loop is inserted into those two-Higgs-doublet
one-loop diagrams, the correspondingly contribution can be
approximated as
\begin{eqnarray}
&&\Delta a_{_\mu}^{\chi,\;2L}\simeq11\times10^{-10}\bigg({\tan\beta\over50}\bigg)
\bigg({100\;{\rm GeV}\over\Lambda_{_{\rm NP}}}\bigg)^2\;{\rm sign}(\mu_{_{\rm H}})\;,
\label{2L-approxi2}
\end{eqnarray}
if all supersymmetric masses are set equal, i.e. $\mu_{_{\rm
H}}=m_{_2}=M_{_A} =\Lambda_{_{\rm NP}}$, and the $U(1)$ gaugino
mass $m_{_1}$ relates to the $SU(2)$ gaugino mass $m_{_2}$ by the
GUT relation $m_{_1} =5m_{_2}/(3s_{_{\rm w}}^2c_{_{\rm w}}^2)$
with the CP conservation assumption. Here $M_{_{A}}$ is the mass
of CP-odd neutral Higgs. Although other contributions in Ref
.\cite{heinemeyer, geng} cannot be neglected also, they cannot be
approximated as the succinct formulaes above.

%%%%%%%%%%%%%%%%%%%%%%%%%%%%%%%%%%%%%%%%%%%%%%%%%%%%%%%%%%%%%%%%%%%
%%%%%%%%%%%%%%%%%%%%%%%%%%%%%%%%%%%%%%%%%%%%%%%%%%%%%%%%%%%%%%%%%%%
Among the two-loop supersymmetric diagrams under investigation,
the corrections to the coefficients of operators in Eq.
(\ref{ops}) originate from three types of graphs: the lepton
self-energy diagrams, where there are two neutralinos and two
sleptons; a chargino, a neutralino and two sleptons or two
charginos and two sleptons as virtual particles in the loop
(Fig.\ref{fig2}). In our previous works \cite{step}, we analyzed
the contributions to the rare decay $b\rightarrow s\gamma$ and
neutron EDM from the same topological two-loop diagrams which are
composed of gluino, chargino (neutralino), and squarks. Certainly,
Figs.\ref{fig2} does not include all diagrams with internal
slepton/neutrnalino (chargino) which contribute to the anomalous
dipole moments of muon. Beside those diagrams in
Fig.\ref{fig2}, there are two-loop diagrams where a neutralino
(chargino) one-loop self-energy composed of lepton-slepton or a
slepton one-loop self-energy composed of lepton-neutralino
(chargino) is inserted into those one-loop diagrams in
Fig.\ref{fig1}. However, those diagrams belong to different
topological classes, and we will analyze the corrections from
those two-loop diagrams in our future works. We will adopt below a
terminology where, for example, the "neutralino-neutralino
contribution" means the sum of those triangle diagrams (indeed two
triangles bound together), which have two neutralinos and two
sleptons with a photon being attached in all possible ways to the
internal lines. Because the sum of the "triangle" diagrams
corresponding to each "self-energy" obviously respects the Ward
identity requested by QED gauge symmetry, we can calculate the
contributions of all the "self-energies" separately.

Since the two-loop analysis is more subtle than the analysis at
one-loop level, we show here in some detail how to evaluate all
the processes, which  contribute at two-loop level to the
theoretical prediction of the lepton MDMs and EDMs. Taking the
same steps, which we did in our earlier works \cite{step}, we
obtain the following expressions for the relevant effective
Lagrangian from the "neutralino-neutralino" self energy diagram:
\begin{eqnarray}
%%%%%%%%%%%%%%%%%%%%%%%%%%%%%%%%%%%%%%%%%%%%%%%%%%%%%%%%%%%%%%%%%%%%%%%%%%%%%%%%%%%
&&{\cal L}_{_{\chi_\alpha^0\chi_\beta^0}}=-{e^4(4\pi)^2\over4(s_{_{\rm w}}
c_{_{\rm w}})^4}\int{d^Dq_1\over(2\pi)^D}{d^Dq_2\over(2\pi)^D}
{1\over{\cal D}_{_{\chi_\alpha^0\chi_\beta^0}}}
\nonumber\\
&&\hspace{1.8cm}\times
\Bigg\{(\xi_{_N}^I)_{_{j\beta}}(\eta_{_N}^J)_{_{i\beta}}
(\eta_{_N}^J)_{_{\alpha j}}^\dagger(\xi_{_N}^I)_{_{\alpha i}}^\dagger
\sum\limits_{\rho=1}^4\Big({\cal N}_{_{\chi_\alpha^0\chi_\beta^0}}^a\Big)_\rho
{\cal O}_{_\rho}^-
\nonumber\\
&&\hspace{1.8cm}
+(\eta_{_N}^I)_{_{j\beta}}(\xi_{_N}^J)_{_{i\beta}}
(\xi_{_N}^J)_{_{\alpha j}}^\dagger(\eta_{_N}^I)_{_{\alpha i}}^\dagger
\sum\limits_{\rho=1}^4\Big({\cal N}_{_{\chi_\alpha^0\chi_\beta^0}}^a\Big)_\rho
{\cal O}_{_\rho}^+
\nonumber\\
&&\hspace{1.8cm}
+m_{_{\chi_\alpha^0}}m_{_{\chi_\beta^0}}(\xi_{_N}^I)_{_{j\beta}}
(\xi_{_N}^J)_{_{i\beta}}(\xi_{_N}^J)_{_{\alpha j}}^\dagger
(\xi_{_N}^I)_{_{\alpha i}}^\dagger
\sum\limits_{\rho=1}^4\Big({\cal N}_{_{\chi_\alpha^0\chi_\beta^0}}^b\Big)_\rho
{\cal O}_{_\rho}^-
\nonumber\\
&&\hspace{1.8cm}
+m_{_{\chi_\alpha^0}}m_{_{\chi_\beta^0}}(\eta_{_N}^I)_{_{j\beta}}
(\eta_{_N}^J)_{_{i\beta}}(\eta_{_N}^J)_{_{\alpha j}}^\dagger
(\eta_{_N}^I)_{_{\alpha i}}^\dagger
\sum\limits_{\rho=1}^4\Big({\cal N}_{_{\chi_\alpha^0\chi_\beta^0}}^b\Big)_\rho
{\cal O}_{_\rho}^+
\nonumber\\
&&\hspace{1.8cm}
-{m_{_{\chi_\alpha^0}}\over m_{_{l^I}}}(\eta_{_N}^I)_{_{j\beta}}
(\xi_{_N}^J)_{_{i\beta}}(\xi_{_N}^J)_{_{\alpha j}}^\dagger
(\xi_{_N}^I)_{_{\alpha i}}^\dagger
\sum\limits_{\rho=5}^6\Big({\cal N}_{_{\chi_\alpha^0\chi_\beta^0}}^c\Big)_\rho
{\cal O}_{_\rho}^-
\nonumber\\
&&\hspace{1.8cm}
-{m_{_{\chi_\alpha^0}}\over m_{_{l^I}}}(\xi_{_N}^I)_{_{j\beta}}
(\eta_{_N}^J)_{_{i\beta}}(\eta_{_N}^J)_{_{\alpha j}}^\dagger
(\eta_{_N}^I)_{_{\alpha i}}^\dagger
\sum\limits_{\rho=5}^6\Big({\cal N}_{_{\chi_\alpha^0\chi_\beta^0}}^c\Big)_\rho
{\cal O}_{_\rho}^+
\nonumber\\
&&\hspace{1.8cm}
-{m_{_{\chi_\beta^0}}\over m_{_{l^I}}}(\eta_{_N}^I)_{_{j\beta}}
(\eta_{_N}^J)_{_{i\beta}}(\eta_{_N}^J)_{_{\alpha j}}^\dagger
(\xi_{_N}^I)_{_{\alpha i}}^\dagger
\sum\limits_{\rho=5}^6\Big({\cal N}_{_{\chi_\alpha^0\chi_\beta^0}}^d\Big)_\rho
{\cal O}_{_\rho}^-
\nonumber\\
&&\hspace{1.8cm}
-{m_{_{\chi_\beta^0}}\over m_{_{l^I}}}(\xi_{_N}^I)_{_{j\beta}}
(\xi_{_N}^J)_{_{i\beta}}(\xi_{_N}^J)_{_{\alpha j}}^\dagger
(\eta_{_N}^I)_{_{\alpha i}}^\dagger
\sum\limits_{\rho=5}^6\Big({\cal N}_{_{\chi_\alpha^0\chi_\beta^0}}^d\Big)_\rho
{\cal O}_{_\rho}^+\Bigg\}
%%%%%%%%%%%%%%%%%%%%%%%%%%%%%%%%%%%%%%%%%%%%%%%%%%%%%%%%%%%%%%%%%%%%%%%%%%%%%%%%%%%
\label{2lnn}
\end{eqnarray}
with ${\cal
D}_{_{\chi_\alpha^0\chi_\beta^0}}=((q_2-q_1)^2-m_{_{l^J}}^2)
(q_1^2-m_{_{\chi_\beta^0}}^2)(q_1^2-m_{_{\tilde{E}_j}}^2)
(q_2^2-m_{_{\tilde{E}_i}}^2)(q_2^2-m_{_{\chi_\alpha^0}}^2)$. Those
tedious expressions of the form factors $\Big({\cal
N}_{_{\chi_\alpha^0\chi_\beta^0}}^{a,\;b}\Big)_\rho\;(\rho=1,\;\cdots,\;4)$
and $\Big({\cal
N}_{_{\chi_\alpha^0\chi_\beta^0}}^{c,\;d}\Big)_\rho\;(\rho=5,\;6)$
are listed in appendix \ref{ap2}. In order to express the sum of
those corresponding triangle diagram amplitudes which satisfy the
Ward identity  required by the QED gauge symmetry explicitly, here
we use the identities given in appendix\ref{ap1}. In a similar
way, we can rigorously verify the following equations with those
identities:
\begin{eqnarray}
%%%%%%%%%%%%%%%%%%%%%%%%%%%%%%%%%%%%%%%%%%%%%%%%%%%%%%%%%%%%%%%%%%%%%%%%%%%%%%%%%%%
&&\int{d^Dq_1\over(2\pi)^D}{d^Dq_2\over(2\pi)^D}
{\Big({\cal N}_{_{\chi_\alpha^0\chi_\beta^0}}^{a,\;b}\Big)_3
\over{\cal D}_{_{\chi_\alpha^0\chi_\beta^0}}}
(q_1\leftrightarrow q_2,\;
\alpha\leftrightarrow \beta,\;i\leftrightarrow j)
\nonumber\\
&&\hspace{-0.6cm}
\equiv\int{d^Dq_1\over(2\pi)^D}{d^Dq_2\over(2\pi)^D}
{\Big({\cal N}_{_{\chi_\alpha^0\chi_\beta^0}}^{a,\;b}\Big)_2
\over{\cal D}_{_{\chi_\alpha^0\chi_\beta^0}}}
\;,\nonumber\\
%%%%%%%%%%%%%%%%%%%%%%%%%%%%%%%%%%%%%%%%%%%%%%%%%%%%%%%%%%%%%%%%%%%%%%%%%%%%%%%%%%%
&&\int{d^Dq_1\over(2\pi)^D}{d^Dq_2\over(2\pi)^D}
{\Big({\cal N}_{_{\chi_\alpha^0\chi_\beta^0}}^d\Big)_6
\over{\cal D}_{_{\chi_\alpha^0\chi_\beta^0}}}
(q_1\leftrightarrow q_2,\;
\alpha\leftrightarrow \beta,\;i\leftrightarrow j)
\nonumber\\
&&\hspace{-0.6cm}
\equiv\int{d^Dq_1\over(2\pi)^D}{d^Dq_2\over(2\pi)^D}
{\Big({\cal N}_{_{\chi_\alpha^0\chi_\beta^0}}^c\Big)_6
\over{\cal D}_{_{\chi_\alpha^0\chi_\beta^0}}}\;.
%%%%%%%%%%%%%%%%%%%%%%%%%%%%%%%%%%%%%%%%%%%%%%%%%%%%%%%%%%%%%%%%%%%%%%%%%%%%%%%%%%%
\label{trnn}
\end{eqnarray}
In fact, this is the direct consequence of the CPT invariance in
the fundamental Lagrangian.

%%%%%%%%%%%%%%%%%%%%%%%%%%%%%%%%%%%%%%%%%%%%%%%%%%%%%%%%%%%%%%%%%%%
\begin{figure}[t]
\setlength{\unitlength}{1mm}
\begin{center}
\begin{picture}(0,100)(0,0)
\put(-82,-100){\includegraphics{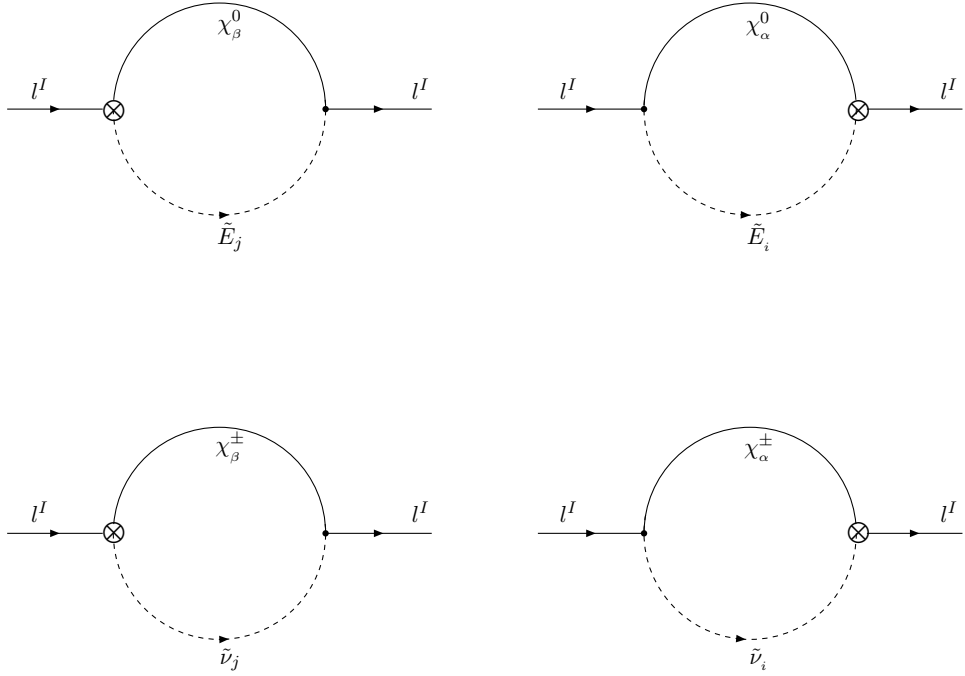}}
\end{picture}
\caption[]{The diagrams which cancel the ultra-violet divergence
of the two-loop diagrams,  $\otimes$ represents the counter terms
which  originate from the corresponding one-loop diagrams. Here,
the corresponding triangle diagrams are obtained by attaching a
photon to the internal charged slepton or chargino.} \label{fig3}
\end{center}
\end{figure}
%%%%%%%%%%%%%%%%%%%%%%%%%%%%%%%%%%%%%%%%%%%%%%%%%%%%%%%%%%%%%%%%%%%

The coefficients of high dimensional operators in Eq. (\ref{2lnn})
contain ultra-violet divergence that is caused by those divergent
sub-diagrams. In order to obtain physical predictions on lepton
MDMs and EDMs, it is necessary to adopt a concrete renormalization
scheme removing the ultra-violet divergence. In literature, the
ultra-violet divergence is removed in either the on-mass-shell
renormalization scheme \cite{czarnecki,heinemeyer}, or the simpler
$\overline{MS}$ renormalization scheme \cite{edm2,geng,step}. As
an over-subtracted renormalization scheme, on-shell scheme looks
more physical than $\overline{MS}$ renormalization scheme.
However, there are at least two external legs of the divergent
sub-diagram being the internal lines of the whole two-loop
diagrams. This signifies that an artistic on-shell scheme is not
more superior to the simpler $\overline{MS}$ scheme in our case.
Certainly, theoretical predictions on lepton anomalous dipole
moments depend on the concrete renormalization scheme. Here, we
present firstly the renormalized results which are obtained in the
$\overline{MS}$ renormalization scheme. We put the relatively
complicated results which are obtained by the on-mass-shell
scheme, in the appendix.

For the two-loop "neutralino-neutralino" diagrams, the bare effective
Lagrangian contains the following ultra-violet divergence
\begin{eqnarray}
%%%%%%%%%%%%%%%%%%%%%%%%%%%%%%%%%%%%%%%%%%%%%%%%%%%%%%%%%%%%%%%%%%%%%%%%%%%%%%%%%%%
&&{\cal L}_{_{\chi_\alpha^0\chi_\beta^0}}\in-{e^4\over96s_{_{\rm w}}^4
c_{_{\rm w}}^4}{1\over(4\pi)^2\Lambda_{_{\rm NP}}^2}{\Gamma^2(1+\epsilon)\over\epsilon}
(4\pi)^{2\epsilon}\Big(1+\epsilon(3+2\ln x_{_{\rm R}})\Big)
\nonumber\\
&&\hspace{1.2cm}\times
\Bigg\{\Bigg[\rho_1(x_{_{\chi_\beta^0}},x_{_{\tilde{E}_j}})
+\rho_1(x_{_{\chi_\alpha^0}},x_{_{\tilde{E}_i}})
+\epsilon\Bigg(\varphi_1(x_{_{\chi_\beta^0}},x_{_{\tilde{E}_j}})
\nonumber\\
&&\hspace{1.2cm}
+\varphi_2(x_{_{\chi_\alpha^0}},x_{_{\tilde{E}_i}})\Bigg)\Bigg]
\Bigg[(\xi_{_N}^I)_{_{j\beta}}(\eta_{_N}^J)_{_{i\beta}}
(\eta_{_N}^J)_{_{\alpha j}}^\dagger(\xi_{_N}^I)_{_{\alpha i}}^\dagger
\Bigg({\cal O}_{_2}^-+{\cal O}_{_3}^-\Bigg)
\nonumber\\
&&\hspace{1.2cm}
+(\eta_{_N}^I)_{_{j\beta}}(\xi_{_N}^J)_{_{i\beta}}
(\xi_{_N}^J)_{_{\alpha j}}^\dagger(\eta_{_N}^I)_{_{\alpha i}}^\dagger
\Bigg({\cal O}_{_2}^++{\cal O}_{_3}^+\Bigg)\Bigg]
\nonumber\\
&&\hspace{1.2cm}
+6{m_{_{\chi_\alpha^0}}\over m_{_{l^I}}}\Bigg[\rho_2(x_{_{\chi_\alpha^0}},
x_{_{\tilde{E}_i}})
+\epsilon\varphi_3(x_{_{\chi_\alpha^0}},x_{_{\tilde{E}_i}})\Bigg]
\nonumber\\
&&\hspace{1.2cm}\times
\Bigg[(\eta_{_N}^I)_{_{j\beta}}
(\xi_{_N}^J)_{_{i\beta}}(\xi_{_N}^J)_{_{\alpha j}}^\dagger
(\xi_{_N}^I)_{_{\alpha i}}^\dagger{\cal O}_{_6}^-
\nonumber\\
&&\hspace{1.2cm}
+(\xi_{_N}^I)_{_{j\beta}}
(\eta_{_N}^J)_{_{i\beta}}(\eta_{_N}^J)_{_{\alpha j}}^\dagger
(\eta_{_N}^I)_{_{\alpha i}}^\dagger{\cal O}_{_6}^+\Bigg]
\nonumber\\
&&\hspace{1.2cm}
+6{m_{_{\chi_\beta^0}}\over m_{_{l^I}}}\Bigg[
\rho_2(x_{_{\chi_\beta^0}},x_{_{\tilde{E}_j}})
+\epsilon\varphi_3(x_{_{\chi_\beta^0}},x_{_{\tilde{E}_j}})\Bigg]
\nonumber\\
&&\hspace{1.2cm}\times
\Bigg[(\eta_{_N}^I)_{_{j\beta}}
(\eta_{_N}^J)_{_{i\beta}}(\eta_{_N}^J)_{_{\alpha j}}^\dagger
(\xi_{_N}^I)_{_{\alpha i}}^\dagger{\cal O}_{_6}^-
\nonumber\\
&&\hspace{1.2cm}
+(\xi_{_N}^I)_{_{j\beta}}
(\xi_{_N}^J)_{_{i\beta}}(\xi_{_N}^J)_{_{\alpha j}}^\dagger
(\eta_{_N}^I)_{_{\alpha i}}^\dagger{\cal O}_{_6}^+\Bigg]
+\cdots\Bigg\}\;,
%%%%%%%%%%%%%%%%%%%%%%%%%%%%%%%%%%%%%%%%%%%%%%%%%%%%%%%%%%%%%%%%%%%%%%%%%%%%%%%%%%%
\label{dvnn}
\end{eqnarray}
with $\epsilon=2-D/2$, where $D$ denotes the dimension of time-space.
Generally, the renormalization scale $\Lambda_{_{\rm RE}}$ and the
new physics scale $\Lambda_{_{\rm NP}}$ should be of the same
order in quantity, but there does not exist a compelling reason to
make them equal. Here we keep the ratio $x_{_{\rm R}}=
\Lambda_{_{\rm RE}}^2/\Lambda_{_{\rm NP}}^2$ as a free parameter
in the expressions. In Eq. (\ref{dvnn}), we only retain the
operators ${\cal O}_{_{2,3,6}}^\pm$ that contribute to the lepton
anomalous dipole moments, and the ellipsis represents the
convergent parts of those coefficients. Certainly, the
ultra-violet divergence contained in the amplitude of counter
diagrams (the first two diagrams in Fig. \ref{fig3}) will exactly
cancel these divergences:
\begin{eqnarray}
%%%%%%%%%%%%%%%%%%%%%%%%%%%%%%%%%%%%%%%%%%%%%%%%%%%%%%%%%%%%%%%%%%%%%%%%%%%%%%%%%%%
&&{\cal L}_{_{\chi_\alpha^0\chi_\beta^0}}^C=
{1\over(4\pi)^2\Lambda_{_{\rm NP}}^2}{e^4\over96s_{_{\rm w}}^4c_{_{\rm w}}^4}
{\Gamma^2(1+\epsilon)\over\epsilon}(4\pi)^{2\epsilon}
\nonumber\\
&&\hspace{1.6cm}\times
\Bigg\{\Bigg[\Bigg(1+\epsilon(1+\ln x_{_{\rm R}})\Bigg)
\Bigg(\rho_1(x_{_{\chi_\alpha^0}},x_{_{\tilde{E}_i}})
+\rho_1(x_{_{\chi_\beta^0}},x_{_{\tilde{E}_j}})\Bigg)
\nonumber\\
&&\hspace{1.6cm}
+{\epsilon\over2}\Bigg(x_{_{\tilde{E}_i}}
{\partial^3\over\partial^3x_{_{\tilde{E}_i}}}
\varrho_{_{2,2}}(x_{_{\chi_\alpha^0}},x_{_{\tilde{E}_i}})
+x_{_{\tilde{E}_j}}{\partial^3\over\partial^3x_{_{\tilde{E}_j}}}
\varrho_{_{2,2}}(x_{_{\chi_\beta^0}},x_{_{\tilde{E}_j}})\Bigg)\Bigg]
\nonumber\\
&&\hspace{1.6cm}\times
\Bigg[(\xi_{_N}^I)_{_{j\beta}}(\eta_{_N}^J)_{_{i\beta}}
(\eta_{_N}^J)_{_{\alpha j}}^\dagger(\xi_{_N}^I)_{_{\alpha i}}^\dagger
\Bigg({\cal O}_{_2}^-+{\cal O}_{_3}^-\Bigg)
\nonumber\\
&&\hspace{1.6cm}
+(\eta_{_N}^I)_{_{j\beta}}(\xi_{_N}^J)_{_{i\beta}}
(\xi_{_N}^J)_{_{\alpha j}}^\dagger(\eta_{_N}^I)_{_{\alpha i}}^\dagger
\Bigg({\cal O}_{_2}^++{\cal O}_{_3}^+\Bigg)\Bigg]
\nonumber\\
&&\hspace{1.6cm}
+6{m_{_{\chi_\beta^0}}\over m_{_{l^I}}}
\Bigg[\Bigg(1+\epsilon(1+\ln x_{_{\rm R}})\Bigg)\rho_2(x_{_{\chi_\beta^0}}
,x_{_{\tilde{E}_j}})
\nonumber\\
&&\hspace{1.6cm}
-{\epsilon\over2}\cdot x_{_{\tilde{E}_j}}
{\partial^2\over\partial^2x_{_{\tilde{E}_j}}}
\varrho_{_{1,2}}(x_{_{\chi_\beta^0}},x_{_{\tilde{E}_j}})\Bigg]
\nonumber\\
&&\hspace{1.6cm}\times
\Bigg[(\eta_{_N}^I)_{_{j\beta}}(\eta_{_N}^J)_{_{i\beta}}
(\eta_{_N}^J)_{_{\alpha j}}^\dagger(\xi_{_N}^I)_{_{\alpha i}}^\dagger
{\cal O}_{_6}^-
\nonumber\\
&&\hspace{1.6cm}
+(\xi_{_N}^I)_{_{j\beta}}(\xi_{_N}^J)_{_{i\beta}}
(\xi_{_N}^J)_{_{\alpha j}}^\dagger(\eta_{_N}^I)_{_{\alpha i}}^\dagger
{\cal O}_{_6}^+\Bigg]
\nonumber\\
&&\hspace{1.6cm}
+6{m_{_{\chi_\alpha^0}}\over m_{_{l^I}}}
\Bigg[\Bigg(1+\epsilon(1+\ln x_{_{\rm R}})\Bigg)
\rho_2(x_{_{\chi_\alpha^0}}, x_{_{\tilde{E}_i}})
\nonumber\\
&&\hspace{1.6cm}
-{\epsilon\over2}\cdot x_{_{\tilde{E}_i}}
{\partial^2\over\partial^2x_{_{\tilde{E}_i}}}
\varrho_{_{1,2}}(x_{_{\chi_\alpha^0}},x_{_{\tilde{E}_i}})\Bigg]
\nonumber\\
&&\hspace{1.6cm}\times
\Bigg[(\eta_{_N}^I)_{_{j\beta}}(\xi_{_N}^J)_{_{i\beta}}
(\xi_{_N}^J)_{_{\alpha j}}^\dagger(\xi_{_N}^I)_{_{\alpha i}}^\dagger
{\cal O}_{_6}^-
\nonumber\\
&&\hspace{1.6cm}
+(\xi_{_N}^I)_{_{j\beta}}(\eta_{_N}^J)_{_{i\beta}}
(\eta_{_N}^J)_{_{\alpha j}}^\dagger(\eta_{_N}^I)_{_{\alpha i}}^\dagger
{\cal O}_{_6}^+\Bigg]+\cdots\Bigg\}\;.
%%%%%%%%%%%%%%%%%%%%%%%%%%%%%%%%%%%%%%%%%%%%%%%%%%%%%%%%%%%%%%%%%%%%%%%%%%%%%%%%%%%
\label{cnn}
\end{eqnarray}

Adding Eq. (\ref{cnn}) and Eq. (\ref{2lnn}), we get the two-loop "neutralino-neutralino"
corrections to the lepton MDMs:
\begin{eqnarray}
%%%%%%%%%%%%%%%%%%%%%%%%%%%%%%%%%%%%%%%%%%%%%%%%%%%%%%%%%%%%%%%%%%%%%%%%%%%%%%%%%%%
&&\Delta a_{_{l^I}}^{2L,\;\chi_\alpha^0\chi_\beta^0}=
-{e^4\over(4\pi)^2(s_{_{\rm w}}c_{_{\rm w}})^4}
\Bigg\{x_{_{l^I}}\Omega_{_{N,1}}(x_{_{l^J}};x_{_{\tilde{E}_i}},x_{_{\chi_\alpha^0}};
x_{_{\tilde{E}_j}},x_{_{\chi_\beta^0}})
\nonumber\\
&&\hspace{1.8cm}\times
\bigg[{\bf Re}\Big((\xi_{_N}^I)_{_{j\beta}}(\eta_{_N}^J)_{_{i\beta}}
(\eta_{_N}^J)_{_{\alpha j}}^\dagger(\xi_{_N}^I)_{_{\alpha i}}^\dagger\Big)
+{\bf Re}\Big((\eta_{_N}^I)_{_{j\beta}}(\xi_{_N}^J)_{_{i\beta}}
(\xi_{_N}^J)_{_{\alpha j}}^\dagger(\eta_{_N}^I)_{_{\alpha i}}^\dagger\Big)\bigg]
\nonumber\\
&&\hspace{1.8cm}
+x_{_{l^I}}(x_{_{\chi_\alpha^0}}x_{_{\chi_\beta^0}})^{1/2}
{\rm F}_{_{N,2}}(x_{_{l^J}};x_{_{\tilde{E}_i}},x_{_{\chi_\alpha^0}};
x_{_{\tilde{E}_j}},x_{_{\chi_\beta^0}})
\nonumber\\
&&\hspace{1.8cm}\times
\bigg[{\bf Re}\Big((\xi_{_N}^I)_{_{j\beta}}
(\xi_{_N}^J)_{_{i\beta}}(\xi_{_N}^J)_{_{\alpha j}}^\dagger
(\xi_{_N}^I)_{_{\alpha i}}^\dagger\Big)
+{\bf Re}\Big((\eta_{_N}^I)_{_{j\beta}}
(\eta_{_N}^J)_{_{i\beta}}(\eta_{_N}^J)_{_{\alpha j}}^\dagger
(\eta_{_N}^I)_{_{\alpha i}}^\dagger\Big)\bigg]
\nonumber\\
&&\hspace{1.8cm}
-(x_{_{l^I}}x_{_{\chi_\alpha^0}})^{1/2}
\Omega_{_{N,3}}(x_{_{l^J}};x_{_{\tilde{E}_i}},x_{_{\chi_\alpha^0}};
x_{_{\tilde{E}_j}},x_{_{\chi_\beta^0}})
{\bf Re}\Big((\eta_{_N}^I)_{_{j\beta}}
(\xi_{_N}^J)_{_{i\beta}}(\xi_{_N}^J)_{_{\alpha j}}^\dagger
(\xi_{_N}^I)_{_{\alpha i}}^\dagger\Big)
\nonumber\\
&&\hspace{1.8cm}
-(x_{_{l^I}}x_{_{\chi_\beta^0}})^{1/2}
\Omega_{_{N,4}}(x_{_{l^J}};x_{_{\tilde{E}_i}},x_{_{\chi_\alpha^0}};
x_{_{\tilde{E}_j}},x_{_{\chi_\beta^0}})
{\bf Re}\Big((\eta_{_N}^I)_{_{j\beta}}
(\eta_{_N}^J)_{_{i\beta}}(\eta_{_N}^J)_{_{\alpha j}}^\dagger
(\xi_{_N}^I)_{_{\alpha i}}^\dagger\Big)\Bigg\}\;,
%%%%%%%%%%%%%%%%%%%%%%%%%%%%%%%%%%%%%%%%%%%%%%%%%%%%%%%%%%%%%%%%%%%%%%%%%%%%%%%%%%%
\label{mdm2nn}
\end{eqnarray}
together with the lepton EDMs
\begin{eqnarray}
%%%%%%%%%%%%%%%%%%%%%%%%%%%%%%%%%%%%%%%%%%%%%%%%%%%%%%%%%%%%%%%%%%%%%%%%%%%%%%%%%%%
&&\Delta d_{_{l^I}}^{2L,\;\chi_\alpha^0\chi_\beta^0}=
-{e^5\over2(4\pi)^2(s_{_{\rm w}}c_{_{\rm w}})^4\Lambda_{_{\rm NP}}}
\Bigg\{(x_{_{l^I}})^{1/2}\Omega_{_{N,1}}(x_{_{l^J}};x_{_{\tilde{E}_i}}
,x_{_{\chi_\alpha^0}};x_{_{\tilde{E}_j}},x_{_{\chi_\beta^0}})
\nonumber\\
&&\hspace{1.8cm}\times
\bigg[{\bf Im}\Big((\xi_{_N}^I)_{_{j\beta}}(\eta_{_N}^J)_{_{i\beta}}
(\eta_{_N}^J)_{_{\alpha j}}^\dagger(\xi_{_N}^I)_{_{\alpha i}}^\dagger\Big)
-{\bf Im}\Big((\eta_{_N}^I)_{_{j\beta}}(\xi_{_N}^J)_{_{i\beta}}
(\xi_{_N}^J)_{_{\alpha j}}^\dagger(\eta_{_N}^I)_{_{\alpha i}}^\dagger\Big)\bigg]
\nonumber\\
&&\hspace{1.8cm}
+(x_{_{l^I}}x_{_{\chi_\alpha^0}}x_{_{\chi_\beta^0}})^{1/2}
{\rm F}_{_{N,2}}(x_{_{l^J}};x_{_{\tilde{E}_i}}
,x_{_{\chi_\alpha^0}};x_{_{\tilde{E}_j}},x_{_{\chi_\beta^0}})
\nonumber\\
&&\hspace{1.8cm}\times
\bigg[{\bf Im}\Big((\xi_{_N}^I)_{_{j\beta}}
(\xi_{_N}^J)_{_{i\beta}}(\xi_{_N}^J)_{_{\alpha j}}^\dagger
(\xi_{_N}^I)_{_{\alpha i}}^\dagger\Big)
-{\bf Im}\Big((\eta_{_N}^I)_{_{j\beta}}
(\eta_{_N}^J)_{_{i\beta}}(\eta_{_N}^J)_{_{\alpha j}}^\dagger
(\eta_{_N}^I)_{_{\alpha i}}^\dagger\Big)\bigg]
\nonumber\\
&&\hspace{1.8cm}
-x_{_{\chi_\alpha^0}}^{1/2}
\Omega_{_{N,3}}(x_{_{l^J}};x_{_{\tilde{E}_i}}
,x_{_{\chi_\alpha^0}};x_{_{\tilde{E}_j}},x_{_{\chi_\beta^0}})
{\bf Im}\Big((\eta_{_N}^I)_{_{j\beta}}
(\xi_{_N}^J)_{_{i\beta}}(\xi_{_N}^J)_{_{\alpha j}}^\dagger
(\xi_{_N}^I)_{_{\alpha i}}^\dagger\Big)
\nonumber\\
&&\hspace{1.8cm}
-x_{_{\chi_\beta^0}}^{1/2}
\Omega_{_{N,3}}(x_{_{l^J}};x_{_{\tilde{E}_i}}
,x_{_{\chi_\alpha^0}};x_{_{\tilde{E}_j}},x_{_{\chi_\beta^0}})
{\bf Im}\Big((\eta_{_N}^I)_{_{j\beta}}(\eta_{_N}^J)_{_{i\beta}}
(\eta_{_N}^J)_{_{\alpha j}}^\dagger(\xi_{_N}^I)_{_{\alpha i}}^\dagger\Big)\Bigg\}\;.
%%%%%%%%%%%%%%%%%%%%%%%%%%%%%%%%%%%%%%%%%%%%%%%%%%%%%%%%%%%%%%%%%%%%%%%%%%%%%%%%%%%
\label{edm2nn}
\end{eqnarray}
The form factors are expressed as
\begin{eqnarray}
%%%%%%%%%%%%%%%%%%%%%%%%%%%%%%%%%%%%%%%%%%%%%%%%%%%%%%%%%%%%%%%%%%%%%%%%%%%%%%%%%%%
&&\Omega_{_{N,1}}(x_{_{l^J}};x_{_{\tilde{E}_i}},x_{_{\chi_\alpha^0}};
x_{_{\tilde{E}_j}},x_{_{\chi_\beta^0}})
\nonumber\\
&&\hspace{-0.4cm}=
{1\over24}\Bigg\{\Big(2+\ln x_{_{\rm R}}\Big)
\Big[\rho_1(x_{_{\chi_\alpha^0}},x_{_{\tilde{E}_i}})
+\rho_1(x_{_{\chi_\beta^0}},x_{_{\tilde{E}_j}})\Big]
+\varphi_1(x_{_{\chi_\alpha^0}},x_{_{\tilde{E}_i}})
\nonumber\\&&
+\varphi_2(x_{_{\chi_\beta^0}},x_{_{\tilde{E}_j}})
-{1\over2}\Big[x_{_{\tilde{E}_i}}
{\partial^3\over\partial^3x_{_{\tilde{E}_i}}}
\varrho_{_{2,2}}(x_{_{\chi_\alpha^0}},x_{_{\tilde{E}_i}})
+x_{_{\tilde{E}_j}}{\partial^3\over\partial^3x_{_{\tilde{E}_j}}}
\varrho_{_{2,2}}(x_{_{\chi_\beta^0}},x_{_{\tilde{E}_j}})\Big]\Bigg\}
\nonumber\\&&
+{\rm F}_{_{N,1}}(x_{_{l^J}};x_{_{\tilde{E}_i}},x_{_{\chi_\alpha^0}};
x_{_{\tilde{E}_j}},x_{_{\chi_\beta^0}})
\;,\nonumber\\
%%%%%%%%%%%%%%%%%%%%%%%%%%%%%%%%%%%%%%%%%%%%%%%%%%%%%%%%%%%%%%%%%%%%%%%%%%%%%%%%%%%
&&\Omega_{_{N,3}}(x_{_{l^J}};x_{_{\tilde{E}_i}},x_{_{\chi_\alpha^0}};
x_{_{\tilde{E}_j}},x_{_{\chi_\beta^0}})
\nonumber\\
&&\hspace{-0.4cm}=
-{1\over4}\Bigg\{\Big(2+\ln x_{_{\rm R}}\Big)\rho_2(x_{_{\chi_\alpha^0}}
,x_{_{\tilde{E}_i}})
+\varphi_3(x_{_{\chi_\alpha^0}},x_{_{\tilde{E}_i}})
+{1\over2}x_{_{\tilde{E}_i}}
{\partial^2\over\partial^2x_{_{\tilde{E}_i}}}
\varrho_{_{1,2}}(x_{_{\chi_\alpha^0}},x_{_{\tilde{E}_i}})\Bigg\}
\nonumber\\&&
+{\rm F}_{_{N,3}}(x_{_{l^J}};x_{_{\tilde{E}_i}},x_{_{\chi_\alpha^0}};
x_{_{\tilde{E}_j}},x_{_{\chi_\beta^0}})
\;,\nonumber\\
%%%%%%%%%%%%%%%%%%%%%%%%%%%%%%%%%%%%%%%%%%%%%%%%%%%%%%%%%%%%%%%%%%%%%%%%%%%%%%%%%%%
&&\Omega_{_{N,4}}(x_{_{l^J}};x_{_{\tilde{E}_i}},x_{_{\chi_\alpha^0}};
x_{_{\tilde{E}_j}},x_{_{\chi_\beta^0}})
\nonumber\\
&&\hspace{-0.4cm}= -{1\over4}\Bigg\{\Big(2+\ln x_{_{\rm
R}}\Big)\rho_2(x_{_{\chi_\beta^0}} ,x_{_{\tilde{E}_j}})
+\varphi_3(x_{_{\chi_\beta^0}},x_{_{\tilde{E}_j}})
+{1\over2}x_{_{\tilde{E}_j}}
{\partial^2\over\partial^2x_{_{\tilde{E}_j}}}
\varrho_{_{1,2}}(x_{_{\chi_\beta^0}},x_{_{\tilde{E}_j}})\Bigg\}
\nonumber\\&& +{\rm
F}_{_{N,4}}(x_{_{l^J}};x_{_{\tilde{E}_i}},x_{_{\chi_\alpha^0}};
x_{_{\tilde{E}_j}},x_{_{\chi_\beta^0}})\;,
%%%%%%%%%%%%%%%%%%%%%%%%%%%%%%%%%%%%%%%%%%%%%%%%%%%%%%%%%%%%%%%%%%%%%%%%%%%%%%%%%%%
\label{form2nn}
\end{eqnarray}
where
\begin{eqnarray}
%%%%%%%%%%%%%%%%%%%%%%%%%%%%%%%%%%%%%%%%%%%%%%%%%%%%%%%%%%%%%%%%%%%%%%%%%%%%%%%%%%%
&&\varrho_{_{m,n}}(x_{_1}, x_{_2})={x_{_1}^m\ln^n x_{_1}-x_{_2}^m\ln^n x_{_2}
\over x_{_1}-x_{_2}}
%%%%%%%%%%%%%%%%%%%%%%%%%%%%%%%%%%%%%%%%%%%%%%%%%%%%%%%%%%%%%%%%%%%%%%%%%%%%%%%%%%%
\label{fvarphi}
\end{eqnarray}
and other functions $\rho_{1,2},\;\varphi_{1,2,3},\;{\rm
F}_{_{N,i}}\;(i=1,\;\cdots\;,4)$ are defined in appendix
\ref{ap3}. In Eq.(\ref{form2nn}), all terms in the brackets are
correct only for the naive dimensional regularization scheme and
$\overline{MS}$ renormalization scheme.

As for the two-loop "neutralino-chargino" corrections, we have
\begin{eqnarray}
%%%%%%%%%%%%%%%%%%%%%%%%%%%%%%%%%%%%%%%%%%%%%%%%%%%%%%%%%%%%%%%%%%%%%%%%%%%%%%%%%%%
&&\Delta a_{_{l^I}}^{2L,\;\chi^0\chi^\pm}=-{e^4\over2(4\pi)^2s_{_{\rm w}}^4
c_{_{\rm w}}^2}
\Bigg\{{2\sqrt{2}x_{_{l^I}}m_{_{l^I}}\over m_{_{\rm w}}c_{_\beta}}
{\bf Re}\Big((\lambda_{_N}^J)_{_{\beta i}}^\dagger
(\zeta_{_C}^J)_{_{\alpha j}}^\dagger(\eta_{_C}^I)_{_{i\alpha}}^*
(\eta_{_N}^I)_{_{j\beta}}\Big)
\nonumber\\
&&\hspace{2.4cm}\times
\Omega_{_{M,1}}(0;x_{_{\tilde{\nu}_i}},x_{_{\chi_\alpha^\pm}};
x_{_{\tilde{E}_j}},x_{_{\chi_\beta^0}})
\nonumber\\
&&\hspace{2.4cm}
+4x_{_{l^I}}(x_{_{\chi_\alpha^\pm}}x_{_{\chi_\beta^0}})^{1/2}
{\bf Re}\Big((\lambda_{_N}^J)_{_{\beta i}}^\dagger
(\zeta_{_C}^J)_{_{\alpha j}}^\dagger
(\xi_{_C}^I)_{_{\alpha i}}^\dagger(\xi_{_N}^I)_{_{j\beta}}\Big)
\nonumber\\
&&\hspace{2.4cm}\times
\Big[{\rm F}_{_{M,3}}+{\rm F}_{_{M,4}}\Big]
(0;x_{_{\tilde{\nu}_i}},x_{_{\chi_\alpha^\pm}};
x_{_{\tilde{E}_j}},x_{_{\chi_\beta^0}})
\nonumber\\
&&\hspace{2.4cm}
-2(x_{_{l^I}}x_{_{\chi_\alpha^\pm}})^{1/2}
{\bf Re}\Big((\lambda_{_N}^J)_{_{\beta i}}^\dagger
(\zeta_{_C}^J)_{_{\alpha j}}^\dagger
(\xi_{_C}^I)_{_{\alpha i}}^\dagger(\eta_{_N}^I)_{_{j\beta}}\Big)
\nonumber\\
&&\hspace{2.4cm}\times
\Omega_{_{M,3}}(0;x_{_{\tilde{\nu}_i}},x_{_{\chi_\alpha^\pm}};
x_{_{\tilde{E}_j}},x_{_{\chi_\beta^0}})
\nonumber\\
&&\hspace{2.4cm}
-{(2x_{_{l^I}}x_{_{\chi_\beta^0}})^{1/2}m_{_{l^I}}\over m_{_{\rm w}}c_{_\beta}}
{\bf Re}\Big((\lambda_{_N}^J)_{_{\beta i}}^\dagger
(\zeta_{_C}^J)_{_{\alpha j}}^\dagger
(\eta_{_C}^I)_{_{\alpha i}}^\dagger(\xi_{_N}^I)_{_{j\beta}}\Big)
\nonumber\\
&&\hspace{2.4cm}\times
\Omega_{_{M,4}}(0;x_{_{\tilde{\nu}_i}},x_{_{\chi_\alpha^\pm}};
x_{_{\tilde{E}_j}},x_{_{\chi_\beta^0}})
\Bigg\}\;,
%%%%%%%%%%%%%%%%%%%%%%%%%%%%%%%%%%%%%%%%%%%%%%%%%%%%%%%%%%%%%%%%%%%%%%%%%%%%%%%%%%%
\label{mdm2nc}
\end{eqnarray}
as well as
\begin{eqnarray}
%%%%%%%%%%%%%%%%%%%%%%%%%%%%%%%%%%%%%%%%%%%%%%%%%%%%%%%%%%%%%%%%%%%%%%%%%%%%%%%%%%%
&&\Delta d_{_{l^I}}^{2L,\;\chi^0\chi^\pm}=-{e^5\over2(4\pi)^2s_{_{\rm w}}^4
c_{_{\rm w}}^2\Lambda_{_{\rm NP}}}
\Bigg\{{\sqrt{2x_{_{l^I}}}m_{_{l^I}}\over m_{_{\rm w}}c_{_\beta}}
{\bf Im}\Big((\lambda_{_N}^J)_{_{\beta i}}^\dagger
(\zeta_{_C}^J)_{_{\alpha j}}^\dagger(\eta_{_C}^I)_{_{i\alpha}}^*
(\eta_{_N}^I)_{_{j\beta}}\Big)
\nonumber\\
&&\hspace{2.4cm}\times
\Big[{\rm F}_{_{M,1}}-{\rm F}_{_{M,2}}\Big]
(0;x_{_{\tilde{\nu}_i}},x_{_{\chi_\alpha^\pm}};
x_{_{\tilde{E}_j}},x_{_{\chi_\beta^0}})
\nonumber\\
&&\hspace{2.4cm}
+2(x_{_{l^I}}x_{_{\chi_\alpha^\pm}}x_{_{\chi_\beta^0}})^{1/2}
{\bf Im}\Big((\lambda_{_N}^J)_{_{\beta i}}^\dagger
(\zeta_{_C}^J)_{_{\alpha j}}^\dagger
(\xi_{_C}^I)_{_{\alpha i}}^\dagger(\xi_{_N}^I)_{_{j\beta}}\Big)
\nonumber\\
&&\hspace{2.4cm}\times
\Big[{\rm F}_{_{M,3}}-{\rm F}_{_{M,4}}\Big]
(0;x_{_{\tilde{\nu}_i}},x_{_{\chi_\alpha^\pm}};
x_{_{\tilde{E}_j}},x_{_{\chi_\beta^0}})
\nonumber\\
&&\hspace{2.4cm}
-(x_{_{\chi_\alpha^\pm}})^{1/2}
{\bf Im}\Big((\lambda_{_N}^J)_{_{\beta i}}^\dagger
(\zeta_{_C}^J)_{_{\alpha j}}^\dagger
(\xi_{_C}^I)_{_{\alpha i}}^\dagger(\eta_{_N}^I)_{_{j\beta}}\Big)
\nonumber\\
&&\hspace{2.4cm}\times
\Omega_{_{M,3}}(0;x_{_{\tilde{\nu}_i}},x_{_{\chi_\alpha^\pm}};
x_{_{\tilde{E}_j}},x_{_{\chi_\beta^0}})
\nonumber\\
&&\hspace{2.4cm}
+{(x_{_{\chi_\beta^0}})^{1/2}m_{_{l^I}}\over\sqrt{2}m_{_{\rm w}}c_{_\beta}}
{\bf Im}\Big((\lambda_{_N}^J)_{_{\beta i}}^\dagger
(\zeta_{_C}^J)_{_{\alpha j}}^\dagger
(\eta_{_C}^I)_{_{\alpha i}}^\dagger(\xi_{_N}^I)_{_{j\beta}}\Big)
\nonumber\\
&&\hspace{2.4cm}\times
\Omega_{_{M,4}}(0;x_{_{\tilde{\nu}_i}},x_{_{\chi_\alpha^\pm}};
x_{_{\tilde{E}_j}},x_{_{\chi_\beta^0}}) \Bigg\}\;,
%%%%%%%%%%%%%%%%%%%%%%%%%%%%%%%%%%%%%%%%%%%%%%%%%%%%%%%%%%%%%%%%%%%%%%%%%%%%%%%%%%%
\label{edm2nc}
\end{eqnarray}
where the couplings are defined as
\begin{eqnarray}
%%%%%%%%%%%%%%%%%%%%%%%%%%%%%%%%%%%%%%%%%%%%%%%%%%%%%%%%%%%%%%%%%%%%%%%%%%%%%%%%%%%
&&(\zeta_{_C}^I)_{_{i\alpha}}=
\Big((R_{_{\tilde E}})_{Ii}({\cal V})
_{_{\alpha1}}-{m_{_l}\over \sqrt{2}m_{_{\rm w}}c_{_\beta}}
(R_{_{\tilde E}})_{(3+I)i}({\cal U})_{_{\alpha2}}\Big)
\;,\nonumber\\
&&(\lambda_{_N}^I)_{_{i\alpha}}=
(R_{_{\tilde\nu}})_{Ii}\Big(({\cal N})_{_{1\alpha}}s_{_{\rm w}}
-({\cal N})_{_{2\alpha}}c_{_{\rm w}}\Big)\;.
%%%%%%%%%%%%%%%%%%%%%%%%%%%%%%%%%%%%%%%%%%%%%%%%%%%%%%%%%%%%%%%%%%%%%%%%%%%%%%%%%%%
\label{coup2}
\end{eqnarray}

With the naive dimensional regularization scheme and $\overline{MS}$
renormalization scheme, the form factors are written as
\begin{eqnarray}
%%%%%%%%%%%%%%%%%%%%%%%%%%%%%%%%%%%%%%%%%%%%%%%%%%%%%%%%%%%%%%%%%%%%%%%%%%%%%%%%%%%
&&\Omega_{_{M,1}}(0;x_{_{\tilde{\nu}_i}},x_{_{\chi_\alpha^\pm}};
x_{_{\tilde{E}_j}},x_{_{\chi_\beta^0}})
\nonumber\\
&&\hspace{-0.4cm}=
-{1\over12}\Bigg\{\Big(2+\ln x_{_{\rm R}}\Big)
\Big[\rho_1(x_{_{\tilde{\nu}_i}},x_{_{\chi_\alpha^\pm}})
-\rho_1(x_{_{\chi_\beta^0}},x_{_{\tilde{E}_j}})\Big]
+\varphi_1(x_{_{\tilde{\nu}_i}},x_{_{\chi_\alpha^\pm}})
\nonumber\\&&
-\varphi_1(x_{_{\chi_\beta^0}},x_{_{\tilde{E}_j}})
-\Big[x_{_{\chi_\alpha^\pm}}
{\partial^3\over\partial^3x_{_{\chi_\alpha^\pm}}}
\varrho_{_{2,2}}(x_{_{\chi_\alpha^\pm}},x_{_{\tilde{\nu}_i}})
-x_{_{\tilde{E}_j}}{\partial^3\over\partial^3x_{_{\tilde{E}_j}}}
\varrho_{_{2,2}}(x_{_{\chi_\beta^0}},x_{_{\tilde{E}_j}})\Big]\Bigg\}
\nonumber\\&&
+\Big[{\rm F}_{_{M,1}}+{\rm F}_{_{M,2}}\Big](x_{_{l^J}};
x_{_{\tilde{E}_i}},x_{_{\chi_\alpha^0}};x_{_{\tilde{E}_j}},
x_{_{\chi_\beta^0}})
\;,\nonumber\\
%%%%%%%%%%%%%%%%%%%%%%%%%%%%%%%%%%%%%%%%%%%%%%%%%%%%%%%%%%%%%%%%%%%%%%%%%%%%%%%%%%%
&&\Omega_{_{M,3}}(0;x_{_{\tilde{\nu}_i}},x_{_{\chi_\alpha^\pm}};
x_{_{\tilde{E}_j}},x_{_{\chi_\beta^0}})
\nonumber\\
&&\hspace{-0.4cm}=
{1\over4}\Bigg\{2\Big(2+\ln x_{_{\rm R}}\Big)\varphi_3(x_{_{\tilde{\nu}_i}}
,x_{_{\chi_\alpha^\pm}})
+\varphi_3(x_{_{\tilde{\nu}_i}},x_{_{\chi_\alpha^\pm}})
-{1\over2}{\partial^2\over\partial^2x_{_{\chi_\alpha^\pm}}}
\varrho_{_{2,2}}(x_{_{\chi_\alpha^\pm}},x_{_{\tilde{\nu}_i}})\Bigg\}
\nonumber\\&&
+{\rm F}_{_{M,5}}(0;x_{_{\tilde{\nu}_i}},x_{_{\chi_\alpha^\pm}};
x_{_{\tilde{E}_j}},x_{_{\chi_\beta^0}})
\;,\nonumber\\
%%%%%%%%%%%%%%%%%%%%%%%%%%%%%%%%%%%%%%%%%%%%%%%%%%%%%%%%%%%%%%%%%%%%%%%%%%%%%%%%%%%
&&\Omega_{_{M,4}}(0;x_{_{\tilde{\nu}_i}},x_{_{\chi_\alpha^\pm}};
x_{_{\tilde{E}_j}},x_{_{\chi_\beta^0}})
\nonumber\\
&&\hspace{-0.4cm}=
-{1\over4}\Bigg\{\Big(2+\ln x_{_{\rm R}}\Big)\rho_2(x_{_{\tilde{E}_j}}
,x_{_{\chi_\beta^0}})
+\varphi_3(x_{_{\tilde{E}_j}},x_{_{\chi_\beta^0}})
+{1\over2}x_{_{\tilde{E}_j}}
{\partial^2\over\partial^2x_{_{\tilde{E}_j}}}
\varrho_{_{1,2}}(x_{_{\chi_\beta^0}},x_{_{\tilde{E}_j}})\Bigg\}
\nonumber\\&&
+{\rm F}_{_{M,6}}(0;x_{_{\tilde{\nu}_i}},x_{_{\chi_\alpha^\pm}};
x_{_{\tilde{E}_j}},x_{_{\chi_\beta^0}})\;.
%%%%%%%%%%%%%%%%%%%%%%%%%%%%%%%%%%%%%%%%%%%%%%%%%%%%%%%%%%%%%%%%%%%%%%%%%%%%%%%%%%%
\label{form2nc}
\end{eqnarray}

The tedious expressions of functions ${\rm F}_{_{M,i}}\;(i=1,\;\cdots,\;6)$
are put in appendix \ref{ap3}.
With the naive dimensional regularization scheme and $\overline{MS}$
renormalization scheme, the resulting theoretical predictions
on lepton anomalous dipole moments from two-loop "chargino-chargino"
diagrams are similarly expressed as
\begin{eqnarray}
%%%%%%%%%%%%%%%%%%%%%%%%%%%%%%%%%%%%%%%%%%%%%%%%%%%%%%%%%%%%%%%%%%%%%%%%%%%%%%%%%%%
&&\Delta a_{_{l^I}}^{2L,\;\chi^\pm\chi^\pm}=
{e^4\over(4\pi)^2s_{_{\rm w}}^4}x_{_{l^I}}
\Bigg\{\bigg[{2m_{_{l^I}}^2\over m_{_{\rm w}}^2c_{_\beta}^2}
{\bf Re}\Big((\eta_{_C}^I)_{_{j\beta}}(\xi_{_C}^J)_{_{i\beta}}
(\xi_{_C}^J)_{_{\alpha j}}^\dagger(\eta_{_C}^I)_{_{\alpha i}}^\dagger\Big)
\nonumber\\
&&\hspace{2.5cm}
+{2m_{_{l^J}}^2\over m_{_{\rm w}}^2c_{_\beta}^2}{\bf Re}\Big(
(\xi_{_C}^I)_{_{j\beta}}(\eta_{_C}^J)_{_{i\beta}}
(\eta_{_C}^J)_{_{\alpha j}}^\dagger(\xi_{_C}^I)_{_{\alpha i}}^\dagger\Big)\bigg]
\Omega_{_{C,1}}(x_{_{l^J}};x_{_{\tilde{\nu}_i}},x_{_{\chi_\alpha^\pm}};
x_{_{\tilde{\nu}_j}},x_{_{\chi_\beta^\pm}})
\nonumber\\
&&\hspace{2.5cm}
+4(x_{_{\chi_\alpha^\pm}}x_{_{\chi_\beta^\pm}})^{1/2}\bigg[
{m_{_{l^J}}^2m_{_{l^I}}^2\over m_{_{\rm w}}^4c_{_\beta}^4}
{\bf Re}\Big((\eta_{_C}^I)_{_{j\beta}}(\eta_{_C}^J)_{_{i\beta}}
(\eta_{_C}^J)_{_{\alpha j}}^\dagger(\eta_{_C}^I)_{_{\alpha i}}^\dagger\Big)
\nonumber\\
&&\hspace{2.5cm}
+{\bf Re}\Big((\xi_{_C}^I)_{_{j\beta}}(\xi_{_C}^J)_{_{i\beta}}
(\xi_{_C}^J)_{_{\alpha j}}^\dagger(\xi_{_C}^I)_{_{\alpha i}}^\dagger\Big)
\bigg]{\rm F}_{_{C,2}}(x_{_{l^J}};x_{_{\tilde{\nu}_i}},x_{_{\chi_\alpha^\pm}};
x_{_{\tilde{\nu}_j}},x_{_{\chi_\beta^\pm}})
\nonumber\\
&&\hspace{2.5cm}
-{m_{_{l^J}}^2m_{_{\chi_\beta^\pm}}\over\sqrt{2}m_{_{\rm w}}^3c_{_\beta}^3}
{\bf Re}\Big((\eta_{_C}^I)_{_{j\beta}}(\eta_{_C}^J)_{_{i\beta}}
(\eta_{_C}^J)_{_{\alpha j}}^\dagger(\xi_{_C}^I)_{_{\alpha i}}^\dagger\Big)
\Omega_{_{C,4}}(x_{_{l^J}};x_{_{\tilde{\nu}_i}},x_{_{\chi_\alpha^\pm}};
x_{_{\tilde{\nu}_j}},x_{_{\chi_\beta^\pm}})
\nonumber\\
&&\hspace{2.5cm}
-{\sqrt{2}m_{_{\chi_\alpha^\pm}}\over m_{_{\rm w}}c_{_\beta}}
{\bf Re}\Big((\eta_{_C}^I)_{_{j\beta}}(\xi_{_C}^J)_{_{i\beta}}
(\xi_{_C}^J)_{_{\alpha j}}^\dagger(\xi_{_C}^I)_{_{\alpha i}}^\dagger\Big)
\Omega_{_{C,3}}(x_{_{l^J}};x_{_{\tilde{\nu}_i}},x_{_{\chi_\alpha^\pm}};
x_{_{\tilde{\nu}_j}},x_{_{\chi_\beta^\pm}})
\Bigg\}\;,
%%%%%%%%%%%%%%%%%%%%%%%%%%%%%%%%%%%%%%%%%%%%%%%%%%%%%%%%%%%%%%%%%%%%%%%%%%%%%%%%%%%
\label{mdm2cc}
\end{eqnarray}
and
\begin{eqnarray}
%%%%%%%%%%%%%%%%%%%%%%%%%%%%%%%%%%%%%%%%%%%%%%%%%%%%%%%%%%%%%%%%%%%%%%%%%%%%%%%%%%%
&&\Delta d_{_{l^I}}^{2L,\;\chi^\pm\chi^\pm}=
-{e^5\over(4\pi)^2s_{_{\rm w}}^4\Lambda_{_{\rm NP}}}(x_{_{l^I}})^{1/2}
\Bigg\{\bigg[{m_{_{l^I}}^2\over m_{_{\rm w}}^2c_{_\beta}^2}
{\bf Im}\Big((\eta_{_C}^I)_{_{j\beta}}(\xi_{_C}^J)_{_{i\beta}}
(\xi_{_C}^J)_{_{\alpha j}}^\dagger(\eta_{_C}^I)_{_{\alpha i}}^\dagger\Big)
\nonumber\\
&&\hspace{2.5cm}
-{m_{_{l^J}}^2\over m_{_{\rm w}}^2c_{_\beta}^2}
{\bf Im}\Big((\xi_{_C}^I)_{_{j\beta}}(\eta_{_C}^J)_{_{i\beta}}
(\eta_{_C}^J)_{_{\alpha j}}^\dagger(\xi_{_C}^I)_{_{\alpha i}}^\dagger\Big)\bigg]
\Omega_{_{C,1}}(x_{_{l^J}};x_{_{\tilde{\nu}_i}},x_{_{\chi_\alpha^\pm}};
x_{_{\tilde{\nu}_j}},x_{_{\chi_\beta^\pm}})
\nonumber\\
&&\hspace{2.5cm}
+2(x_{_{\chi_\alpha^\pm}}x_{_{\chi_\beta^\pm}})^{1/2}\bigg[{m_{_{l^J}}^2m_{_{l^I}}^2\over
4m_{_{\rm w}}^4c_{_\beta}^4}
{\bf Im}\Big((\eta_{_C}^I)_{_{j\beta}}(\eta_{_C}^J)_{_{i\beta}}
(\eta_{_C}^J)_{_{\alpha j}}^\dagger(\eta_{_C}^I)_{_{\alpha i}}^\dagger\Big)
\nonumber\\
&&\hspace{2.5cm}
-{\bf Im}\Big((\xi_{_C}^I)_{_{j\beta}}(\xi_{_C}^J)_{_{i\beta}}
(\xi_{_C}^J)_{_{\alpha j}}^\dagger(\xi_{_C}^I)_{_{\alpha i}}^\dagger\Big)\bigg]
{\rm F}_{_{C,2}}(x_{_{l^J}};x_{_{\tilde{\nu}_i}},x_{_{\chi_\alpha^\pm}};
x_{_{\tilde{\nu}_j}},x_{_{\chi_\beta^\pm}})
\nonumber\\
&&\hspace{2.5cm}
+{m_{_{l^J}}^2m_{_{\chi_\beta^\pm}}\over2\sqrt{2}m_{_{\rm w}}^3c_{_\beta}^3}
{\bf Im}\Big((\eta_{_C}^I)_{_{j\beta}}(\eta_{_C}^J)_{_{i\beta}}
(\eta_{_C}^J)_{_{\alpha j}}^\dagger(\xi_{_C}^I)_{_{\alpha i}}^\dagger\Big)
\Omega_{_{C,4}}(x_{_{l^J}};x_{_{\tilde{\nu}_i}},x_{_{\chi_\alpha^\pm}};
x_{_{\tilde{\nu}_j}},x_{_{\chi_\beta^\pm}})
\nonumber\\
&&\hspace{2.5cm}
+{m_{_{\chi_\alpha^\pm}}\over\sqrt{2}m_{_{\rm w}}c_{_\beta}}
{\bf Im}\Big((\eta_{_C}^I)_{_{j\beta}}(\xi_{_C}^J)_{_{i\beta}}
(\xi_{_C}^J)_{_{\alpha j}}^\dagger(\xi_{_C}^I)_{_{\alpha i}}^\dagger\Big)
\Omega_{_{C,3}}(x_{_{l^J}};x_{_{\tilde{\nu}_i}},x_{_{\chi_\alpha^\pm}};
x_{_{\tilde{\nu}_j}},x_{_{\chi_\beta^\pm}})
\Bigg\}\;.
%%%%%%%%%%%%%%%%%%%%%%%%%%%%%%%%%%%%%%%%%%%%%%%%%%%%%%%%%%%%%%%%%%%%%%%%%%%%%%%%%%%
\label{edm2cc}
\end{eqnarray}

Here,
\begin{eqnarray}
%%%%%%%%%%%%%%%%%%%%%%%%%%%%%%%%%%%%%%%%%%%%%%%%%%%%%%%%%%%%%%%%%%%%%%%%%%%%%%%%%%%
&&\Omega_{_{C,1}}(x_{_{l^J}};x_{_{\tilde{\nu}_i}},x_{_{\chi_\alpha^\pm}};
x_{_{\tilde{\nu}_j}},x_{_{\chi_\beta^\pm}})
\nonumber\\
&&\hspace{-0.4cm}=
{1\over24}\Bigg\{\Big(2+\ln x_{_{\rm R}}\Big)
\Big[\rho_1(x_{_{\tilde{\nu}_i}},x_{_{\chi_\alpha^\pm}})
+\rho_1(x_{_{\tilde{\nu}_j}},x_{_{\chi_\beta^\pm}})\Big]
+\varphi_1(x_{_{\tilde{\nu}_i}},x_{_{\chi_\alpha^\pm}})
\nonumber\\&&
+\varphi_2(x_{_{\tilde{\nu}_j}},x_{_{\chi_\beta^\pm}})
-{1\over2}\Big[x_{_{\chi_\alpha^\pm}}
{\partial^3\over\partial^3x_{_{\chi_\alpha^\pm}}}
\varrho_{_{2,2}}(x_{_{\chi_\alpha^\pm}},x_{_{\tilde{\nu}_i}})
+x_{_{\chi_\beta^\pm}}{\partial^3\over\partial^3x_{_{\chi_\beta^\pm}}}
\varrho_{_{2,2}}(x_{_{\chi_\beta^\pm}},x_{_{\tilde{\nu}_j}})\Big]\Bigg\}
\nonumber\\&&
+{\rm F}_{_{C,1}}(x_{_{l^J}};x_{_{\tilde{\nu}_i}},x_{_{\chi_\alpha^\pm}};
x_{_{\tilde{\nu}_j}},x_{_{\chi_\beta^\pm}})
\;,\nonumber\\
%%%%%%%%%%%%%%%%%%%%%%%%%%%%%%%%%%%%%%%%%%%%%%%%%%%%%%%%%%%%%%%%%%%%%%%%%%%%%%%%%%%
&&\Omega_{_{C,3}}(x_{_{l^J}};x_{_{\tilde{\nu}_i}},x_{_{\chi_\alpha^\pm}};
x_{_{\tilde{\nu}_j}},x_{_{\chi_\beta^\pm}})
\nonumber\\
&&\hspace{-0.4cm}=
{1\over2}\Bigg\{\Big(2+\ln x_{_{\rm R}}\Big)\varphi_3(x_{_{\chi_\alpha^\pm}}
,x_{_{\tilde{\nu}_i}})
+{1\over2}\varphi_3(x_{_{\chi_\alpha^\pm}},x_{_{\tilde{\nu}_i}})
+{1\over8}{\partial^2\over\partial^2x_{_{\chi_\alpha^\pm}}}
\varrho_{_{2,2}}(x_{_{\chi_\alpha^\pm}},x_{_{\tilde{\nu}_i}})\Bigg\}
\nonumber\\&&
+{\rm F}_{_{C,3}}(x_{_{l^J}};x_{_{\tilde{\nu}_i}},x_{_{\chi_\alpha^\pm}};
x_{_{\tilde{\nu}_j}},x_{_{\chi_\beta^\pm}})
\;,\nonumber\\
%%%%%%%%%%%%%%%%%%%%%%%%%%%%%%%%%%%%%%%%%%%%%%%%%%%%%%%%%%%%%%%%%%%%%%%%%%%%%%%%%%%
&&\Omega_{_{C,4}}(x_{_{l^J}};x_{_{\tilde{\nu}_i}},x_{_{\chi_\alpha^\pm}};
x_{_{\tilde{\nu}_j}},x_{_{\chi_\beta^\pm}})
\nonumber\\
&&\hspace{-0.4cm}= {1\over2}\Bigg\{\Big(2+\ln x_{_{\rm
R}}\Big)\varphi_3(x_{_{\chi_\beta^\pm}} ,x_{_{\tilde{\nu}_j}})
+{1\over2}\varphi_3(x_{_{\chi_\beta^\pm}},x_{_{\tilde{\nu}_j}})
+{1\over8}{\partial^2\over\partial^2x_{_{\chi_\beta^\pm}}}
\varrho_{_{2,2}}(x_{_{\chi_\beta^\pm}},x_{_{\tilde{\nu}_j}})\Bigg\}
\nonumber\\&& +{\rm
F}_{_{C,4}}(x_{_{l^J}};x_{_{\tilde{\nu}_i}},x_{_{\chi_\alpha^\pm}};
x_{_{\tilde{\nu}_j}},x_{_{\chi_\beta^\pm}})\;,
%%%%%%%%%%%%%%%%%%%%%%%%%%%%%%%%%%%%%%%%%%%%%%%%%%%%%%%%%%%%%%%%%%%%%%%%%%%%%%%%%%%
\label{form2cc}
\end{eqnarray}
where the definitions of the functions ${\rm
F}_{_{C,i}},\;(i=1,\;\cdots,\;4)$ can be found in appendix
\ref{ap3}.

Thus, we obtain the MDMs and EDMs of leptons in the $\overline{MS}$
renormalization scheme. However, the on-shell renormalization
scheme is also adopted frequently to remove the ultra-violet
divergence which appears in the radiative electroweak corrections
\cite{onshell}. As an over-subtract scheme, the counter terms
include some finite terms which originate from those renormalization conditions
in the on-shell scheme beside the ultra-violet divergence to cancel
the corresponding ultra-violet divergence in amplitude. In the concrete
calculation performed here, we need the following counter terms
to cancel the ultra-violet divergence in the one-loop corrections
to the vertex $\tilde{E}^*_i\overline{\chi^0_\alpha}\;l^I$
\begin{eqnarray}
&&\delta C_{_{\tilde{E}^*_i\overline{\chi^0_\alpha}\;l^I}}
=\Bigg\{{e\over\sqrt{2}s_{_{\rm w}}c_{_{\rm w}}}\Bigg[
\Bigg({\delta e\over e}\delta_{IJ}+{1\over2}(\delta Z_l^L)_{JI}\Bigg)
\delta_{\alpha\beta}\delta_{ij}+{1\over2}(\delta Z_{_{\tilde E}}^\dagger)
_{ij}\delta_{IJ}\delta_{\alpha\beta}
\nonumber\\
&&\hspace{2.2cm}
+{1\over2}(\delta Z_{\chi^0})_{\beta\alpha}\delta_{IJ}\delta_{ij}
\Bigg](R_{_{\tilde E}}^\dagger)_{jJ}\Bigg({\cal N}_{1\beta}s_{_{\rm w}}
+{\cal N}_{2\beta}c_{_{\rm w}}\Bigg)
\nonumber\\
&&\hspace{2.2cm}
-{e\over\sqrt{2}c_{_{\rm w}}^2}\delta c_{_{\rm w}}(R_{_{\tilde E}}^\dagger)_{jJ}
{\cal N}_{1\beta}\delta_{IJ}\delta_{ij}\delta_{\alpha\beta}
-{e\over\sqrt{2}s_{_{\rm w}}^2}\delta s_{_{\rm w}}(R_{_{\tilde E}}^\dagger)_{jJ}
{\cal N}_{2\beta}\delta_{IJ}\delta_{ij}\delta_{\alpha\beta}
\nonumber\\
&&\hspace{2.2cm}
-{em_{_{l^J}}\over\sqrt{2}m_{_{\rm w}}s_{_{\rm w}}c_{_\beta}}
\Bigg[\Bigg({\delta e\over e}+{\delta m_{_{l^J}}\over m_{_{l^J}}}
+{\delta m_{_{\rm w}}\over m_{_{\rm w}}}-{\delta s_{_{\rm w}}\over s_{_{\rm w}}}
-{\delta c_{_\beta}\over c_{_\beta}}\Bigg)\delta_{IJ}\delta_{ij}\delta_{\alpha\beta}
\nonumber\\
&&\hspace{2.2cm}
+{1\over2}(\delta Z_l^L)_{JI}\delta_{ij}\delta_{\alpha\beta}
+{1\over2}(\delta Z_{_{\tilde E}}^\dagger)_{ij}\delta_{IJ}\delta_{\alpha\beta}
+{1\over2}(\delta Z_{\chi^0})_{\beta\alpha}\delta_{IJ}\delta_{ij}
\Bigg](R_{_{\tilde E}}^\dagger)_{j(3+J)}{\cal N}_{3\beta}\Bigg\}\omega_-
\nonumber\\
&&\hspace{2.2cm}
-\Bigg\{{\sqrt{2}e\over c_{_{\rm w}}}\Bigg[\Bigg({\delta e\over e}
-{\delta c_{_{\rm w}}\over c_{_{\rm w}}}\Bigg)\delta_{IJ}\delta_{ij}\delta_{\alpha\beta}
+{1\over2}(\delta Z_l^R)_{JI}\delta_{ij}\delta_{\alpha\beta}
+{1\over2}(\delta Z_{_{\tilde E}}^\dagger)_{ij}\delta_{IJ}\delta_{\alpha\beta}
\nonumber\\
&&\hspace{2.2cm}
+{1\over2}(\delta Z_{\chi^0}^*)_{\beta\alpha}\delta_{IJ}\delta_{ij}
\Bigg](R_{_{\tilde E}}^\dagger)_{j(3+J)}{\cal N}_{1\beta}^*
\nonumber\\
&&\hspace{2.2cm}
-{em_{_{l^J}}\over\sqrt{2}m_{_{\rm w}}s_{_{\rm w}}c_{_\beta}}
\Bigg[\Bigg({\delta e\over e}+{\delta m_{_{l^J}}\over m_{_{l^J}}}
+{\delta m_{_{\rm w}}\over m_{_{\rm w}}}-{\delta s_{_{\rm w}}\over s_{_{\rm w}}}
-{\delta c_{_\beta}\over c_{_\beta}}\Bigg)\delta_{IJ}\delta_{ij}\delta_{\alpha\beta}
\nonumber\\
&&\hspace{2.2cm}
+{1\over2}(\delta Z_l^R)_{JI}\delta_{ij}\delta_{\alpha\beta}
+{1\over2}(\delta Z_{_{\tilde E}}^\dagger)_{ij}\delta_{IJ}\delta_{\alpha\beta}
+{1\over2}(\delta Z_{\chi^0}^*)_{\beta\alpha}\delta_{IJ}\delta_{ij}
\Bigg](R_{_{\tilde E}}^\dagger)_{jJ}{\cal N}_{3\beta}^*\Bigg\}\omega_+\;.
\label{counter1}
\end{eqnarray}
Here, $\delta e$ represents the renormalization correction to
electrical charge, $\delta m_{_{\rm w}}$ and $\delta m_{_{l^J}}$
stand the renormalization corrections to the W-boson and lepton
masses respectively, $\delta c_{_{\rm w}},\; \delta s_{_{\rm w}}$
as well as $\delta c_{_\beta}$ are the renormalization corrections
to parameters $c_{_{\rm w}}\;, s_{_{\rm w}}$ and $c_{_\beta}$, and
$(\delta Z_l^{L,R})_{JI},\;(\delta Z_{_{\tilde E}})_{ij},\;(\delta
Z_{\chi^0})_{\alpha\beta}$ separately denote the wave function
renormalization constants of leptons, sleptons, and neutralinos.
In the on-shell scheme, we can fix those renormalization
parameters by the mass-shell renormalization conditions
\cite{os1,os2}. They include UV divergence which cancel the
corresponding UV divergence in amplitudes and the finite
contributions which are determined by the on-shell condition, to
the resultant expression of the finite amplitudes. In a similar
way, we can write the counter terms for the one-loop corrections
to the vertex $\tilde{\nu}_{_i}^* \overline{\chi_\alpha^-}\;l^I$
\begin{eqnarray}
&&\delta C_{_{\tilde{\nu}_{_i}^*\overline{\chi_\alpha^-}\;l^I}}
=-{e\over s_{_{\rm w}}}\Bigg[\Bigg({\delta e\over e}-{\delta s_{_{\rm w}}
\over s_{_{\rm w}}}\Bigg)\delta_{IJ}\delta_{ij}\delta_{\alpha\beta}
+{1\over2}(\delta Z_{_{\tilde\nu}}^\dagger)_{ij}\delta_{IJ}\delta_{\alpha\beta}
+{1\over2}(\delta Z_{\chi^+})_{\beta\alpha}\delta_{IJ}\delta_{ij}
\nonumber\\
&&\hspace{2.2cm}
+{1\over2}(\delta Z_l^L)_{JI}\delta_{ij}\delta_{\alpha\beta}
\Bigg](R_{_{\tilde \nu}}^\dagger)_{jJ}{\cal V}^\dagger_{1\beta}\omega_-
-{em_{_{l^J}}\over\sqrt{2}m_{_{\rm w}}s_{_{\rm w}}c_{_\beta}}
\Bigg[\Bigg({\delta e\over e}+{\delta m_{_{l^J}}\over m_{_{l^J}}}
\nonumber\\
&&\hspace{2.2cm}
+{\delta m_{_{\rm w}}\over m_{_{\rm w}}}-{\delta s_{_{\rm w}}\over s_{_{\rm w}}}
-{\delta c_{_\beta}\over c_{_\beta}}\Bigg)\delta_{IJ}\delta_{ij}\delta_{\alpha\beta}
+{1\over2}(\delta Z_l^R)_{JI}\delta_{ij}\delta_{\alpha\beta}
+{1\over2}(\delta Z_{_{\tilde\nu}}^\dagger)_{ij}\delta_{IJ}\delta_{\alpha\beta}
\nonumber\\
&&\hspace{2.2cm}
+{1\over2}(\delta Z_{\chi^-}^*)_{\beta\alpha}\delta_{IJ}\delta_{ij}
\Bigg](R_{_{\tilde\nu}}^\dagger)_{jJ}{\cal U}_{\beta2}\Bigg\}\omega_+\;,
\label{counter2}
\end{eqnarray}
with $(\delta Z_{_{\tilde\nu}})_{ij},\;(\delta
Z_{\chi^-})_{\alpha\beta}$ are the wave function renormalization
constants of sneutrino and chargino respectively. In order to
shorten the length of text, we put the expressions of the
theoretical predictions on the MDMs and EDMs of leptons in terms
of the on-shell renormalization scheme in the appendix.

So far, we have obtained all the corrections from two-loop
supersymmetric diagrams shown in Fig.\ref{fig2}. Beside the two
loop diagrams discussed here, it is well known that the two-loop
Bar-Zee type diagrams also lead to significant contributions to
the fermion MDMs and EDMs in the supersymmetric theory
\cite{edm2}. The Bar-Zee diagram corrections to muon MDM are
discussed in \cite{geng}, the contributions of Bar-Zee diagrams to
muon EDM are analyzed in \cite{muonedm}. Beside those two-loop
diagrams that have been analyzed in literature and the diagrams
presented in this work, there are still large amounts of two-loop
diagrams that have concrete contributions to muon MDM and EDM. The
present status of two-loop calculations cannot be considered as a
complete analysis on MDM and EDM of muon in the framework of supersymmetry.
In the following section, we will only consider the corrections
from those two-loop diagrams in Fig. \ref{fig2} to the one-loop
supersymmetric theoretical predictions on lepton MDM and EDM
through numerical method with some assumptions on the parameter
space of MSSM.

\section{Numerical results \label{sec4}}
\indent\indent With the theoretical formulation derived in
previous sections, we numerically analyze the dependence of the
muon MDM and EDM on the supersymmetric parameters in this section.
Especially, we will present the dependence of the muon MDM and EDM
on some supersymmetric $CP$ phases in some detail here. Within
three standard error deviations, the present experimental data can
tolerate new physics corrections to the muon MDM as
$-10\times10^{-10}<\Delta a_\mu <52\times10^{-10}$. Since the
scalar leptons $\tilde{\nu}_{_\mu},\;\tilde{\mu}_{1,\;2}$ appear
as the internal intermediate particles in the two-loop diagrams
which are investigated in this work, the corrections of these
diagrams will be suppressed strongly when slepton masses are much
higher than the electroweak scale. To investigate if those
diagrams can result in concrete corrections to the muon MDM and
EDM, we choose a suitable supersymmetric parameter  region where
the masses of the second generation sleptons are lying in the
range $M_{_{\tilde\mu}}<1\;{\rm TeV}$. In this work, we neglect
all other possible sources of flavor violation except those due to
the CKM matrix, and try to avoid ambiguities of the unification
conditions of the soft-breaking parameters at the grand
unification scale in the mSUGRA scheme. The MSSM Lagrangian
contains several sources of CP violating phases: the phases of the
$\mu$ parameter in the superpotential and the corresponding
bilinear coupling of the soft breaking terms, three phases of the
gaugino mass terms, and the phases of the trilinear sfermion
Yukawa couplings in the soft Lagrangian. As we are not considering
the spontaneous CP violation in this work, the CP phase of the
soft bilinear coupling vanishes due to the tree level neutral
Higgs tadpole condition. Moreover, for the model we employ here,
the mass of the lightest Higgs boson sets a strong constraint on
the parameter space of the new physics. As indicated in the
literature \cite{Pilaftsis}, the CP violation would cause changes
to the neutral-Higgs-quark coupling, neutral Higgs-gauge-boson
coupling and self-coupling of Higgs boson. The present
experimental lower bound on the mass of the lightest Higgs bosons
is relaxed to 60 GeV. In our numerical analysis we will take this
constraint for the parameter space into account.

As a cross check, we have compared our one-loop supersymmetric
prediction on muon MDM in CP conservation framework with that
obtained with corresponding Fortran subroutine on muon MDM in the
code {\it FeynHiggs} \cite{feynhiggs}, and find a perfect
agreement. In the two-loop sector, we check our Fortran subroutine
on two-loop vacuum integrals with the corresponding programs in
the package {\it FeynHiggs}, and also find that they agree with
each other very well. For guaranteeing validity of the results, we
also independently develop certain programs for those two-loop
integrals to check our two-loop integrals in Fortran code.

\begin{figure}[t]
\setlength{\unitlength}{1mm}
\begin{center}
\begin{picture}(0,100)(0,0)
\put(-82,-20){\includegraphics{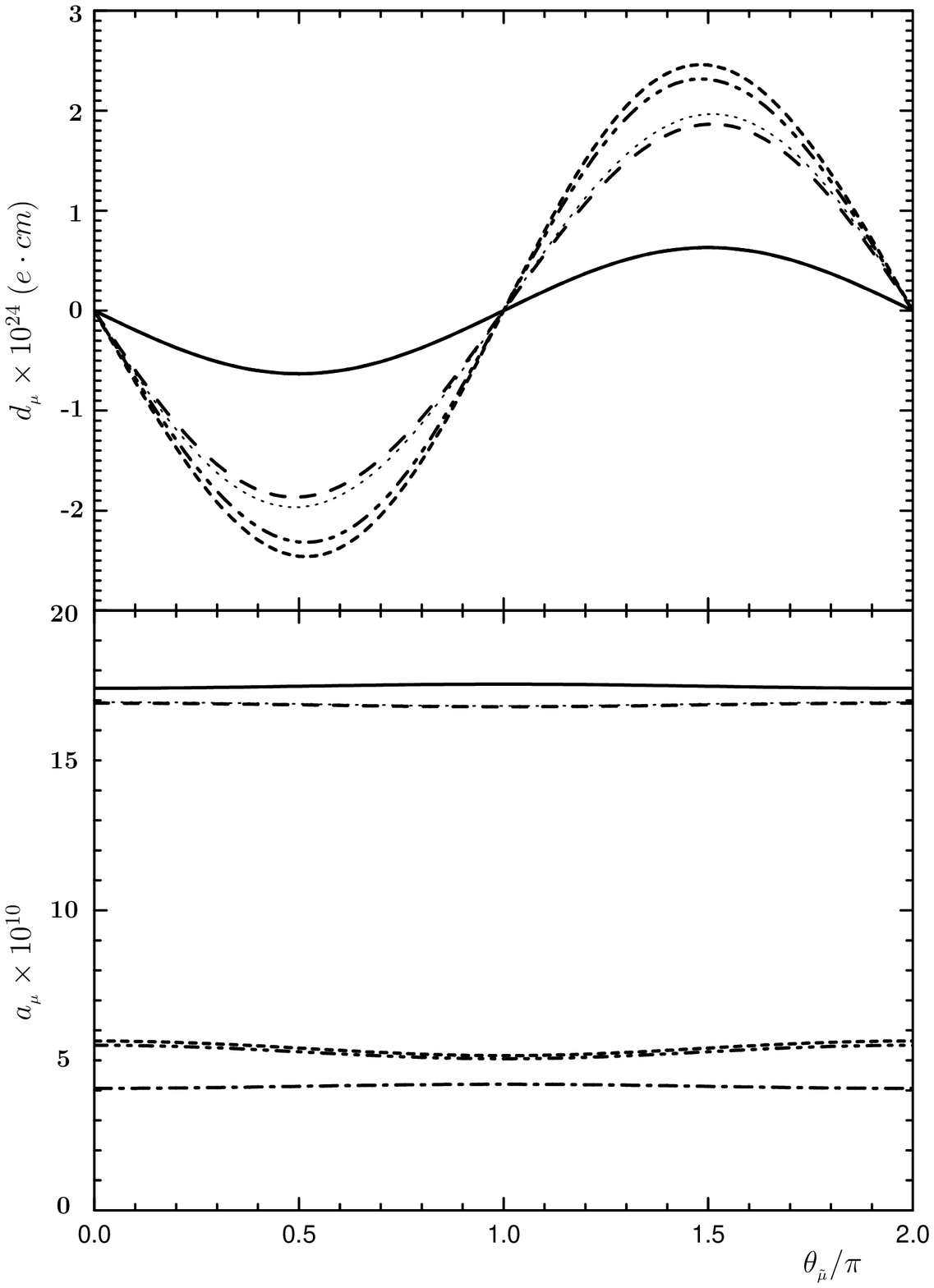}}
\end{picture}
\caption[]{The supersymmetric corrections to the MDM and EDM of
muon vary with the CP violating phase
$\theta_{_{\tilde\mu}}=\arg({A_{_{\tilde \mu}}})$ when $\mu_{_{\rm
H}}=300\;{\rm GeV}$ and $\tan\beta=5$ or $\tan\beta=20$, where the
dash-dot lines stand for the results of one loop with
$\tan\beta=5$, the dash-dot-dot lines stand for the results of two
loop in $\overline{MS}$ scheme with $\tan\beta=5$, the short dash
lines stand for the results of two loop in mass shell scheme with
$\tan\beta=5$, the solid lines stand for the results of one loop
with $\tan\beta=20$, the dash lines stand for the results of two
loop in $\overline{MS}$ scheme with $\tan\beta=20$, and the dot
lines stand for the results of two loop in mass shell scheme with
$\tan\beta=20$.} \label{fig4}
\end{center}
\end{figure}

Without losing too much generality, we will fix the following
values for the supersymmetric parameters: $M_{_{{\tilde
\mu}_L}}=M_{_{{\tilde \mu}_R}} =|A_{_{\tilde\mu}}|=500\;{\rm GeV}$
, $|m_{_1}|=|m_{_2}| = 300\;{\rm GeV}$. Taking $\mu_{_{\rm
H}}=300\;{\rm GeV}$, we plot the MDM and EDM of muon versus the CP
phase $\theta_{_{\tilde\mu}}=\arg({A_{_{\tilde \mu}}})$ for
$\tan\beta=5$ or $\tan\beta=20$ in Fig.\ref{fig4}. As
$\tan\beta=5$, one-loop supersymmetric correction to the MDM of
muon (Dash-Dot line) reaches $4\times10^{-10}$. With our choice
for the parameter space, the two-loop supersymmetric corrections
can approximately be as large as 30\% of the one-loop results. For
$\tan\beta=20$, one-loop supersymmetric prediction on the MDM of
muon (Sold line) is about $17.4\times10^{-10}$, whereas the
relative correction from two-loop supersymmetric contribution to
one-loop result is approximated as 3\%. In other words, the
corrections of these two-loop diagrams turn more and more
insignificantly along with the increase of $\tan\beta$. Actually,
the third and fourth terms of neutralino-neutralino contribution
(Eq. \ref{mdm2nn}) and the third term of neutralino-chargino
contribution (Eq. \ref{mdm2nc}) dominate the corrections to muon
MDM from the two-loop diagrams in Fig. \ref{fig2}, and those terms
depend on the parameter $\tan\beta$ very mildly. Because the CP
phase $\theta_{_{\tilde\mu}}$ affects the anomalous dipole moments
of muon through the mixing matrix $R_{_{\tilde E}}$ for sleptons
of second generation , this leads to that the variation of the
muon MDM versus $\theta_{_{\tilde\mu}}$ is very gentle. Taking
$\theta_{_{\tilde\mu}}=\pm\pi/2$, the theoretical prediction on
muon EDM approximates as $1.8\times10^{-24}\;(e\cdot cm)$ for
$\tan\beta=5$, and $2.5\times10^{-24}\;(e\cdot cm)$ for
$\tan\beta=20$. The muon EDM of this order can be detected
hopefully in near future experiments with experimental precision
$10^{-24}\;(e\cdot cm)$ \cite{nexp}.

%%%%%%%%%%%%%%%%%%%%%%%%%%%%%%%%%%%%%%%%%%%%%%%%%%%%%%%%%%%%%%%%%%%
%%%%%%%%%%%%%%%%%%%%%%%%%%%%%%%%%%%%%%%%%%%%%%%%%%%%%%%%%%%%%%%%%%%

%%%%%%%%%%%%%%%%%%%%%%%%%%%%%%%%%%%%%%%%%%%%%%%%%%%%%%%%%%%%%%%%%%%
\begin{figure}[t]
\setlength{\unitlength}{1mm}
\begin{center}
\begin{picture}(0,100)(0,0)
\put(-82,-20){\includegraphics{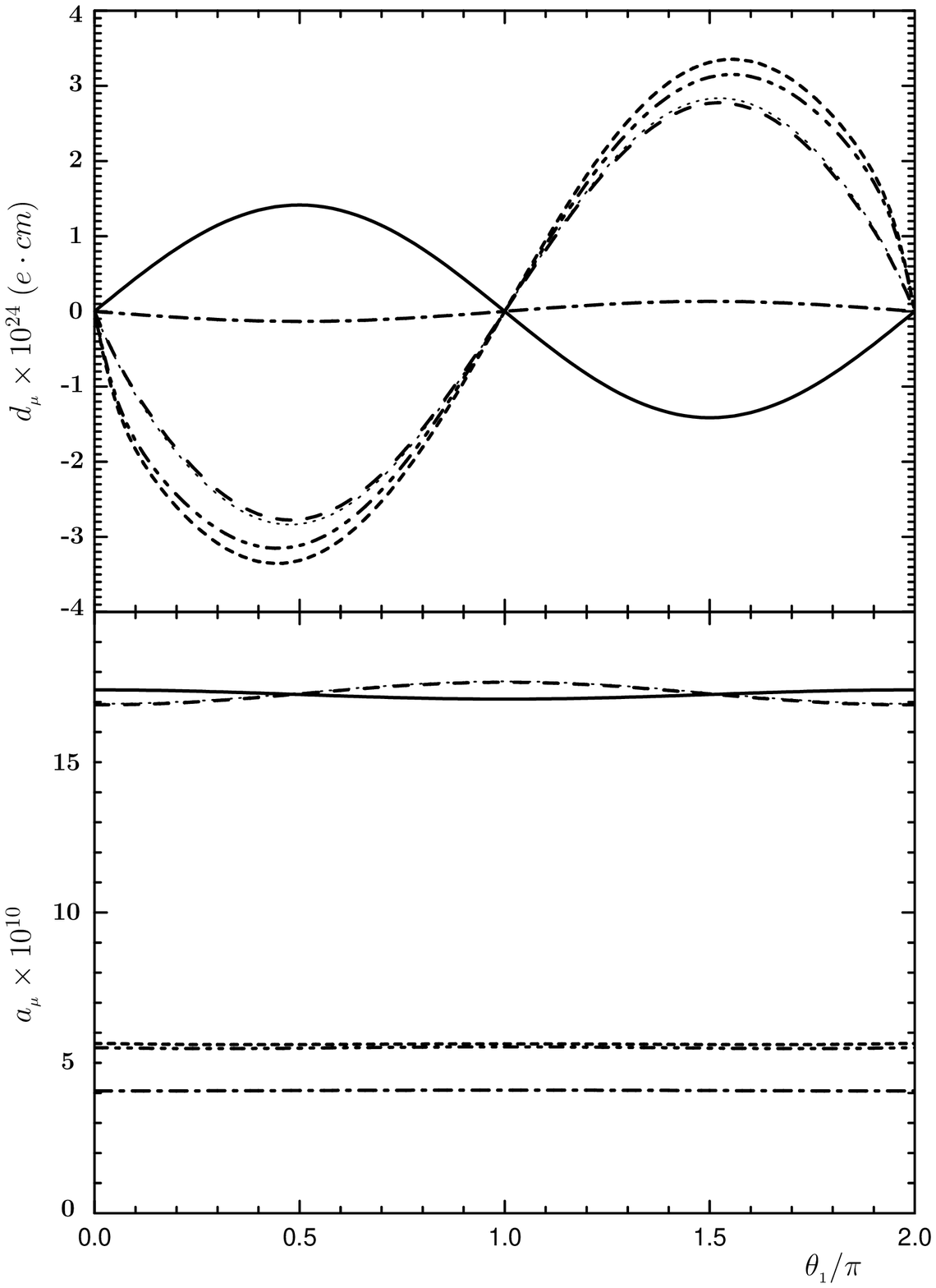}}
\end{picture}
\caption[]{The supersymmetric corrections to the MDM and EDM of
muon vary with the CP violating phase $\theta_{_{1}}=\arg(m_{_1})$
when $\mu_{_{\rm H}}=300\;{\rm GeV}$ and $\tan\beta=5$ or
$\tan\beta=20$, where the dash-dot lines stand for the results of
one loop with $\tan\beta=5$, the dash-dot-dot lines stand for the
results of two loop in $\overline{MS}$ scheme with $\tan\beta=5$,
the short dash lines stand for the results of two loop in mass
shell scheme with $\tan\beta=5$, the solid lines stand for the
results of one loop with $\tan\beta=20$, the dash lines stand for
the results of two loop in $\overline{MS}$ scheme with
$\tan\beta=20$, and the dot lines stand for the results of two
loop in mass shell scheme with $\tan\beta=20$.} \label{fig5}
\end{center}
\end{figure}
%%%%%%%%%%%%%%%%%%%%%%%%%%%%%%%%%%%%%%%%%%%%%%%%%%%%%%%%%%%%%%%%%%%

Now, we analyze the variation of supersymmetric corrections to the
muon anomalous dipole moments with the CP phase
$\theta_{_1}=\arg(m_{_1})$. Taking $\mu_{_{\rm H}}=300\;{\rm
GeV}$, we plot the MDM and EDM of muon versus the CP phase
$\theta_{_{1}}$ for $\tan\beta=5$ or $\tan\beta=20$ in
Fig.\ref{fig5}. Here, we find that muon MDM depends on the CP
phase $\theta_{_{1}}$ very gently. For $\tan\beta=5$, the one-loop
correction to the muon MDM is about $4\times10^{-10}$. With our
choice for the supersymmetric parameters, muon MDM is approximated
as $5.5\times10^{-10}$ when we include the corrections from those
two-loop diagrams in Fig. \ref{fig2}. As $\tan\beta=5$, the
one-loop contribution to the muon EDM originates from the
"neutralino-slepton" diagram, and two-loop contribution mainly
originates from the "neutralino-neutralino" diagrams. Because the
concrete dependence of one-loop result on the CP phase
$\theta_{_{1}}$ differs from that of two-loop result on
$\theta_{_{1}}$ drastically, it is easy to understand why the
correction to muon EDM from two-loop diagrams becomes dominant.
When $\tan\beta=20$, the one-loop correction to muon MDM is
enhanced, this leads to that $\Delta a_\mu$ can reach
$17.4\times10^{-10}$, the correction from those two-loop diagrams
to muon MDM turn insignificant now. However, the theoretical
prediction on muon EDM exceeds the precision of future experiment
already.

%%%%%%%%%%%%%%%%%%%%%%%%%%%%%%%%%%%%%%%%%%%%%%%%%%%%%%%%%%%%%%%%%%%
\begin{figure}[t]
\setlength{\unitlength}{1mm}
\begin{center}
\begin{picture}(0,100)(0,0)
\put(-82,-20){\includegraphics{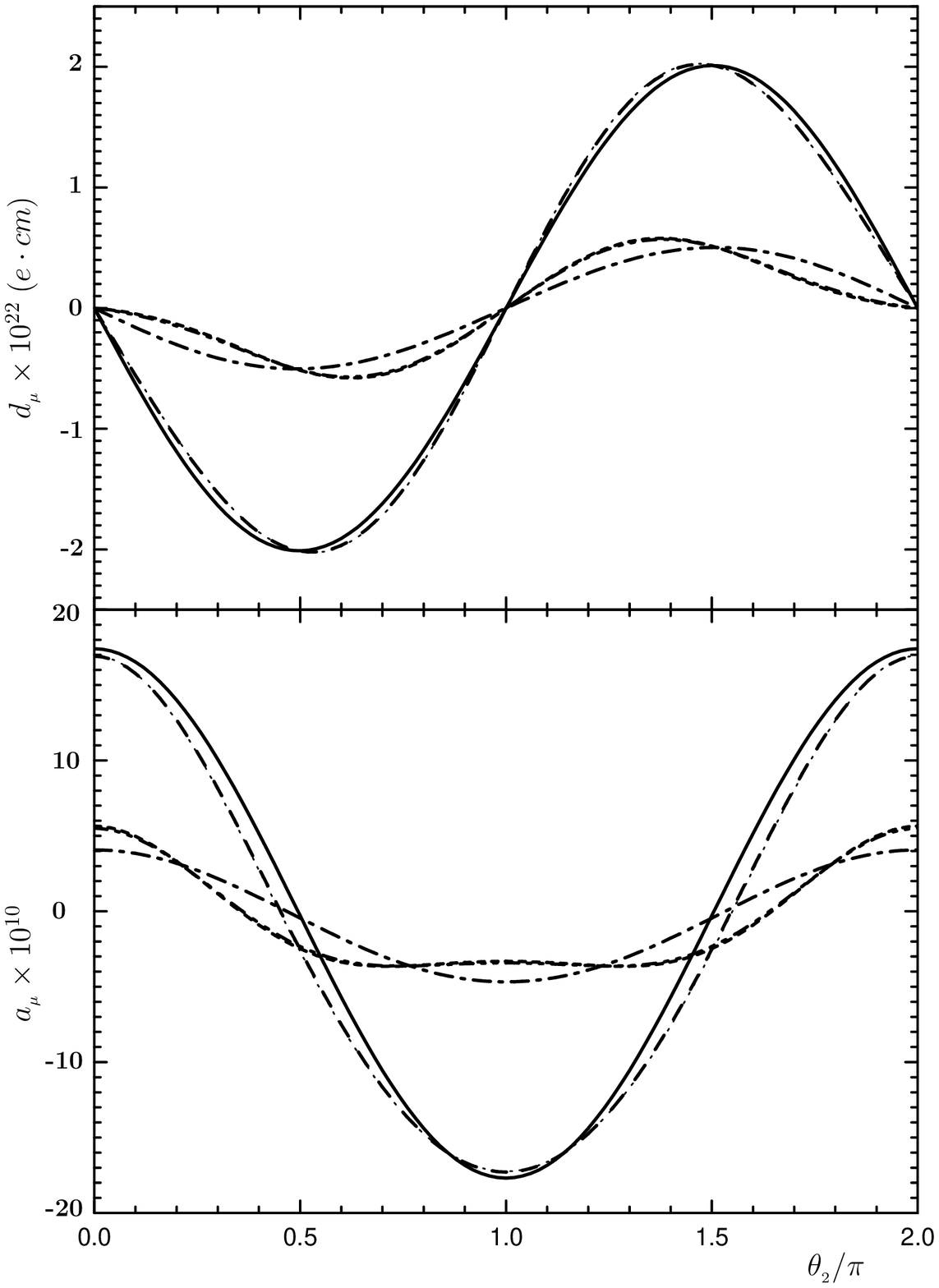}}
\end{picture}
\caption[]{The supersymmetric corrections to the MDM and EDM of
muon vary with the CP violating phase $\theta_{_{2}}=\arg(m_{_2})$
when $\mu_{_{\rm H}}=300\;{\rm GeV}$ and $\tan\beta=5$ or
$\tan\beta=20$, where the dash-dot lines stand for the results of
one loop with $\tan\beta=5$, the dash-dot-dot lines stand for the
results of two loop in $\overline{MS}$ scheme with $\tan\beta=5$,
the short dash lines stand for the results of two loop in mass
shell scheme with $\tan\beta=5$, the solid lines stand for the
results of one loop with $\tan\beta=20$, the dash lines stand for
the results of two loop in $\overline{MS}$ scheme with
$\tan\beta=20$, and the dot lines stand for the results of two
loop in mass shell scheme with $\tan\beta=20$.} \label{fig6}
\end{center}
\end{figure}
%%%%%%%%%%%%%%%%%%%%%%%%%%%%%%%%%%%%%%%%%%%%%%%%%%%%%%%%%%%%%%%%%%%

%%%%%%%%%%%%%%%%%%%%%%%%%%%%%%%%%%%%%%%%%%%%%%%%%%%%%%%%%%%%%%%%%%%
\begin{figure}[t]
\setlength{\unitlength}{1mm}
\begin{center}
\begin{picture}(0,60)(0,0)
\put(-82,-60){\includegraphics{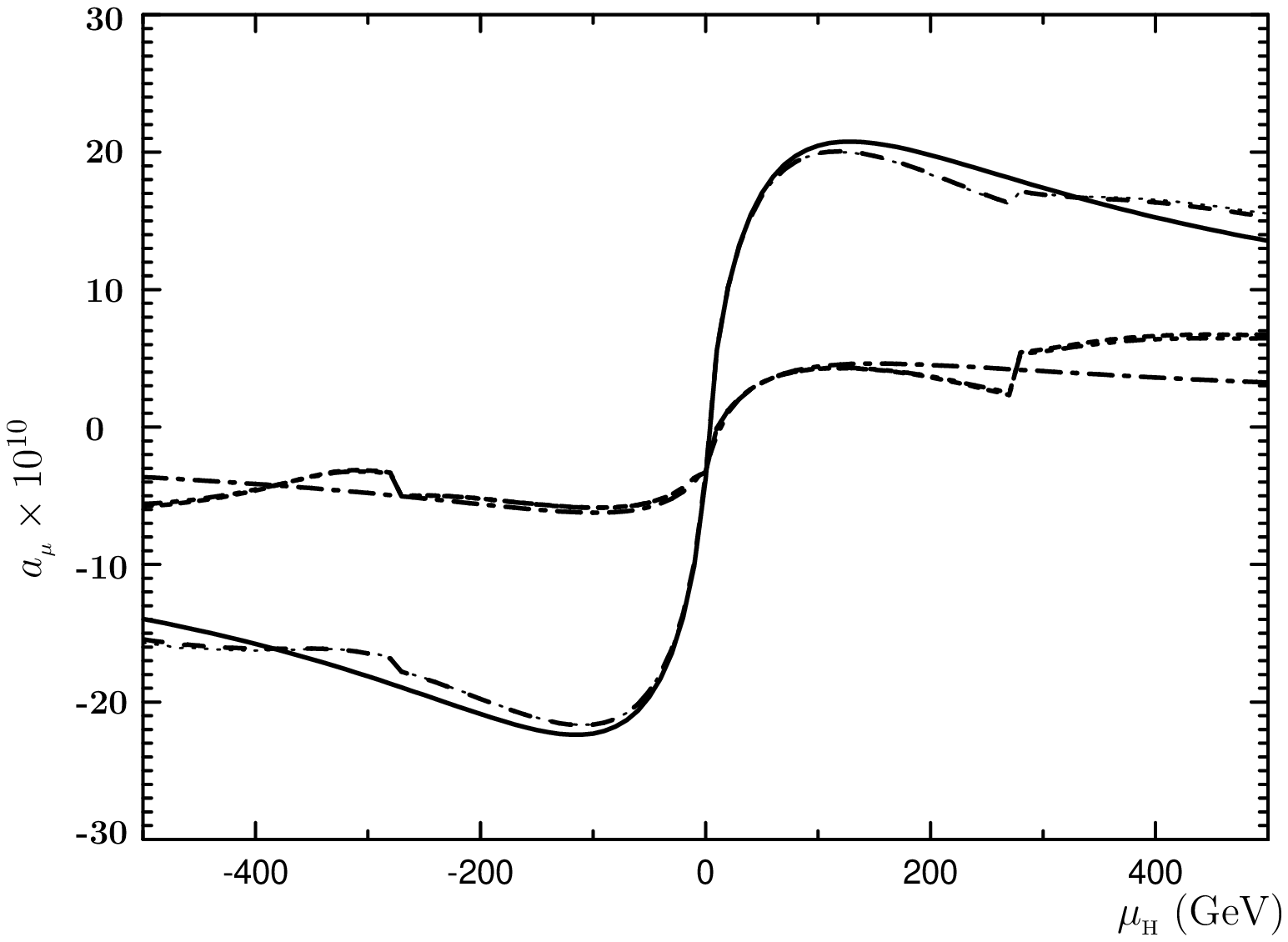}}
\end{picture}
\caption[]{The supersymmetric corrections to the MDM and EDM of
muon vary with $\mu_{_{\rm H}}$ for $\tan\beta=5$ or
$\tan\beta=20$ in CP conservation framework, where the dash-dot
line stands for the results of one loop with $\tan\beta=5$, the
dash-dot-dot line stands for the results of two loop in
$\overline{MS}$ scheme with $\tan\beta=5$, the short dash line
stands for the results of two loop in mass shell scheme with
$\tan\beta=5$, the solid line stands for the results of one loop
with $\tan\beta=20$, the dash line stands for the results of two
loop in $\overline{MS}$ scheme with $\tan\beta=20$, and the dot
line stands for the results of two loop in mass shell scheme with
$\tan\beta=20$.} \label{fig7}
\end{center}
\end{figure}
%%%%%%%%%%%%%%%%%%%%%%%%%%%%%%%%%%%%%%%%%%%%%%%%%%%%%%%%%%%%%%%%%%%

Perhaps the most interesting subject to study is the variation of
muon MDM and EDM versus the CP phase $\theta_{_2}=\arg(m_{_2})$.
Taking $\mu_{_{\rm H}}=300\;{\rm GeV}$, we plot
the MDM and EDM of muon versus the CP phase
$\theta_{_{2}}=\arg(m_{_2})$ for large $\tan\beta=5$ or
$\tan\beta=20$ in Fig.\ref{fig6}.
Generally, the two-loop correction to the muon MDM is 30\%
approximately for $\tan\beta=5$. In the largest CP violation
($\theta_{_2}=\pm\pi/2$) case, the EDM of muon is large enough and
can be experimentally tested with the experimental precision in
near future: $10^{-24}\;(e\cdot cm)$. For $\tan\beta=20$, the
one-loop supersymmetric corrections to the MDM and EDM are enhanced
drastically. Especially for the muon EDM, it reaches
$2\times10^{-22}\;(e\cdot cm)$, which can be detected easily in
the future experiment.

\begin{figure}
\setlength{\unitlength}{1mm}
\begin{center}
\begin{picture}(0,60)(0,0)
\put(-82,-60){\includegraphics{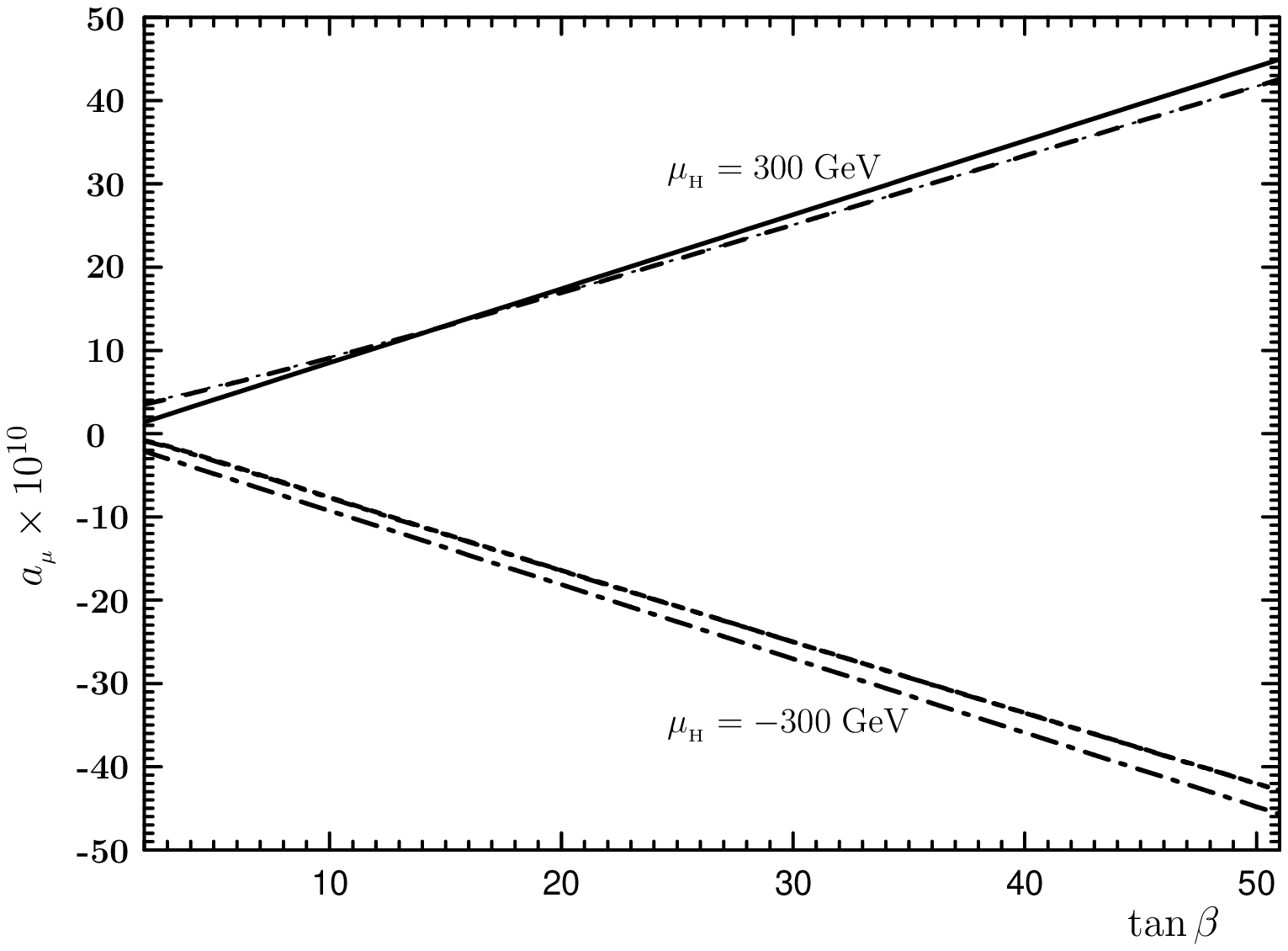}}
\end{picture}
\caption[]{The supersymmetric corrections to the MDM and EDM of muon
vary with $\tan\beta$ for $\mu_{_{\rm H}}=\pm300$ GeV in
CP conservation framework, Where the dash-dot line stands
for the results of one loop with $\mu_{_{\rm H}}=-300$ GeV, the dash-dot-dot line
stands for the results of two loop in
$\overline{MS}$ scheme with $\mu_{_{\rm H}}=-300$ GeV, the short dash line
stands for the results of two loop in mass shell scheme with $\mu_{_{\rm H}}=-300$ GeV,
the solid line stands for the results of one loop with $\mu_{_{\rm H}}=300$ GeV,
the dash line stands for the results of two loop in
$\overline{MS}$ scheme with $\mu_{_{\rm H}}=300$ GeV, and the dot line stands for the results of
two loop in mass shell scheme with $\mu_{_{\rm H}}=300$ GeV.}
\label{fig8}
\end{center}
\end{figure}

In the above analysis, we always suppose $\mu_{_{\rm H}}>0$. It is
well known that the sign of the one-loop contribution depends on
the relative sign of $\mu_{_{\rm H}}$ and $m_{_2}$. Assuming CP
conservation with $\theta_{_1}=\theta_{_2}
=\theta_{_{\tilde\nu}}=0$, we plot the muon MDM versus $\mu_{_{\rm
H}}$ as $\tan\beta=5$ or $\tan\beta=20$ in Fig. \ref{fig7}. The
plot shows that the correction from the two-loop diagrams can be
neglected safely when $|\mu_{_{\rm H}}|\le100$ GeV. Except the
parameter $\mu_{_{\rm H}}$, another parameter $\tan\beta$ plays an
important role in our analysis. Assuming CP conservation and
setting $\mu_{_{\rm H}}=\pm300$ GeV, we plot the theoretical
predictions on muon MDM versus $\tan\beta$ in Fig. \ref{fig8}.
Since the dominant contribution from the two-loop diagrams depends
on $\tan\beta$ weakly, the variation of two-loop correction is not
very obvious  with the increase of $\tan\beta$. In other words,
the correction from those two-loop diagrams turns insignificant in
large $\tan\beta$ case because the one-loop supersymmetric
prediction on muon MDM is proportional to $\tan\beta$.

All the numerical results are obtained under the assumption
$M_{_{\tilde\mu}}<1\;({\rm TeV})$. Since the
supersymmetric corrections to the muon MDM and EDM contain a
suppression factor $\Lambda_{_{\rm EW}}^2/ \Lambda_{_{\rm
SUSY}}^2$, the contributions from the considered diagrams
Fig .\ref{fig2} become insignificant
if the supersymmetric scale $\Lambda_{_{\rm SUSY}}\gg
\Lambda_{_{\rm EW}}$.

\section{Conclusions\label{sec5}}
\indent\indent In this work, we analyze some two-loop
supersymmetric corrections to the anomalous dipole moments of muon
by the effective Lagrangian method. In our calculation, we keep
all dimension 6 operators. We remove the ultra-violet divergence
caused by the divergent sub-diagrams  in the $\overline{MS}$ and
on-shell renormalization schemes respectively. After applying the
equation of motion for muon, we derive the muon MDM and EDM.
Numerically, we analyze the dependence of muon anomalous dipole
moments on  CP violation phases. There is an experimentally
allowed supersymmetric parameter space where those two-loop
corrections on the muon MDM are significant and cannot be
neglected, meanwhile the EDM of muon can be large enough to be
experimentally detected with the experimental precision of near
future.

\begin{acknowledgments}

The work is supported by the Academy of Finland under the contract
no.\ 104915 and 107293, the ABRL Grant No. R14-2003-012-01001-0 of
Korea, and also partly  by the National Natural Science Foundation
of China(NNSFC). The author (T.F.F.) also thanks Prof. Lee, C.-K.
for useful comments.
\end{acknowledgments}
\vspace{1.6cm}
\appendix

\section{The identities for two-loop integrals\label{ap1}}

\begin{eqnarray}
%%%%%%%%%%%%%%%%%%%%%%%%%%%%%%%%%%%%%%%%%%%%%%%%%%%%%%%%%%%%%%%%%%%%%%
&&\int{d^Dq_1\over(2\pi)^D}{d^Dq_2\over(2\pi)^D}{1
\over{\cal D}_{_0}}\bigg\{{q_1^2(q_1\cdot q_2)^2
-(q_1^2)^2q_1\cdot q_2\over (q_2-q_1)^2-m_{_0}^2}
+{q_1^2(q_1\cdot q_2)^2\over q_2^2-m_{_2}^2}
-{(q_1^2)^2\over2}
\bigg\}\equiv0\;,
\nonumber\\
&&\int{d^Dq_1\over(2\pi)^D}{d^Dq_2\over(2\pi)^D}{1\over{\cal D}_{_0}}
\bigg\{{(q_1^2)^2q_2^2-(q_1^2)^2q_1\cdot q_2\over(q_2-q_1)^2-m_{_0}^2}
+{(q_1^2)^2q_2^2\over q_2^2-m_{_2}^2}-{D(q_1^2)^2\over2}
\bigg\}\equiv0\;,
\nonumber\\
%%%%%%%%%%%%%%%%%%%%%%%%%%%%%%%%%%%%%%%%%%%%%%%%%%%%%%%%%%%%%%%%%%%%%%
&&\int{d^Dq_1\over(2\pi)^D}{d^Dq_2\over(2\pi)^D}{1\over{\cal D}_{_0}}
\bigg\{{q_1^2(q_1\cdot q_2)^2-(q_1^2)^2q_2^2\over(q_2-q_1)^2-m_{_0}^2}
+{D-1\over2}q_1^2q_1\cdot q_2\bigg\}\equiv0\;,
\nonumber\\
&&\int{d^Dq_1\over(2\pi)^D}{d^Dq_2\over(2\pi)^D}{1\over{\cal D}_{_0}}
\bigg\{{q_1^2q_1\cdot q_2q_2^2-q_1^2(q_1\cdot q_2)^2\over
(q_2-q_1)^2-m_{_0}^2}+{q_1^2q_1\cdot q_2q_2^2\over q_2^2-m_{_2}^2}
-{D+1\over2}q_1^2q_1\cdot q_2\bigg\}\equiv0\;,
\nonumber\\
&&\int{d^Dq_1\over(2\pi)^D}{d^Dq_2\over(2\pi)^D}{1\over{\cal D}_{_0}}
\bigg\{{(q_1\cdot q_2)^3-q_1^2(q_1\cdot q_2)^2\over(q_2-q_1)^2-m_{_0}^2}
+{(q_1\cdot q_2)^3\over q_2^2-m_{_2}^2}-q_1^2q_1\cdot q_2\bigg\}\equiv0
\;,\nonumber\\
&&\int{d^Dq_1\over(2\pi)^D}{d^Dq_2\over(2\pi)^D}{1\over{\cal D}_{_0}}
\bigg\{{q_1^2q_1\cdot q_2q_2^2-(q_1^2)^2q_2^2\over(q_2-q_1)^2-m_{_0}^2}
+{q_1^2q_1\cdot q_2q_2^2\over q_2^2-m_{_2}^2}
-q_1^2q_1\cdot q_2\bigg\}\equiv0\;,
\nonumber\\
&&\int{d^Dq_1\over(2\pi)^D}{d^Dq_2\over(2\pi)^D}{1\over{\cal D}_{_0}}
\bigg\{{(q_1\cdot q_2)^3-(q_1^2)^2q_2^2\over(q_2-q_1)^2-m_{_0}^2}
+{(q_1\cdot q_2)^3\over q_2^2-m_{_2}^2}
+{D-3\over2}q_1^2q_1\cdot q_2\bigg\}\equiv0\;,
\nonumber\\
%%%%%%%%%%%%%%%%%%%%%%%%%%%%%%%%%%%%%%%%%%%%%%%%%%%%%%%%%%%%%%%%%%%%%%%%%%%%%%%%%%%
&&\int{d^Dq_1\over(2\pi)^D}{d^Dq_2\over(2\pi)^D}{1\over{\cal D}_{_0}}
\bigg\{{q_1^2q_1\cdot q_2q_2^2-(q_1\cdot q_2)^3\over(q_2-q_1)^2-m_{_0}^2}
-{D(q_1\cdot q_2)^2-q_1^2q_2^2\over2}\bigg\}\equiv0
\;,\nonumber\\
&&\int{d^Dq_1\over(2\pi)^D}{d^Dq_2\over(2\pi)^D}{1\over{\cal D}_{_0}}
\bigg\{{(q_1\cdot q_2)^2q_2^2-q_1^2q_1\cdot q_2q_2^2\over(q_2-q_1)^2-m_{_0}^2}
+{(q_1\cdot q_2)^2q_2^2\over q_2^2-m_{_2}^2}
-{2(q_1\cdot q_2)^2+q_1^2q_2^2\over2}\bigg\}\equiv0
\;,\nonumber\\
&&\int{d^Dq_1\over(2\pi)^D}{d^Dq_2\over(2\pi)^D}{1\over{\cal D}_{_0}}
\bigg\{{q_1^2(q_2^2)^2-q_1^2q_1\cdot q_2q_2^2\over(q_2-q_1)^2-m_{_0}^2}
+{q_1^2(q_2^2)^2\over q_2^2-m_{_2}^2}-{D+2\over2}q_1^2q_2^2\bigg\}\equiv0
\;,\nonumber\\
&&\int{d^Dq_1\over(2\pi)^D}{d^Dq_2\over(2\pi)^D}{1\over{\cal D}_{_0}}
\bigg\{{(q_1\cdot q_2)^2q_2^2-(q_1\cdot q_2)^3\over(q_2-q_1)^2-m_{_0}^2}
+{(q_1\cdot q_2)^2q_2^2\over q_2^2-m_{_2}^2}
-{D+2\over2}(q_1\cdot q_2)^2\bigg\}\equiv0
\;,\nonumber\\
&&\int{d^Dq_1\over(2\pi)^D}{d^Dq_2\over(2\pi)^D}{1\over{\cal D}_{_0}}
\bigg\{{q_1^2(q_2^2)^2-(q_1\cdot q_2)^3\over(q_2-q_1)^2-m_{_0}^2}
+{q_1^2(q_2^2)^2\over q_2^2-m_{_2}^2}
-{D(q_1\cdot q_2)^2+(D+1)q_1^2q_2^2\over2}\bigg\}\equiv0
\;,\nonumber\\
%%%%%%%%%%%%%%%%%%%%%%%%%%%%%%%%%%%%%%%%%%%%%%%%%%%%%%%%%%%%%%%%%%%%%%%%%%%%%%%%%%%
&&\int{d^Dq_1\over(2\pi)^D}{d^Dq_2\over(2\pi)^D}
{1\over{\cal D}_{_0}}\bigg\{{q_1\cdot q_2(q_2^2)^2-(q_1\cdot q_2)^2q_2^2
\over (q_2-q_1)^2-m_{_0}^2}
+{q_1\cdot q_2(q_2^2)^2\over q_2^2-m_{_2}^2}
-{D+3\over2}q_1\cdot q_2q_2^2\bigg\}\equiv0
\;,\nonumber\\
&&\int{d^Dq_1\over(2\pi)^D}{d^Dq_2\over(2\pi)^D}
{1\over{\cal D}_{_0}}\bigg\{{q_1\cdot q_2(q_2^2)^2-q_1^2(q_2^2)^2
\over (q_2-q_1)^2-m_{_0}^2}
+{q_1\cdot q_2(q_2^2)^2\over q_2^2-m_{_2}^2}
-2q_1\cdot q_2q_2^2\bigg\}\equiv0
\;,\nonumber\\
%%%%%%%%%%%%%%%%%%%%%%%%%%%%%%%%%%%%%%%%%%%%%%%%%%%%%%%%%%%%%%%%%%%%%%%%%%%%%%%%%%%
&&\int{d^Dq_1\over(2\pi)^D}{d^Dq_2\over(2\pi)^D}{1\over{\cal D}_{_0}}
\bigg\{{q_1^2(q_1\cdot q_2)^2-(q_1^2)^2q_2^2\over q_2^2-m_{_2}^2}
+{D-1\over2}\Big[(q_1^2)^2-q_1^2q_1\cdot q_2\Big]\bigg\}\equiv0
\;,\nonumber\\
&&\int{d^Dq_1\over(2\pi)^D}{d^Dq_2\over(2\pi)^D}{1\over{\cal D}_{_0}}
\bigg\{{(q_1^2)^2q_1\cdot q_2\over q_1^2-m_{_1}^2}
+{q_1^2(q_1\cdot q_2)^2\over q_2^2-m_{_2}^2}
-{D+3\over2}q_1^2q_1\cdot q_2-{(q_1^2)^2\over2}\bigg\}\equiv0\;,
\nonumber\\
&&\int{d^Dq_1\over(2\pi)^D}{d^Dq_2\over(2\pi)^D}{1\over{\cal D}_{_0}}
\bigg\{{(q_1^2)^2q_1\cdot q_2\over q_1^2-m_{_1}^2}
+{(q_1^2)^2q_2^2\over q_2^2-m_{_2}^2}
-2q_1^2q_1\cdot q_2-{D\over2}(q_1^2)^2\bigg\}\equiv0
\;,\nonumber\\
%%%%%%%%%%%%%%%%%%%%%%%%%%%%%%%%%%%%%%%%%%%%%%%%%%%%%%%%%%%%%%%%%%%%%%%%%%%%%%%%%%%
&&\int{d^Dq_1\over(2\pi)^D}{d^Dq_2\over(2\pi)^D}{1\over{\cal D}_{_0}}
\bigg\{{q_1^2q_1\cdot q_2q_2^2-(q_1\cdot q_2)^3\over q_2^2-m_{_2}^2}
-{D-1\over2}q_1^2q_1\cdot q_2+{D(q_1\cdot q_2)^2
-q_1^2q_2^2\over2}\bigg\}\equiv0
\;,\nonumber\\
&&\int{d^Dq_1\over(2\pi)^D}{d^Dq_2\over(2\pi)^D}{1\over{\cal D}_{_0}}
\bigg\{{q_1^2(q_1\cdot q_2)^2\over q_1^2-m_{_1}^2}
+{q_1^2q_1\cdot q_2q_2^2\over q_2^2-m_{_2}^2}
-{D+1\over2}q_1^2q_1\cdot q_2
\nonumber\\
&&\hspace{0.4cm}-(q_1\cdot q_2)^2-{q_1^2q_2^2\over2}\bigg\}\equiv0
\;,\nonumber\\
&&\int{d^Dq_1\over(2\pi)^D}{d^Dq_2\over(2\pi)^D}{1\over{\cal D}_{_0}}
\bigg\{{(q_1^2)^2q_2^2\over q_1^2-m_{_1}^2}
+{q_1^2q_1\cdot q_2q_2^2\over q_2^2-m_{_2}^2}
-q_1^2q_1\cdot q_2-{D+2\over2}q_1^2q_2^2
\bigg\}\equiv0
\;,\nonumber\\
&&\int{d^Dq_1\over(2\pi)^D}{d^Dq_2\over(2\pi)^D}{1\over{\cal D}_{_0}}
\bigg\{{q_1^2(q_1\cdot q_2)^2\over q_1^2-m_{_1}^2}
+{(q_1\cdot q_2)^3\over q_2^2-m_{_2}^2}
-q_1^2q_1\cdot q_2-{D+2\over2}(q_1\cdot q_2)^2\bigg\}\equiv0
\;,\nonumber\\
&&\int{d^Dq_1\over(2\pi)^D}{d^Dq_2\over(2\pi)^D}{1\over{\cal D}_{_0}}
\bigg\{{(q_1^2)^2q_2^2\over q_1^2-m_{_1}^2}
+{(q_1\cdot q_2)^3\over q_2^2-m_{_2}^2}
+{D-3\over2}q_1^2q_1\cdot q_2
\nonumber\\
&&\hspace{0.4cm}-{D\over2}(q_1\cdot q_2)^2-{D+1\over2}q_1^2q_2^2\bigg\}\equiv0
\;,\nonumber\\
%%%%%%%%%%%%%%%%%%%%%%%%%%%%%%%%%%%%%%%%%%%%%%%%%%%%%%%%%%%%%%%%%%%%%%%%%%%%%%%%%%%
&&\int{d^Dq_1\over(2\pi)^D}{d^Dq_2\over(2\pi)^D}{1\over{\cal D}_{_0}}
\bigg\{{q_1^2q_1\cdot q_2-(q_1^2)^2\over(q_2-q_1)^2-m_{_0}^2}
+{q_1^2q_1\cdot q_2\over q_2^2-m_{_2}^2}\bigg\}\equiv0
\;,\nonumber\\
&&\int{d^Dq_1\over(2\pi)^D}{d^Dq_2\over(2\pi)^D}{1\over{\cal D}_{_0}}
\bigg\{{q_1^2q_2^2-q_1^2q_1\cdot q_2\over(q_2-q_1)^2-m_{_0}^2}
+{q_1^2q_2^2\over q_2^2-m_{_2}^2}-{D\over2}q_1^2\bigg\}\equiv0
\;,\nonumber\\
&&\int{d^Dq_1\over(2\pi)^D}{d^Dq_2\over(2\pi)^D}{1\over{\cal D}_{_0}}
\bigg\{{(q_1\cdot q_2)^2-q_1^2q_1\cdot q_2\over(q_2-q_1)^2-m_{_0}^2}
+{(q_1\cdot q_2)^2\over q_2^2-m_{_2}^2}-{q_1^2\over2}\bigg\}\equiv0
\;,\nonumber\\
&&\int{d^Dq_1\over(2\pi)^D}{d^Dq_2\over(2\pi)^D}{1\over{\cal D}_{_0}}
\bigg\{{q_1\cdot q_2q_2^2-(q_1\cdot q_2)^2\over(q_2-q_1)^2-m_{_0}^2}
+{q_1\cdot q_2q_2^2\over q_2^2-m_{_2}^2}-{D+1\over2}q_1\cdot q_2\bigg\}\equiv0
\;,\nonumber\\
&&\int{d^Dq_1\over(2\pi)^D}{d^Dq_2\over(2\pi)^D}{1\over{\cal D}_{_0}}
\bigg\{{q_1\cdot q_2q_2^2-q_1^2q_2^2\over(q_2-q_1)^2-m_{_0}^2}
+{q_1\cdot q_2q_2^2\over q_2^2-m_{_2}^2}-q_1\cdot q_2\bigg\}\equiv0
\;,\nonumber\\
&&\int{d^Dq_1\over(2\pi)^D}{d^Dq_2\over(2\pi)^D}{1\over{\cal D}_{_0}}
\bigg\{{(q_2^2)^2-q_1\cdot q_2q_2^2\over(q_2-q_1)^2-m_{_0}^2}
+{(q_2^2)^2\over q_2^2-m_{_2}^2}-{D+2\over2}q_2^2\bigg\}\equiv0
\;,\nonumber\\
%%%%%%%%%%%%%%%%%%%%%%%%%%%%%%%%%%%%%%%%%%%%%%%%%%%%%%%%%%%%%%%%%%%%%%%%%%%%%%%%%%%
&&\int{d^Dq_1\over(2\pi)^D}{d^Dq_2\over(2\pi)^D}{1\over{\cal D}_{_0}}
\bigg\{{(q_1^2)^2\over q_1^2-m_{_1}^2}
+{q_1^2q_1\cdot q_2\over q_2^2-m_{_2}^2}-{D+2\over2}q_1^2\bigg\}\equiv0
\;,\nonumber\\
&&\int{d^Dq_1\over(2\pi)^D}{d^Dq_2\over(2\pi)^D}{1\over{\cal D}_{_0}}
\bigg\{{q_1^2q_1\cdot q_2\over q_1^2-m_{_1}^2}
+{(q_1\cdot q_2)^2\over q_2^2-m_{_2}^2}-{D+1\over2}q_1\cdot q_2
-{q_1^2\over2}\bigg\}\equiv0
\;,\nonumber\\
&&\int{d^Dq_1\over(2\pi)^D}{d^Dq_2\over(2\pi)^D}{1\over{\cal D}_{_0}}
\bigg\{{q_1^2q_1\cdot q_2\over q_1^2-m_{_1}^2}
+{q_1^2q_2^2\over q_2^2-m_{_2}^2}-q_1\cdot q_2
-{D\over2}q_1^2\bigg\}\equiv0
\;,\nonumber\\
%%%%%%%%%%%%%%%%%%%%%%%%%%%%%%%%%%%%%%%%%%%%%%%%%%%%%%%%%%%%%%%%%%%%%%%%%%%%%%%%%%%
&&\int{d^Dq_1\over(2\pi)^D}{d^Dq_2\over(2\pi)^D}{1\over{\cal D}_{_0}}
\bigg\{{q_1\cdot q_2-q_1^2\over(q_2-q_1)^2-m_{_0}^2}
+{q_1\cdot q_2\over q_2^2-m_{_2}^2}\bigg\}\equiv0
\;,\nonumber\\
&&\int{d^Dq_1\over(2\pi)^D}{d^Dq_2\over(2\pi)^D}{1\over{\cal D}_{_0}}
\bigg\{{q_2^2-q_1\cdot q_2\over(q_2-q_1)^2-m_{_0}^2}
+{q_2^2\over q_2^2-m_{_2}^2}-{D\over2}\bigg\}\equiv0
\;,\nonumber\\
%%%%%%%%%%%%%%%%%%%%%%%%%%%%%%%%%%%%%%%%%%%%%%%%%%%%%%%%%%%%%%%%%%%%%%%%%%%%%%%%%%%
&&\int{d^Dq_1\over(2\pi)^D}{d^Dq_2\over(2\pi)^D}{1\over{\cal D}_{_0}}
\bigg\{{q_1^2\over q_1^2-m_{_1}^2}
+{q_1\cdot q_2\over q_2^2-m_{_2}^2}-{D\over2}\bigg\}\equiv0
\;,\nonumber\\
&&\int{d^Dq_1\over(2\pi)^D}{d^Dq_2\over(2\pi)^D}{1\over{\cal D}_{_0}}
\bigg\{{q_1\cdot q_2\over q_1^2-m_{_1}^2}
+{q_2^2\over q_2^2-m_{_2}^2}-{D\over2}\bigg\}\equiv0
\;.\nonumber\\
%%%%%%%%%%%%%%%%%%%%%%%%%%%%%%%%%%%%%%%%%%%%%%%%%%%%%%%%%%%%%%%%%%%%%%%%%%%%%%%%%%%
\label{iden2}
\end{eqnarray}
with ${\cal D}_{_0}=((q_2-q_1)^2-m_{_0}^2)(q_1^2-m_{_1}^2)(q_2^2-m_{_2}^2)$.

\section{The form factors\label{ap2}}
\begin{eqnarray}
%%%%%%%%%%%%%%%%%%%%%%%%%%%%%%%%%%%%%%%%%%%%%%%%%%%%%%%%%%%%%%%%%%%%%%
&&\Big({\cal N}_{_{\chi_\alpha^0\chi_\beta^0}}^a\Big)_1=
-{24\over D(D+2)}{(q_1^2)^2q_1\cdot q_2-(q_1^2)^2q_2^2
\over(q_1^2-m_{_{\tilde{E}_j}}^2)^3}
+{4\over D}{(q_1^2)^2-q_1^2q_2^2\over(q_1^2-m_{_{\tilde{E}_j}}^2)^2}
\nonumber\\
&&\hspace{1.6cm}
+{8\over D(D+2)}{3q_1^2q_1\cdot q_2q_2^2-2q_1^2(q_1\cdot q_2)^2-(q_1^2)^2q_2^2
\over(q_1^2-m_{_{\tilde{E}_j}}^2)^2(q_2^2-m_{_{\chi_\alpha^0}}^2)}
\nonumber\\
&&\hspace{1.6cm}
+{8\over D(D+2)}{2(q_1\cdot q_2)^2q_2^2-3q_1^2q_1\cdot q_2q_2^2+q_1^2(q_2^2)^2
\over(q_1^2-m_{_{\tilde{E}_j}}^2)(q_2^2-m_{_{\chi_\alpha^0}}^2)^2}
\nonumber\\
&&\hspace{1.6cm}
+{2\over D}{3q_1^2q_1\cdot q_2-2(q_1\cdot q_2)^2-q_1\cdot q_2q_2^2
\over(q_1^2-m_{_{\tilde{E}_j}}^2)(q_2^2-m_{_{\chi_\alpha^0}}^2)}
\nonumber\\
&&\hspace{1.6cm}
-{q_1^2-q_1\cdot q_2\over q_1^2-m_{_{\tilde{E}_j}}^2}
+{24\over D(D+2)}{q_1\cdot q_2(q_2^2)^2-q_1^2(q_2^2)^2
\over(q_2^2-m_{_{\chi_\alpha^0}}^2)^3}
\nonumber\\
&&\hspace{1.6cm}
+{8\over D}{q_1^2q_2^2-q_1\cdot q_2q_2^2\over(q_2^2-m_{_{\chi_\alpha^0}}^2)^2}
-{q_1^2-q_1\cdot q_2\over q_2^2-m_{_{\chi_\alpha^0}}^2}
\;,\nonumber\\
%%%%%%%%%%%%%%%%%%%%%%%%%%%%%%%%%%%%%%%%%%%%%%%%%%%%%%%%%%%%%%%%%%%%%%
&&\Big({\cal N}_{_{\chi_\alpha^0\chi_\beta^0}}^a\Big)_2=
{6\over D(D+2)}{(q_1^2)^2q_1\cdot q_2-(q_1^2)^2q_2^2\over(q_1^2-m_{_{\tilde{E}_j}}^2)^3}
+{1\over D}{q_1^2q_2^2-q_1^2q_1\cdot q_2\over(q_1^2-m_{_{\tilde{E}_j}}^2)^2}
\nonumber\\
&&\hspace{1.6cm}
+{6\over(q_1^2-m_{_{\tilde{E}_j}}^2)^2(q_2^2-m_{_{\chi_\alpha^0}}^2)}
\Big[{(q_1^2)^2q_2^2-q_1^2q_1\cdot q_2q_2^2\over D(D+2)}
\nonumber\\
&&\hspace{1.6cm}
+{q_1^2(q_1\cdot q_2)^2-(q_1^2)^2q_2^2\over (D-1)(D+2)}\Big]
-{2\over(q_1^2-m_{_{\tilde{E}_j}}^2)(q_2^2-m_{_{\chi_\alpha^0}}^2)^2}
\nonumber\\
&&\hspace{1.6cm}\times
\Big[3\cdot{q_1^2(q_2^2)^2-q_1^2q_1\cdot q_2q_2^2\over D(D+2)}
+(D-4)\cdot{(q_1\cdot q_2)^2q_2^2-q_1^2(q_2^2)^2\over D(D-1)(D+2)}\Big]
\nonumber\\
&&\hspace{1.6cm}
-{1\over(q_1^2-m_{_{\tilde{E}_j}}^2)(q_2^2-m_{_{\chi_\alpha^0}}^2)}
\Big[3\cdot{q_1^2q_1\cdot q_2\over 2D}
-{q_1^2q_2^2\over D}-(D-4)\cdot{(q_1\cdot q_2)^2-q_1^2q_2^2\over D(D-1)}\Big]
\nonumber\\
&&\hspace{1.6cm}
+{1\over2D}{q_1^2q_1\cdot q_2-q_1^2q_2^2\over(q_1^2-m_{_{\tilde{E}_j}}^2)
(q_2^2-m_{_{\tilde{E}_i}}^2)}
+{D-2\over2D}{q_1^2\over q_1^2-m_{_{\tilde{E}_j}}^2}
-{1\over D}{2q_1^2q_2^2-q_1\cdot q_2q_2^2\over(q_2^2-m_{_{\chi_\alpha^0}}^2)^2}
\nonumber\\
&&\hspace{1.6cm}
-{6\over D(D+2)}{q_1\cdot q_2(q_2^2)^2-q_1^2(q_2^2)^2
\over(q_2^2-m_{_{\chi_\alpha^0}}^2)^3}
-{1\over2D}{q_1\cdot q_2q_2^2\over(q_2^2-m_{_{\chi_\alpha^0}}^2)
(q_2^2-m_{_{\tilde{E}_i}}^2)}
\nonumber\\
&&\hspace{1.6cm}
+{1\over4D}{Dq_1^2+2q_1\cdot q_2\over q_2^2-m_{_{\chi_\alpha^0}}^2}
\;,\nonumber\\
%%%%%%%%%%%%%%%%%%%%%%%%%%%%%%%%%%%%%%%%%%%%%%%%%%%%%%%%%%%%%%%%%%%%%%
&&\Big({\cal N}_{_{\chi_\alpha^0\chi_\beta^0}}^a\Big)_3=
{6\over D(D+2)}{(q_1^2)^2q_1\cdot q_2-(q_1^2)^2q_2^2\over(q_1^2-m_{_{\tilde{E}_j}}^2)^3}
-{1\over D}{2(q_1^2)^2-q_1^2q_1\cdot q_2-q_1^2q_2^2\over(q_1^2-m_{_{\tilde{E}_j}}^2)^2}
\nonumber\\
&&\hspace{1.6cm}
+{2\over(q_1^2-m_{_{\tilde{E}_j}}^2)^2(q_2^2-m_{_{\chi_\alpha^0}}^2)}
\Big[3\cdot{(q_1^2)^2q_2^2-q_1^2q_1\cdot q_2q_2^2\over D(D+2)}
\nonumber\\
&&\hspace{1.6cm}
+(D-4)\cdot{q_1^2(q_1\cdot q_2)^2-(q_1^2)^2q_2^2\over D(D-1)(D+2)}\Big]
\nonumber\\
&&\hspace{1.6cm}
-{6\over D(D+2)}{q_1^2(q_2^2)^2-q_1^2q_1\cdot q_2q_2^2\over(q_1^2-m_{_{\tilde{E}_j}}^2)
(q_2^2-m_{_{\chi_\alpha^0}}^2)^2}
-{1\over2D}{3q_1^2q_1\cdot q_2-2q_1\cdot q_2q_2^2\over(q_1^2-m_{_{\tilde{E}_j}}^2)
(q_2^2-m_{_{\chi_\alpha^0}}^2)}
\nonumber\\
&&\hspace{1.6cm}
-{1\over(q_1^2-m_{_{\tilde{E}_j}}^2)(q_2^2-m_{_{\tilde{E}_i}}^2)}
\Big[{q_1^2q_1\cdot q_2\over 2D}+{(q_1\cdot q_2)^2-q_1^2q_2^2\over D-1}\Big]
\nonumber\\
&&\hspace{1.6cm}
-{6\over D(D+2)}{q_1\cdot q_2(q_2^2)^2-q_1^2(q_2^2)^2\over(q_2^2-m_{_{\chi_\alpha^0}}^2)^3}
-{1\over D}{2q_1^2q_2^2-q_1\cdot q_2q_2^2\over(q_2^2-m_{_{\chi_\alpha^0}}^2)^2}
\nonumber\\
&&\hspace{1.6cm}
-{1\over2D}{q_1\cdot q_2q_2^2\over(q_2^2-m_{_{\chi_\alpha^0}}^2)(q_2^2-m_{_{\tilde{E}_i}}^2)}
+{q_1^2+2q_1\cdot q_2\over4(q_2^2-m_{_{\chi_\alpha^0}}^2)}
+{D-2\over2D}{q_1\cdot q_2\over q_2^2-m_{_{\tilde{E}_i}}^2}
\;,\nonumber\\
%%%%%%%%%%%%%%%%%%%%%%%%%%%%%%%%%%%%%%%%%%%%%%%%%%%%%%%%%%%%%%%%%%%%%%
&&\Big({\cal N}_{_{\chi_\alpha^0\chi_\beta^0}}^a\Big)_4=
{4\over D(D+2)}{(q_1^2)^2q_1\cdot q_2-(q_1^2)^2q_2^2\over(q_1^2-m_{_{\tilde{E}_j}}^2)^3}
-{4\over D(D+2)}{q_1\cdot q_2(q_2^2)^2-q_1^2(q_2^2)^2\over(q_2^2-m_{_{\chi_\alpha^0}}^2)^3}
\nonumber\\
&&\hspace{1.6cm}
+{4\over(q_1^2-m_{_{\tilde{E}_j}}^2)^2(q_2^2-m_{_{\chi_\alpha^0}}^2)}
\Big[{(q_1^2)^2q_2^2-q_1^2q_1\cdot q_2q_2^2\over D(D+2)}
+{q_1^2(q_1\cdot q_2)^2-(q_1^2)^2q_2^2\over (D-1)(D+2)}\Big]
\nonumber\\
&&\hspace{1.6cm}
-{4\over(q_1^2-m_{_{\tilde{E}_j}}^2)(q_2^2-m_{_{\chi_\alpha^0}}^2)^2}
\Big[{q_1^2(q_2^2)^2-q_1^2q_1\cdot q_2q_2^2\over D(D+2)}
+{(q_1\cdot q_2)^2q_2^2-q_1^2(q_2^2)^2\over (D-1)(D+2)}\Big]
\nonumber\\
&&\hspace{1.6cm}
-{1\over(q_1^2-m_{_{\tilde{E}_j}}^2)(q_2^2-m_{_{\chi_\alpha^0}}^2)}
\Big[{q_1^2q_1\cdot q_2-2(q_1\cdot q_2)^2\over D}
-2\cdot{(q_1\cdot q_2)^2-q_1^2q_2^2\over D(D-1)}\Big]
\nonumber\\
&&\hspace{1.6cm}
+{1\over D}{q_1^2q_1\cdot q_2\over(q_1^2-m_{_{\tilde{E}_j}}^2)(q_2^2-m_{_{\tilde{E}_i}}^2)}
-{1\over D(D+2)}{q_1^2-q_1\cdot q_2\over q_1^2-m_{_{\tilde{E}_j}}^2}
-{1\over3}{q_1^2-q_1\cdot q_2\over q_2^2-m_{_{\tilde{E}_i}}^2}
\nonumber\\
&&\hspace{1.6cm}
-{2\over 3D}{6(q_1\cdot q_2)^2+8q_1^2q_2^2
-11q_1\cdot q_2q_2^2\over(q_2^2-m_{_{\chi_\alpha^0}}^2)^2}
+{1\over3D}{2q_1^2q_2^2+q_1\cdot q_2q_2^2\over(q_2^2-m_{_{\chi_\alpha^0}}^2)
(q_2^2-m_{_{\tilde{E}_i}}^2)}
\nonumber\\
&&\hspace{1.6cm}
+{1\over q_2^2-m_{_{\chi_\alpha^0}}^2}\Big({q_1^2\over6}
-{2D+3\over3D}q_1\cdot q_2\Big)
+{D-2\over 3D}{q_1\cdot q_2\over(q_2-q_1)^2-m_{_{l^J}}^2}
\;,\nonumber\\
%%%%%%%%%%%%%%%%%%%%%%%%%%%%%%%%%%%%%%%%%%%%%%%%%%%%%%%%%%%%%%%%%%%%%%
%%%%%%%%%%%%%%%%%%%%%%%%%%%%%%%%%%%%%%%%%%%%%%%%%%%%%%%%%%%%%%%%%%%%%%
&&\Big({\cal N}_{_{\chi_\alpha^0\chi_\beta^0}}^b\Big)_1=
{24\over D(D+2)}{(q_1^2)^2-q_1^2q_1\cdot q_2\over(q_1^2-m_{_{\tilde{E}_j}}^2)^3}
+{24\over D(D+2)}{q_1\cdot q_2q_2^2
-(q_2^2)^2\over(q_2^2-m_{_{\chi_\alpha^0}}^2)^3}
\nonumber\\
&&\hspace{1.6cm}
+{8\over D(D+2)}{3q_1^2q_1\cdot q_2-4(q_1\cdot q_2)^2+q_1^2q_2^2
\over(q_1^2-m_{_{\tilde{E}_j}}^2)^2(q_2^2-m_{_{\chi_\alpha^0}}^2)}
\nonumber\\
&&\hspace{1.6cm}
-{4\over D}{q_1^2-q_1\cdot q_2\over(q_1^2-m_{_{\tilde{E}_j}}^2)^2}
+{8\over D(D+2)}{5(q_1\cdot q_2)^2-2q_1^2q_2^2-3q_1\cdot q_2q_2^2
\over(q_1^2-m_{_{\tilde{E}_j}}^2)(q_2^2-m_{_{\chi_\alpha^0}}^2)^2}
\nonumber\\
&&\hspace{1.6cm}
+{2\over D}{q_2^2-q_1^2\over(q_1^2-m_{_{\tilde{E}_j}}^2)(q_2^2-m_{_{\chi_\alpha^0}}^2)}
-{4\over D}{q_1\cdot q_2-q_2^2\over(q_2^2-m_{_{\chi_\alpha^0}}^2)^2}
\;,\nonumber\\
%%%%%%%%%%%%%%%%%%%%%%%%%%%%%%%%%%%%%%%%%%%%%%%%%%%%%%%%%%%%%%%%%%%%%%
&&\Big({\cal N}_{_{\chi_\alpha^0\chi_\beta^0}}^b\Big)_2=
-{6\over D(D+2)}{(q_1^2)^2-q_1^2q_1\cdot q_2\over(q_1^2-m_{_{\tilde{E}_j}}^2)^3}
+{1\over D}{q_1^2-q_1\cdot q_2\over(q_1^2-m_{_{\tilde{E}_j}}^2)^2}
\nonumber\\
&&\hspace{1.6cm}
-{6\over(q_1^2-m_{_{\tilde{E}_j}}^2)^2(q_2^2-m_{_{\chi_\alpha^0}}^2)}
\Big[{q_1^2q_1\cdot q_2-q_1^2q_2^2\over D(D+2)}
-{(q_1\cdot q_2)^2-q_1^2q_2^2\over (D-1)(D+2)}\Big]
\nonumber\\
&&\hspace{1.6cm}
-{2\over(q_1^2-m_{_{\tilde{E}_j}}^2)(q_2^2-m_{_{\chi_\alpha^0}}^2)^2}
\Big[3\cdot{(q_1\cdot q_2)^2-q_1\cdot q_2q_2^2\over D(D+2)}
\nonumber\\
&&\hspace{1.6cm}
-(2D+1)\cdot{(q_1\cdot q_2)^2-q_1^2q_2^2\over D(D-1)(D+2)}\Big]
+{1\over2D}{q_1\cdot q_2\over(q_1^2-m_{_{\tilde{E}_j}}^2)(q_2^2-m_{_{\tilde{E}_i}}^2)}
\nonumber\\
&&\hspace{1.6cm}
+{1\over2D}{2q_1^2-q_1\cdot q_2\over(q_1^2-m_{_{\tilde{E}_j}}^2)
(q_2^2-m_{_{\chi_\alpha^0}}^2)}
-{6\over D(D+2)}{q_1\cdot q_2q_2^2-(q_2^2)^2\over(q_2^2-m_{_{\chi_\alpha^0}}^2)^3}
\nonumber\\
&&\hspace{1.6cm}
+{1\over D}{q_1\cdot q_2\over(q_2^2-m_{_{\chi_\alpha^0}}^2)^2}
+{1\over2D}{q_1\cdot q_2\over(q_2^2-m_{_{\chi_\alpha^0}}^2)
(q_2^2-m_{_{\tilde{E}_i}}^2)}-{1\over4(q_2^2-m_{_{\chi_\alpha^0}}^2)}
\;,\nonumber\\
%%%%%%%%%%%%%%%%%%%%%%%%%%%%%%%%%%%%%%%%%%%%%%%%%%%%%%%%%%%%%%%%%%%%%%
&&\Big({\cal N}_{_{\chi_\alpha^0\chi_\beta^0}}^b\Big)_3=
-{6\over D(D+2)}{(q_1^2)^2-q_1^2q_1\cdot q_2\over(q_1^2-m_{_{\tilde{E}_j}}^2)^3}
+{1\over D}{q_1^2-q_1\cdot q_2\over(q_1^2-m_{_{\tilde{E}_j}}^2)^2}
\nonumber\\
&&\hspace{1.6cm}
-{6\over(q_1^2-m_{_{\tilde{E}_j}}^2)(q_2^2-m_{_{\chi_\alpha^0}}^2)^2}
\Big[{q_1^2q_2^2-q_1\cdot q_2q_2^2\over D(D+2)}
+{(q_1\cdot q_2)^2-q_1^2q_2^2\over (D-1)(D+2)}\Big]
\nonumber\\
&&\hspace{1.6cm}
-{2\over(q_1^2-m_{_{\tilde{E}_j}}^2)^2(q_2^2-m_{_{\chi_\alpha^0}}^2)}
\Big[3\cdot{q_1^2q_1\cdot q_2-q_1^2q_2^2\over D(D+2)}
\nonumber\\
&&\hspace{1.6cm}
-(D-4)\cdot{(q_1\cdot q_2)^2-q_1^2q_2^2\over D(D-1)(D+2)}\Big]
+{1\over2D}{3q_1\cdot q_2-2q_2^2\over(q_1^2-m_{_{\tilde{E}_j}}^2)
(q_2^2-m_{_{\chi_\alpha^0}}^2)}
\nonumber\\
&&\hspace{1.6cm}
+{1\over2D}{q_1\cdot q_2\over(q_1^2-m_{_{\tilde{E}_j}}^2)
(q_2^2-m_{_{\tilde{E}_i}}^2)}
-{6\over D(D+2)}{q_1\cdot q_2q_2^2-(q_2^2)^2\over(q_2^2-m_{_{\chi_\alpha^0}}^2)^3}
\nonumber\\
&&\hspace{1.6cm}
+{1\over D}{q_1\cdot q_2\over(q_2^2-m_{_{\chi_\alpha^0}}^2)^2}
+{1\over2D}{2q_2^2-q_1\cdot q_2\over(q_2^2-m_{_{\chi_\alpha^0}}^2)
(q_2^2-m_{_{\tilde{E}_i}}^2)}
-{1\over4(q_2^2-m_{_{\chi_\alpha^0}}^2)}
\;,\nonumber\\
%%%%%%%%%%%%%%%%%%%%%%%%%%%%%%%%%%%%%%%%%%%%%%%%%%%%%%%%%%%%%%%%%%%%%%
&&\Big({\cal N}_{_{\chi_\alpha^0\chi_\beta^0}}^b\Big)_4=
-{4\over D(D+2)}{(q_1^2)^2-q_1^2q_1\cdot q_2\over(q_1^2-m_{_{\tilde{E}_j}}^2)^3}
-{1\over D}{q_1\cdot q_2\over(q_1^2-m_{_{\tilde{E}_j}}^2)(q_2^2-m_{_{\chi_\alpha^0}}^2)}
\nonumber\\
&&\hspace{1.6cm}
-{4\over(q_1^2-m_{_{\tilde{E}_j}}^2)^2(q_2^2-m_{_{\chi_\alpha^0}}^2)}
\Big[{q_1^2q_1\cdot q_2-q_1^2q_2^2\over D(D+2)}
-{(q_1\cdot q_2)^2-q_1^2q_2^2\over (D-1)(D+2)}\Big]
\nonumber\\
&&\hspace{1.6cm}
-{4\over(q_1^2-m_{_{\tilde{E}_j}}^2)(q_2^2-m_{_{\chi_\alpha^0}}^2)^2}
\Big[{q_1^2q_2^2-q_1\cdot q_2q_2^2\over D(D+2)}
+{(q_1\cdot q_2)^2-q_1^2q_2^2\over (D-1)(D+2)}\Big]
\nonumber\\
&&\hspace{1.6cm}
-{1\over D}{q_1\cdot q_2\over(q_1^2-m_{_{\tilde{E}_j}}^2)(q_2^2-m_{_{\tilde{E}_i}}^2)}
-{4\over D(D+2)}{q_1\cdot q_2q_2^2-(q_2^2)^2\over(q_2^2-m_{_{\chi_\alpha^0}}^2)^3}
\nonumber\\
&&\hspace{1.6cm}
+{2\over3D}{q_1\cdot q_2-6q_2^2\over(q_2^2-m_{_{\chi_\alpha^0}}^2)^2}
-{1\over3D}{q_1\cdot q_2+6q_2^2\over(q_2^2-m_{_{\chi_\alpha^0}}^2)
(q_2^2-m_{_{\tilde{E}_i}}^2)}
+{2\over3D}{q_1\cdot q_2-3q_2^2\over(q_2^2-m_{_{\tilde{E}_i}}^2)^2}
\nonumber\\
&&\hspace{1.6cm}
+{1\over6}\Big[{2\over(q_2-q_1)^2-m_{_{l^I}}^2}
+{5\over q_2^2-m_{_{\chi_\alpha^0}}^2}+{2\over q_2^2-m_{_{\tilde{E}_i}}^2}\Big]
\;,\nonumber\\
%%%%%%%%%%%%%%%%%%%%%%%%%%%%%%%%%%%%%%%%%%%%%%%%%%%%%%%%%%%%%%%%%%%%%%
%%%%%%%%%%%%%%%%%%%%%%%%%%%%%%%%%%%%%%%%%%%%%%%%%%%%%%%%%%%%%%%%%%%%%%
&&\Big({\cal N}_{_{\chi_\alpha^0\chi_\beta^0}}^c\Big)_5=
{4\over D}{(q_1^2)^2-q_1^2q_1\cdot q_2\over(q_1^2-m_{_{\tilde{E}_j}}^2)^2}
+{4\over D}{q_1^2q_1\cdot q_2-q_1\cdot q_2q_2^2\over(q_1^2-m_{_{\tilde{E}_j}}^2)
(q_2^2-m_{_{\chi_\alpha^0}}^2)}
\nonumber\\
&&\hspace{1.6cm}
+{4\over D}{q_1^2q_2^2-q_1\cdot q_2q_2^2\over(q_2^2-m_{_{\chi_\alpha^0}}^2)^2}
-{q_1^2-q_1\cdot q_2\over q_1^2-m_{_{\tilde{E}_j}}^2}
-{q_1^2-q_1\cdot q_2\over q_2^2-m_{_{\chi_\alpha^0}}^2}
\;,\nonumber\\
%%%%%%%%%%%%%%%%%%%%%%%%%%%%%%%%%%%%%%%%%%%%%%%%%%%%%%%%%%%%%%%%%%%%%%
&&\Big({\cal N}_{_{\chi_\alpha^0\chi_\beta^0}}^c\Big)_6=
-{2\over D}{(q_1^2)^2-q_1^2q_1\cdot q_2\over(q_1^2-m_{_{\tilde{E}_j}}^2)^2}
+{q_1^2-q_1\cdot q_2\over2(q_1^2-m_{_{\tilde{E}_j}}^2)}
-{2\over D}{q_1^2q_2^2-q_1\cdot q_2q_2^2\over(q_2^2-m_{_{\chi_\alpha^0}}^2)^2}
\nonumber\\
&&\hspace{1.6cm}
-{2\over(q_1^2-m_{_{\tilde{E}_j}}^2)(q_2^2-m_{_{\chi_\alpha^0}}^2)}
\Big[{q_1^2q_1\cdot q_2-q_1^2q_2^2\over D}
-{(q_1\cdot q_2)^2-q_1^2q_2^2\over D-1}\Big]
\nonumber\\
&&\hspace{1.6cm}
+{q_1^2-q_1\cdot q_2\over2(q_2^2-m_{_{\chi_\alpha^0}}^2)}
+{D-2\over2D}{q_1^2-q_1\cdot q_2\over(q_2-q_1)^2-m_{_{l^I}}^2}
\;,\nonumber\\
%%%%%%%%%%%%%%%%%%%%%%%%%%%%%%%%%%%%%%%%%%%%%%%%%%%%%%%%%%%%%%%%%%%%%%
%%%%%%%%%%%%%%%%%%%%%%%%%%%%%%%%%%%%%%%%%%%%%%%%%%%%%%%%%%%%%%%%%%%%%%
&&\Big({\cal N}_{_{\chi_\alpha^0\chi_\beta^0}}^d\Big)_5=
-{4\over D}{q_1^2q_1\cdot q_2-q_1^2q_2^2\over(q_1^2-m_{_{\tilde{E}_j}}^2)^2}
-{4\over D}{(q_1\cdot q_2)^2-q_1\cdot q_2q_2^2
\over(q_1^2-m_{_{\tilde{E}_j}}^2)(q_2^2-m_{_{\chi_\alpha^0}}^2)}
\nonumber\\
&&\hspace{1.6cm}
+{q_1\cdot q_2-q_2^2\over q_1^2-m_{_{\tilde{E}_j}}^2}
+{2\over D}{q_1^2-q_1\cdot q_2\over q_1^2-m_{_{\tilde{E}_j}}^2}
-{4\over D}{q_1\cdot q_2q_2^2-(q_2^2)^2\over(q_2^2-m_{_{\chi_\alpha^0}}^2)^2}
\nonumber\\
&&\hspace{1.6cm}
+{2+D\over D}{q_1\cdot q_2-q_2^2\over q_2^2-m_{_{\chi_\alpha^0}}^2}
\;,\nonumber\\
%%%%%%%%%%%%%%%%%%%%%%%%%%%%%%%%%%%%%%%%%%%%%%%%%%%%%%%%%%%%%%%%%%%%%%
&&\Big({\cal N}_{_{\chi_\alpha^0\chi_\beta^0}}^d\Big)_6=
{2\over D}{q_1^2q_1\cdot q_2-q_1^2q_2^2\over(q_1^2-m_{_{\tilde{E}_j}}^2)^2}
+{2\over(q_1^2-m_{_{\tilde{E}_j}}^2)(q_2^2-m_{_{\chi_\alpha^0}}^2)}
\Big[{q_1^2q_2^2-q_1\cdot q_2q_2^2\over D}
\nonumber\\
&&\hspace{1.6cm}
+{(q_1\cdot q_2)^2-q_1^2q_2^2\over D-1}\Big]
-{q_1\cdot q_2-q_2^2\over2(q_1^2-m_{_{\tilde{E}_j}}^2)}
+{2\over D}{q_1\cdot q_2q_2^2-(q_2^2)^2\over(q_2^2-m_{_{\chi_\alpha^0}}^2)^2}
\nonumber\\
&&\hspace{1.6cm} -{2+D\over2D}{q_1\cdot q_2-q_2^2\over
q_2^2-m_{_{\chi_\alpha^0}}^2} +{1\over D}{q_1\cdot q_2-q_2^2\over
q_2^2-m_{_{\tilde{E}_i}}^2} +{2-D\over2D}{q_1\cdot
q_2-q_2^2\over(q_2-q_1)^2-m_{_{l^I}}^2}\;.
%%%%%%%%%%%%%%%%%%%%%%%%%%%%%%%%%%%%%%%%%%%%%%%%%%%%%%%%%%%%%%%%%%%%%%
\label{aeq1}
\end{eqnarray}

\section{The functions \label{ap3}}

\begin{eqnarray}
%%%%%%%%%%%%%%%%%%%%%%%%%%%%%%%%%%%%%%%%%%%%%%%%%%%%%%%%%%%%%%%%%%%%%%%%%%%%%%%%%%%
&&\rho_1(x_{_1}, x_{_2})=-6x_{_1}^2x_{_2}{\ln x_{_1}-\ln x_{_2}\over(x_{_1}-x_{_2})^4}
+{2x_{_1}^2+5x_{_1}x_{_2}-x_{_2}^2\over(x_{_1}-x_{_2})^3}
\;,\nonumber\\
%%%%%%%%%%%%%%%%%%%%%%%%%%%%%%%%%%%%%%%%%%%%%%%%%%%%%%%%%%%%%%%%%%%%%%%%%%%%%%%%%%%
&&\rho_2(x_{_1}, x_{_2})=2x_{_1}x_{_2}{\ln x_{_1}-\ln x_{_2}\over(x_{_1}-x_{_2})^3}
-{x_{_1}+x_{_2}\over(x_{_1}-x_{_2})^2}
\;,\nonumber\\
%%%%%%%%%%%%%%%%%%%%%%%%%%%%%%%%%%%%%%%%%%%%%%%%%%%%%%%%%%%%%%%%%%%%%%%%%%%%%%%%%%%
%%%%%%%%%%%%%%%%%%%%%%%%%%%%%%%%%%%%%%%%%%%%%%%%%%%%%%%%%%%%%%%%%%%%%%%%%%%%%%%%%%%
&&\varphi_1(x_{_1}, x_{_2})=-(2x_{_1}^3+3x_{_1}^2x_{_2})
{\ln x_{_1}-\ln x_{_2}\over(x_{_1}-x_{_2})^4}
+{28x_{_1}^2+x_{_1}x_{_2}+x_{_2}^2\over6(x_{_1}-x_{_2})^3}
\;,\nonumber\\
%%%%%%%%%%%%%%%%%%%%%%%%%%%%%%%%%%%%%%%%%%%%%%%%%%%%%%%%%%%%%%%%%%%%%%%%%%%%%%%%%%%
&&\varphi_2(x_{_1}, x_{_2})=-(6x_{_1}x_{_2}^2-x_{_2}^3)
{\ln x_{_1}-\ln x_{_2}\over(x_{_1}-x_{_2})^4}
-{8x_{_1}^2-37x_{_1}x_{_2}-x_{_2}^2\over6(x_{_1}-x_{_2})^3}\;,\nonumber\\
%%%%%%%%%%%%%%%%%%%%%%%%%%%%%%%%%%%%%%%%%%%%%%%%%%%%%%%%%%%%%%%%%%%%%%%%%%%%%%%%%%%
&&\varphi_3(x_{_1}, x_{_2})=x_{_2}^2{\ln x_{_1}-\ln x_{_2}\over(x_{_1}-x_{_2})^3}
+{x_{_1}-3x_{_2}\over2(x_{_1}-x_{_2})^2}\;.
%%%%%%%%%%%%%%%%%%%%%%%%%%%%%%%%%%%%%%%%%%%%%%%%%%%%%%%%%%%%%%%%%%%%%%%%%%%%%%%%%%%
\label{fun1}
\end{eqnarray}

\begin{eqnarray}
%%%%%%%%%%%%%%%%%%%%%%%%%%%%%%%%%%%%%%%%%%%%%%%%%%%%%%%%%%%%%%%%%%%%%%%%%%%%%%%%%%%
&&\Psi_{_{3a}}(x_{_0};x_{_A},x_{_B};x_{_\alpha},x_{_\beta})
\nonumber\\
&&\hspace{-0.4cm}=
{1\over24}\bigg\{\Big[2\varrho_{_{3,1}}+\varrho_{_{3,2}}\Big](x_{_A},x_{_B})
+{1\over(x_{_A}-x_{_B})(x_{_\alpha}-x_{_\beta})}\bigg[
x_{_A}^3x_{_\alpha}\ln^2(x_{_A}x_{_\alpha})
\nonumber\\&&
-x_{_A}^3x_{_\beta}\ln^2(x_{_A}x_{_\beta})
-x_{_B}^3x_{_\alpha}\ln^2(x_{_B}x_{_\alpha})+x_{_B}^3x_{_\beta}\ln^2(x_{_B}x_{_\beta})
\nonumber\\&&
+x_{_A}^2\Big(x_{_0}-x_{_A}+x_{_\alpha}\Big)\Phi(x_{_0},x_{_A},x_{_\alpha})
-x_{_A}^2\Big(x_{_0}-x_{_A}+x_{_\beta}\Big)\Phi(x_{_0},x_{_A},x_{_\beta})
\nonumber\\&&
-x_{_B}^2\Big(x_{_0}-x_{_B}+x_{_\alpha}\Big)\Phi(x_{_0},x_{_B},x_{_\alpha})
+x_{_B}^2\Big(x_{_0}-x_{_B}+x_{_\beta}\Big)\Phi(x_{_0},x_{_B},x_{_\beta})\bigg]\bigg\}\;.
%%%%%%%%%%%%%%%%%%%%%%%%%%%%%%%%%%%%%%%%%%%%%%%%%%%%%%%%%%%%%%%%%%%%%%%%%%%%%%%%%%%
\label{f1}
\end{eqnarray}

\begin{eqnarray}
%%%%%%%%%%%%%%%%%%%%%%%%%%%%%%%%%%%%%%%%%%%%%%%%%%%%%%%%%%%%%%%%%%%%%%%%%%%%%%%%%%%
&&\Psi_{_{3b}}(x_{_0};x_{_A},x_{_B};x_{_\alpha},x_{_\beta})
\nonumber\\
&&\hspace{-0.4cm}=
-{1\over8}\bigg\{\Big(x_{_\alpha}+x_{_\beta}\Big)
\Big[2\varrho_{_{2,1}}-\varrho_{_{2,2}}\Big](x_{_A},x_{_B})
+{1\over(x_{_A}-x_{_B})(x_{_\alpha}-x_{_\beta})}\bigg[
(x_{_A}x_{_\alpha})^2\ln^2(x_{_A}x_{_\alpha})
\nonumber\\&&
-(x_{_A}x_{_\beta})^2\ln^2(x_{_A}x_{_\beta})
-(x_{_B}x_{_\alpha})^2\ln^2(x_{_B}x_{_\alpha})
+(x_{_B}x_{_\beta})^2\ln^2(x_{_B}x_{_\beta})
%%%%%%%%%%%%%%%%%%%%%%%%%%%%%%%%%%%%%%%%%%%%%%%%%%%%%%%%%%%%%%%%%%%%%%%%%%%%%%%%%%%
\nonumber\\&&
+x_{_A}x_{_\alpha}\Big(x_{_0}+x_{_A}-x_{_\alpha}\Big)
\Phi(x_{_0},x_{_A},x_{_\alpha})
-x_{_A}x_{_\beta}\Big(x_{_0}+x_{_A}-x_{_\beta}\Big)
\Phi(x_{_0},x_{_A},x_{_\beta})
\nonumber\\&&
-x_{_B}x_{_\alpha}\Big(x_{_0}+x_{_B}-x_{_\alpha}\Big)
\Phi(x_{_0},x_{_B},x_{_\alpha})
+x_{_B}x_{_\beta}\Big(x_{_0}+x_{_B}-x_{_\beta}\Big)
\Phi(x_{_0},x_{_B},x_{_\beta})\bigg)\bigg\}\;.
%%%%%%%%%%%%%%%%%%%%%%%%%%%%%%%%%%%%%%%%%%%%%%%%%%%%%%%%%%%%%%%%%%%%%%%%%%%%%%%%%%%
\label{f2}
\end{eqnarray}

\begin{eqnarray}
%%%%%%%%%%%%%%%%%%%%%%%%%%%%%%%%%%%%%%%%%%%%%%%%%%%%%%%%%%%%%%%%%%%%%%%%%%%%%%%%%%%
&&\Psi_{_{3c}}(x_{_0};x_{_A},x_{_B};x_{_\alpha},x_{_\beta})
\nonumber\\
&&\hspace{-0.4cm}=
{1\over16}\bigg\{\Big(x_{_\alpha}+x_{_\beta}\Big)
\Big[2\varrho_{_{2,1}}-\varrho_{_{2,2}}\Big](x_{_A},x_{_B})
+\Big(x_{_A}^2+x_{_A}x_{_B}+x_{_B}^2\Big)
\Big[2\varrho_{_{1,1}}-\varrho_{_{1,2}}\Big](x_{_\alpha},x_{_\beta})
\nonumber\\&&
+{1\over(x_{_A}-x_{_B})(x_{_\alpha}-x_{_\beta})}\bigg(
(x_{_A}^3x_{_\alpha}+x_{_A}^2x_{_\alpha}^2-x_{_0}x_{_A}^2x_{_\alpha})\ln^2(x_{_A}x_{_\alpha})
\nonumber\\&&
-(x_{_A}^3x_{_\beta}+x_{_A}^2x_{_\beta}^2-x_{_0}x_{_A}^2x_{_\beta})\ln^2(x_{_A}x_{_\beta})
-(x_{_B}^3x_{_\alpha}+x_{_B}^2x_{_\alpha}^2-x_{_0}x_{_B}^2x_{_\alpha})\ln^2(x_{_B}x_{_\alpha})
\nonumber\\&&
+(x_{_B}^3x_{_\beta}+x_{_B}^2x_{_\beta}^2-x_{_0}x_{_B}^2x_{_\beta})\ln^2(x_{_B}x_{_\beta})
%%%%%%%%%%%%%%%%%%%%%%%%%%%%%%%%%%%%%%%%%%%%%%%%%%%%%%%%%%%%%%%%%%%%%%%%%%%%%%%%%%%
-x_{_A}\lambda^2(x_{_0},x_{_A},x_{_\alpha})\Phi(x_{_0},x_{_A},x_{_\alpha})
\nonumber\\&&
+x_{_A}\lambda^2(x_{_0},x_{_A},x_{_\beta})\Phi(x_{_0},x_{_A},x_{_\beta})
+x_{_B}\lambda^2(x_{_0},x_{_B},x_{_\alpha})\Phi(x_{_0},x_{_B},x_{_\alpha})
\nonumber\\&&
-x_{_B}\lambda^2(x_{_0},x_{_B},x_{_\beta})\Phi(x_{_0},x_{_B},x_{_\beta})\bigg)
\bigg\}\;.
%%%%%%%%%%%%%%%%%%%%%%%%%%%%%%%%%%%%%%%%%%%%%%%%%%%%%%%%%%%%%%%%%%%%%%%%%%%%%%%%%%%
\label{f3}
\end{eqnarray}

\begin{eqnarray}
%%%%%%%%%%%%%%%%%%%%%%%%%%%%%%%%%%%%%%%%%%%%%%%%%%%%%%%%%%%%%%%%%%%%%%%%%%%%%%%%%%%
&&\Psi_{_{2a}}(x_{_0};x_{_A},x_{_B};x_{_\alpha},x_{_\beta})
\nonumber\\
&&\hspace{-0.4cm}=
{1\over2}\bigg\{{1\over(x_{_A}-x_{_B})(x_{_\alpha}-x_{_\beta})}
\bigg(x_{_A}^2\Phi(x_{_0},x_{_A},x_{_\alpha})
-x_{_A}^2\Phi(x_{_0},x_{_A},x_{_\beta})
\nonumber\\&&
-x_{_B}^2\Phi(x_{_0},x_{_B},x_{_\alpha})
+x_{_B}^2\Phi(x_{_0},x_{_B},x_{_\beta})\bigg)\bigg\}\;.
%%%%%%%%%%%%%%%%%%%%%%%%%%%%%%%%%%%%%%%%%%%%%%%%%%%%%%%%%%%%%%%%%%%%%%%%%%%%%%%%%%%
\label{f4}
\end{eqnarray}

\begin{eqnarray}
%%%%%%%%%%%%%%%%%%%%%%%%%%%%%%%%%%%%%%%%%%%%%%%%%%%%%%%%%%%%%%%%%%%%%%%%%%%%%%%%%%%
&&\Psi_{_{2b}}(x_{_0};x_{_A},x_{_B};x_{_\alpha},x_{_\beta})
\nonumber\\
&&\hspace{-0.4cm}=
{1\over4}\bigg\{\Big[2\varrho_{_{2,1}}-\varrho_{_{2,2}}\Big](x_{_A},x_{_B})
+\Big(x_{_A}+x_{_B}\Big)\Big[2\varrho_{_{1,1}}-\varrho_{_{1,2}}\Big]
(x_{_\alpha},x_{_\beta})
\nonumber\\&&
%%%%%%%%%%%%%%%%%%%%%%%%%%%%%%%%%%%%%%%%%%%%%%%%%%%%%%%%%%%%%%%%%%%%%%%%%%%%%%%%%%%
+{1\over(x_{_A}-x_{_B})(x_{_\alpha}-x_{_\beta})}\bigg(
x_{_A}^2x_{_\alpha}\ln^2(x_{_A}x_{_\alpha})
-x_{_A}^2x_{_\beta}\ln^2(x_{_A}x_{_\beta})
\nonumber\\&&
-x_{_B}^2x_{_\alpha}\ln^2(x_{_B}x_{_\alpha})
+x_{_B}^2x_{_\beta}\ln^2(x_{_B}x_{_\beta})
+x_{_A}(x_{_0}-x_{_A}-x_{_\alpha})\Phi(x_{_0},x_{_A},x_{_\alpha})
\nonumber\\&&
-x_{_A}(x_{_0}-x_{_A}-x_{_\beta})\Phi(x_{_0},x_{_A},x_{_\beta})
-x_{_B}(x_{_0}-x_{_B}-x_{_\alpha})\Phi(x_{_0},x_{_B},x_{_\alpha})
\nonumber\\&&
+x_{_B}(x_{_0}-x_{_B}-x_{_\beta})\Phi(x_{_0},x_{_B},x_{_\beta})\bigg)\bigg\}\;.
%%%%%%%%%%%%%%%%%%%%%%%%%%%%%%%%%%%%%%%%%%%%%%%%%%%%%%%%%%%%%%%%%%%%%%%%%%%%%%%%%%%
\label{f5}
\end{eqnarray}

\begin{eqnarray}
%%%%%%%%%%%%%%%%%%%%%%%%%%%%%%%%%%%%%%%%%%%%%%%%%%%%%%%%%%%%%%%%%%%%%%%%%%%%%%%%%%%
&&\Psi_{_{2c}}(x_{_0};x_{_A},x_{_B};x_{_\alpha},x_{_\beta})
\nonumber\\
&&\hspace{-0.4cm}=
{1\over8}\bigg\{\Big(x_{_\alpha}+x_{_\beta}\Big)
\Big[2\varrho_{_{1,1}}-\varrho_{_{1,2}}\Big](x_{_A},x_{_B})
+\Big(x_{_A}+x_{_B}\Big)\Big[2\varrho_{_{1,1}}-\varrho_{_{1,2}}\Big]
(x_{_\alpha},x_{_\beta})
\nonumber\\&&
+{1\over(x_{_A}-x_{_B})(x_{_\alpha}-x_{_\beta})}\bigg(
(x_{_A}+x_{_\alpha}-x_{_0})x_{_A}x_{_\alpha}\ln^2(x_{_A}x_{_\alpha})
\nonumber\\&&
-(x_{_A}+x_{_\beta}-x_{_0})x_{_A}x_{_\beta}\ln^2(x_{_A}x_{_\beta})
-(x_{_B}+x_{_\alpha}-x_{_0})x_{_B}x_{_\alpha}\ln^2(x_{_B}x_{_\alpha})
\nonumber\\&&
+(x_{_B}+x_{_\beta}-x_{_0})x_{_B}x_{_\beta}\ln^2(x_{_B}x_{_\beta})
%%%%%%%%%%%%%%%%%%%%%%%%%%%%%%%%%%%%%%%%%%%%%%%%%%%%%%%%%%%%%%%%%%%%%%%%%%%%%%%%%%%
-(x_{_0}-x_{_A}-x_{_\alpha})^2\Phi(x_{_0},x_{_A},x_{_\alpha})
\nonumber\\&&
+(x_{_0}-x_{_A}-x_{_\beta})^2\Phi(x_{_0},x_{_A},x_{_\beta})
+(x_{_0}-x_{_B}-x_{_\alpha})^2\Phi(x_{_0},x_{_B},x_{_\alpha})
\nonumber\\&&
-(x_{_0}-x_{_B}-x_{_\beta})^2\Phi(x_{_0},x_{_B},x_{_\beta})\bigg)\bigg\}\;.
%%%%%%%%%%%%%%%%%%%%%%%%%%%%%%%%%%%%%%%%%%%%%%%%%%%%%%%%%%%%%%%%%%%%%%%%%%%%%%%%%%%
\label{f6}
\end{eqnarray}

\begin{eqnarray}
%%%%%%%%%%%%%%%%%%%%%%%%%%%%%%%%%%%%%%%%%%%%%%%%%%%%%%%%%%%%%%%%%%%%%%%%%%%%%%%%%%%
&&\Psi_{_{2d}}(x_{_0};x_{_A},x_{_B};x_{_\alpha},x_{_\beta})
\nonumber\\
&&\hspace{-0.4cm}=
-{1\over2}\bigg\{\varrho_{_{2,2}}(x_{_A},x_{_B})
+\varrho_{_{2,2}}(x_{_\alpha},x_{_\beta})
+{1\over(x_{_A}-x_{_B})(x_{_\alpha}-x_{_\beta})}\Big(
x_{_A}x_{_\alpha}\Phi(x_{_0},x_{_A},x_{_\alpha})
\nonumber\\&&
-x_{_A}x_{_\beta}\Phi(x_{_0},x_{_A},x_{_\beta})
-x_{_B}x_{_\alpha}\Phi(x_{_0},x_{_B},x_{_\alpha})
+x_{_B}x_{_\beta}\Phi(x_{_0},x_{_B},x_{_\beta})\Big)\bigg\}\;.
%%%%%%%%%%%%%%%%%%%%%%%%%%%%%%%%%%%%%%%%%%%%%%%%%%%%%%%%%%%%%%%%%%%%%%%%%%%%%%%%%%%
\label{f7}
\end{eqnarray}

\begin{eqnarray}
%%%%%%%%%%%%%%%%%%%%%%%%%%%%%%%%%%%%%%%%%%%%%%%%%%%%%%%%%%%%%%%%%%%%%%%%%%%%%%%%%%%
&&\Psi_{_{1a}}(x_{_0};x_{_A},x_{_B};x_{_\alpha},x_{_\beta})
\nonumber\\
&&\hspace{-0.4cm}=
-{1\over2}\bigg\{\varrho_{_{1,2}}(x_{_\alpha},x_{_\beta})
+{1\over(x_{_A}-x_{_B})(x_{_\alpha}-x_{_\beta})}\Big[
x_{_A}\Phi(x_{_0},x_{_A},x_{_\alpha})
\nonumber\\&&
-x_{_A}\Phi(x_{_0},x_{_A},x_{_\beta})-x_{_B}\Phi(x_{_0},x_{_B},x_{_\alpha})
+x_{_B}\Phi(x_{_0},x_{_B},x_{_\beta})\Big]\bigg\}\;.
%%%%%%%%%%%%%%%%%%%%%%%%%%%%%%%%%%%%%%%%%%%%%%%%%%%%%%%%%%%%%%%%%%%%%%%%%%%%%%%%%%%
\label{f8}
\end{eqnarray}

\begin{eqnarray}
%%%%%%%%%%%%%%%%%%%%%%%%%%%%%%%%%%%%%%%%%%%%%%%%%%%%%%%%%%%%%%%%%%%%%%%%%%%%%%%%%%%
&&\Psi_{_{1b}}(x_{_0};x_{_A},x_{_B};x_{_\alpha},x_{_\beta})
\nonumber\\
&&\hspace{-0.4cm}=
{1\over4}\bigg\{\Big[2\varrho_{_{1,1}}-\varrho_{_{1,2}}\Big](x_{_A},x_{_B})
+\Big[2\varrho_{_{1,1}}-\varrho_{_{1,2}}\Big](x_{_\alpha},x_{_\beta})
\nonumber\\&&
+{1\over(x_{_A}-x_{_B})(x_{_\alpha}-x_{_\beta})}\bigg(
x_{_A}x_{_\alpha}\ln^2(x_{_A}x_{_\alpha})
-x_{_A}x_{_\beta}\ln^2(x_{_A}x_{_\beta})
\nonumber\\&&
-x_{_B}x_{_\alpha}\ln^2(x_{_B}x_{_\alpha})
+x_{_B}x_{_\beta}\ln^2(x_{_B}x_{_\beta})
+(x_{_0}-x_{_A}-x_{_\alpha})\Phi(x_{_0},x_{_A},x_{_\alpha})
\nonumber\\&&
-(x_{_0}-x_{_A}-x_{_\beta})\Phi(x_{_0},x_{_A},x_{_\beta})
-(x_{_0}-x_{_B}-x_{_\alpha})\Phi(x_{_0},x_{_B},x_{_\alpha})
\nonumber\\&&
+(x_{_0}-x_{_B}-x_{_\beta})\Phi(x_{_0},x_{_B},x_{_\beta})\bigg)\bigg\}\;.
%%%%%%%%%%%%%%%%%%%%%%%%%%%%%%%%%%%%%%%%%%%%%%%%%%%%%%%%%%%%%%%%%%%%%%%%%%%%%%%%%%%
\label{f9}
\end{eqnarray}

\begin{eqnarray}
%%%%%%%%%%%%%%%%%%%%%%%%%%%%%%%%%%%%%%%%%%%%%%%%%%%%%%%%%%%%%%%%%%%%%%%%%%%%%%%%%%%
&&\Psi_{_{0}}(x_{_0};x_{_A},x_{_B};x_{_\alpha},x_{_\beta})
\nonumber\\
&&\hspace{-0.4cm}=
{1\over(x_{_A}-x_{_B})(x_{_\alpha}-x_{_\beta})}\bigg[
\Phi(x_{_0},x_{_A},x_{_\alpha})-\Phi(x_{_0},x_{_A},x_{_\beta})
\nonumber\\&&
-\Phi(x_{_0},x_{_B},x_{_\alpha})+\Phi(x_{_0},x_{_B},x_{_\beta})\bigg]\;.
%%%%%%%%%%%%%%%%%%%%%%%%%%%%%%%%%%%%%%%%%%%%%%%%%%%%%%%%%%%%%%%%%%%%%%%%%%%%%%%%%%%
\label{f10}
\end{eqnarray}

\begin{eqnarray}
%%%%%%%%%%%%%%%%%%%%%%%%%%%%%%%%%%%%%%%%%%%%%%%%%%%%%%%%%%%%%%%%%%%%%%%%%%%%%%%%%%%
&&{\rm F}_{_{N,1}}(x_{_{l^J}};x_{_{\tilde{E}_i}},x_{_{\chi_\alpha^0}};
x_{_{\tilde{E}_j}},x_{_{\chi_\beta^0}})
\nonumber\\
&&\hspace{-0.4cm}=
\Bigg\{{1\over4}{\partial^3\over\partial^3x_{_{\tilde{E}_j}}}
\Psi_{_{3a}}+{1\over8}{\partial^2\over\partial^2x_{_{\tilde{E}_j}}}\Big[
\Psi_{_{2d}}-\Psi_{_{2b}}\Big]+{1\over4}{\partial^3\over\partial^2x_{_{\tilde{E}_j}}
\partial x_{_{\chi_\alpha^0}}}\Psi_{_{3b}}+{1\over3}{\partial^3\over\partial^2x_{_{\tilde{E}_j}}
\partial x_{_{\chi_\alpha^0}}}\Psi_{_{3c}}
\nonumber\\&&
-{3\over8}{\partial^2\over\partial x_{_{\tilde{E}_j}}
\partial x_{_{\chi_\alpha^0}}}\Psi_{_{2b}}
+{1\over4}{\partial^2\over\partial x_{_{\tilde{E}_j}}
\partial x_{_{\chi_\alpha^0}}}\Psi_{_{2d}}
+{1\over8}{\partial^2\over\partial x_{_{\tilde{E}_j}}
\partial x_{_{\tilde{E}_i}}}\Big[\Psi_{_{2b}}-\Psi_{_{2d}}\Big]
+{1\over4}{\partial\over\partial x_{_{\tilde{E}_j}}}\Psi_{_{1a}}
\nonumber\\&&
+{1\over8}{\partial\over\partial x_{_{\chi_\alpha^0}}}\Big[2\Psi_{_{1a}}
+\Psi_{_{1b}}\Big]\Bigg\}(x_{_{l^J}};x_{_{\tilde{E}_j}},
x_{_{\chi_\beta^0}};x_{_{\tilde{E}_i}},x_{_{\chi_\alpha^0}})
\nonumber\\&&
-\Bigg\{{1\over4}{\partial^3\over\partial^2x_{_{\chi_\alpha^0}}
\partial x_{_{\tilde{E}_j}}}\Psi_{_{3b}}
+{1\over4}{\partial^3\over\partial^3x_{_{\chi_\alpha^0}}}\Psi_{_{3a}}
+{1\over8}{\partial^2\over\partial^2x_{_{\chi_\alpha^0}}}
\Big[2\Psi_{_{2d}}-\Psi_{_{2b}}\Big]
\nonumber\\&&
+{1\over8}{\partial^2\over\partial x_{_{\chi_\alpha^0}}
\partial x_{_{\tilde{E}_i}}}\Psi_{_{2b}}
\Bigg\}(x_{_{l^J}};x_{_{\tilde{E}_i}},x_{_{\chi_\alpha^0}};
x_{_{\tilde{E}_j}},x_{_{\chi_\beta^0}})\;.
%%%%%%%%%%%%%%%%%%%%%%%%%%%%%%%%%%%%%%%%%%%%%%%%%%%%%%%%%%%%%%%%%%%%%%%%%%%%%%%%%%%
\label{fnn1}
\end{eqnarray}

\begin{eqnarray}
%%%%%%%%%%%%%%%%%%%%%%%%%%%%%%%%%%%%%%%%%%%%%%%%%%%%%%%%%%%%%%%%%%%%%%%%%%%%%%%%%%%
&&{\rm F}_{_{N,2}}(x_{_{l^J}};x_{_{\tilde{E}_i}},x_{_{\chi_\alpha^0}};
x_{_{\tilde{E}_j}},x_{_{\chi_\beta^0}})
\nonumber\\
&&\hspace{-0.4cm}=
\Bigg\{-{1\over24}{\partial^3\over\partial^3x_{_{\tilde{E}_j}}}
\Big[\Psi_{_{2a}}-\Psi_{_{2b}}\Big]
+{1\over8}{\partial^2\over\partial^2x_{_{\tilde{E}_j}}}
\Big[\Psi_{_{1a}}-\Psi_{_{1b}}\Big]
-{1\over24}{\partial^3\over\partial^2x_{_{\tilde{E}_j}}
\partial x_{_{\chi_\alpha^0}}}\Big[3\Psi_{_{2b}}
\nonumber\\&&
-4\Psi_{_{2c}}+\Psi_{_{2d}}\Big]
+{1\over8}{\partial^2\over\partial x_{_{\tilde{E}_i}}
\partial x_{_{\tilde{E}_j}}}\Psi_{_{1b}}
+{1\over8}{\partial^2\over\partial x_{_{\tilde{E}_j}}
\partial x_{_{\chi_\alpha^0}}}\Big[2\Psi_{_{1a}}-\Psi_{_{1b}}\Big]
\nonumber\\&&
-{1\over4}{\partial\over\partial x_{_{\chi_\alpha^0}}}\Psi_{_{0}}
\Bigg\}(x_{_{l^J}};x_{_{\tilde{E}_j}},
x_{_{\chi_\beta^0}};x_{_{\tilde{E}_i}},x_{_{\chi_\alpha^0}})
\nonumber\\&&
+\Bigg\{{1\over8}{\partial^3\over\partial x_{_{\tilde{E}_j}}
\partial^2x_{_{\chi_\alpha^0}}}\Big[\Psi_{_{2b}}-\Psi_{_{2d}}\Big]
+{1\over24}{\partial^3\over\partial^3x_{_{\chi_\alpha^0}}}
\Big[\Psi_{_{2a}}-\Psi_{_{2b}}\Big]
+{1\over8}{\partial^2\over\partial^2x_{_{\chi_\alpha^0}}}\Psi_{_{1b}}
\nonumber\\&&
+{1\over8}{\partial^2\over\partial x_{_{\chi_\alpha^0}}
\partial x_{_{\tilde{E}_i}}}\Psi_{_{1b}}
\Bigg\}(x_{_{l^J}};x_{_{\tilde{E}_i}},x_{_{\chi_\alpha^0}};
x_{_{\tilde{E}_j}},x_{_{\chi_\beta^0}})\;.
%%%%%%%%%%%%%%%%%%%%%%%%%%%%%%%%%%%%%%%%%%%%%%%%%%%%%%%%%%%%%%%%%%%%%%%%%%%%%%%%%%%
\label{fnn2}
\end{eqnarray}

\begin{eqnarray}
%%%%%%%%%%%%%%%%%%%%%%%%%%%%%%%%%%%%%%%%%%%%%%%%%%%%%%%%%%%%%%%%%%%%%%%%%%%%%%%%%%%
&&{\rm F}_{_{N,3}}(x_{_{l^J}};x_{_{\tilde{E}_i}},x_{_{\chi_\alpha^0}};
x_{_{\tilde{E}_j}},x_{_{\chi_\beta^0}})
\nonumber\\
&&\hspace{-0.4cm}=
\Bigg\{-{1\over4}{\partial^2\over\partial^2x_{_{\tilde{E}_j}}}
\Big[\Psi_{_{2a}}-\Psi_{_{2b}}\Big]
+{1\over2}{\partial\over\partial x_{_{\tilde{E}_j}}}
\Big[\Psi_{_{1a}}-\Psi_{_{1b}}\Big]
\nonumber\\&&
-{1\over6}{\partial^2\over\partial x_{_{\tilde{E}_j}}\partial x_{_{\chi_\alpha^0}}}
\Big[3\Psi_{_{2b}}-4\Psi_{_{2c}}+\Psi_{_{2d}}\Big]
+{1\over2}{\partial\over\partial x_{_{\chi_\alpha^0}}}
\Big[\Psi_{_{1a}}-\Psi_{_{1b}}\Big]
\nonumber\\&&
+{1\over4}{\partial\over\partial x_{_{l^J}}}
\Big[\Psi_{_{1a}}-\Psi_{_{1b}}\Big]
\Bigg\}(x_{_{l^J}};x_{_{\tilde{E}_j}},x_{_{\chi_\beta^0}};
x_{_{\tilde{E}_i}},x_{_{\chi_\alpha^0}})
\nonumber\\&&
+{1\over4}{\partial^2\over\partial^2x_{_{\chi_\alpha^0}}}
\Big[\Psi_{_{2b}}-\Psi_{_{2d}}\Big](x_{_{l^J}};x_{_{\tilde{E}_i}},x_{_{\chi_\alpha^0}};
x_{_{\tilde{E}_j}},x_{_{\chi_\beta^0}})\;.
%%%%%%%%%%%%%%%%%%%%%%%%%%%%%%%%%%%%%%%%%%%%%%%%%%%%%%%%%%%%%%%%%%%%%%%%%%%%%%%%%%%
\label{fnn3}
\end{eqnarray}

\begin{eqnarray}
%%%%%%%%%%%%%%%%%%%%%%%%%%%%%%%%%%%%%%%%%%%%%%%%%%%%%%%%%%%%%%%%%%%%%%%%%%%%%%%%%%%
&&{\rm F}_{_{N,4}}(x_{_{l^J}};x_{_{\tilde{E}_i}},x_{_{\chi_\alpha^0}};
x_{_{\tilde{E}_j}},x_{_{\chi_\beta^0}})
\nonumber\\
&&\hspace{-0.4cm}=
\Bigg\{-{1\over6}{\partial^2\over\partial x_{_{\tilde{E}_j}}
\partial x_{_{\chi_\alpha^0}}}\Big[3\Psi_{_{2b}}-4\Psi_{_{2c}}
+\Psi_{_{2d}}\Big]
-{1\over4}{\partial^2\over\partial^2x_{_{\chi_\alpha^0}}}
\Big[\Psi_{_{2a}}-\Psi_{_{2b}}\Big]
+\Big[{1\over2}{\partial\over\partial x_{_{\tilde{E}_j}}}
\nonumber\\&&
+{3\over4}{\partial\over\partial x_{_{\chi_\alpha^0}}}
-{1\over4}{\partial\over\partial x_{_{\tilde{E}_i}}}
+{1\over4}{\partial\over\partial x_{_{l^J}}}\Big]
\Big[\Psi_{_{1a}}-\Psi_{_{1b}}\Big]
\Bigg\}(x_{_{l^J}};x_{_{\tilde{E}_i}},x_{_{\chi_\alpha^0}};
x_{_{\tilde{E}_j}},x_{_{\chi_\beta^0}})
\nonumber\\&&
+{1\over4}{\partial^2\over\partial^2x_{_{\tilde{E}_j}}}\Big[\Psi_{_{2b}}
-\Psi_{_{2d}}\Big](x_{_{l^J}};x_{_{\tilde{E}_j}},x_{_{\chi_\beta^0}};
x_{_{\tilde{E}_i}},x_{_{\chi_\alpha^0}})\;.
%%%%%%%%%%%%%%%%%%%%%%%%%%%%%%%%%%%%%%%%%%%%%%%%%%%%%%%%%%%%%%%%%%%%%%%%%%%%%%%%%%%
\label{fnn4}
\end{eqnarray}

\begin{eqnarray}
%%%%%%%%%%%%%%%%%%%%%%%%%%%%%%%%%%%%%%%%%%%%%%%%%%%%%%%%%%%%%%%%%%%%%%%%%%%%%%%%%%%
&&{\rm F}_{_{M,1}}(0;x_{_{\tilde{\nu}_i}},x_{_{\chi_\alpha^\pm}};
x_{_{\tilde{E}_j}},x_{_{\chi_\beta^0}})
\nonumber\\
&&\hspace{-0.4cm}=
\Bigg\{{1\over4}{\partial^3\over\partial^3x_{_{\tilde{E}_j}}}\Psi_{_{3a}}
-{1\over8}{\partial^2\over\partial^2x_{_{\tilde{E}_j}}}\Big[\Psi_{_{2b}}
-\Psi_{_{2d}}\Big]
+{\partial^3\over\partial^2x_{_{\tilde{E}_j}}\partial x_{_{\chi_\alpha^\pm}}}
\Big[{1\over4}\Psi_{_{3b}}+{1\over3}\Psi_{_{3c}}\Big]
\nonumber\\&&
-{1\over4}{\partial^2\over\partial x_{_{\tilde{E}_j}}\partial x_{_{\chi_\alpha^\pm}}}
\Big[\Psi_{_{2b}}-\Psi_{_{2d}}\Big]\Bigg\}(0;x_{_{\tilde{E}_j}},x_{_{\chi_\beta^0}};
x_{_{\tilde{\nu}_i}},x_{_{\chi_\alpha^\pm}})
\nonumber\\&&
+\Bigg\{{1\over8}{\partial^2\over\partial^2x_{_{\chi_\alpha^\pm}}}\Big[\Psi_{_{2b}}
-\Psi_{_{2d}}\Big]
-{1\over4}{\partial^3\over\partial x_{_{\tilde{E}_j}}\partial^2x_{_{\chi_\alpha^\pm}}}
\Psi_{_{3b}}
\nonumber\\&&
-{1\over4}{\partial^3\over\partial^3x_{_{\chi_\alpha^\pm}}}\Psi_{_{3a}}
\Bigg\}(0;x_{_{\tilde{\nu}_i}},x_{_{\chi_\alpha^\pm}};
x_{_{\tilde{E}_j}},x_{_{\chi_\beta^0}})\;.
%%%%%%%%%%%%%%%%%%%%%%%%%%%%%%%%%%%%%%%%%%%%%%%%%%%%%%%%%%%%%%%%%%%%%%%%%%%%%%%%%%%
\label{fnc1}
\end{eqnarray}

\begin{eqnarray}
%%%%%%%%%%%%%%%%%%%%%%%%%%%%%%%%%%%%%%%%%%%%%%%%%%%%%%%%%%%%%%%%%%%%%%%%%%%%%%%%%%%
&&{\rm F}_{_{M,2}}(0;x_{_{\tilde{\nu}_i}},x_{_{\chi_\alpha^\pm}};
x_{_{\tilde{E}_j}},x_{_{\chi_\beta^0}})
\nonumber\\
&&\hspace{-0.4cm}=
\Bigg\{{1\over4}{\partial^3\over\partial^3x_{_{\tilde{E}_j}}}\Psi_{_{3a}}
-{1\over8}{\partial^2\over\partial^2x_{_{\tilde{E}_j}}}\Big[2\Psi_{_{2a}}
-\Psi_{_{2b}}-\Psi_{_{2d}}\Big]
\nonumber\\&&
+{1\over4}{\partial^3\over\partial^2x_{_{\tilde{E}_j}}
\partial x_{_{\chi_\alpha^\pm}}}\Psi_{_{3b}}
-{\partial^2\over\partial x_{_{\tilde{E}_j}}\partial x_{_{\chi_\alpha^\pm}}}
\Big[{1\over4}\Psi_{_{2b}}-{1\over3}\Psi_{_{2c}}
\nonumber\\&&
+{1\over3}\Psi_{_{2d}}\Big]
+{1\over2}{\partial\over\partial x_{_{\tilde{E}_j}}}\Big[\Psi_{_{1a}}
-\Psi_{_{1b}}\Big]\Bigg\}(0;x_{_{\tilde{E}_j}},x_{_{\chi_\beta^0}};
x_{_{\tilde{\nu}_i}},x_{_{\chi_\alpha^\pm}})
\nonumber\\&&
+\Bigg\{-{\partial^3\over\partial x_{_{\tilde{E}_j}}
\partial^2x_{_{\chi_\alpha^\pm}}}\Big[{1\over4}\Psi_{_{3b}}
+{1\over3}\Psi_{_{3c}}\Big]
+{1\over4}{\partial^2\over\partial x_{_{\tilde{E}_j}}
\partial x_{_{\chi_\alpha^\pm}}}\Psi_{_{2b}}
\nonumber\\&&
+{1\over8}{\partial^2\over\partial^2x_{_{\chi_\alpha^\pm}}}\Big[\Psi_{_{2b}}
-\Psi_{_{2d}}\Big]
-{1\over4}{\partial^3\over\partial^3x_{_{\chi_\alpha^\pm}}}\Psi_{_{3a}}
\Bigg\}(0;x_{_{\tilde{\nu}_i}},x_{_{\chi_\alpha^\pm}};
x_{_{\tilde{E}_j}},x_{_{\chi_\beta^0}})\;.
%%%%%%%%%%%%%%%%%%%%%%%%%%%%%%%%%%%%%%%%%%%%%%%%%%%%%%%%%%%%%%%%%%%%%%%%%%%%%%%%%%%
\label{fnc2}
\end{eqnarray}

\begin{eqnarray}
%%%%%%%%%%%%%%%%%%%%%%%%%%%%%%%%%%%%%%%%%%%%%%%%%%%%%%%%%%%%%%%%%%%%%%%%%%%%%%%%%%%
&&{\rm F}_{_{M,3}}(0;x_{_{\tilde{\nu}_i}},x_{_{\chi_\alpha^\pm}};
x_{_{\tilde{E}_j}},x_{_{\chi_\beta^0}})
\nonumber\\
&&\hspace{-0.4cm}=
\Bigg\{-{1\over24}{\partial^3\over\partial^3x_{_{\tilde{E}_j}}}
\Big[\Psi_{_{2a}}-\Psi_{_{2b}}\Big]
-{1\over24}{\partial^3\over\partial^2x_{_{\tilde{E}_j}}
\partial x_{_{\chi_\alpha^\pm}}}\Big[3\Psi_{_{2b}}-4\Psi_{_{2c}}
\nonumber\\&&
+\Psi_{_{2d}}\Big]
+{1\over8}{\partial^2\over\partial^2x_{_{\tilde{E}_j}}}
\Big[\Psi_{_{1a}}-\Psi_{_{1b}}\Big]
\Bigg\}(0;x_{_{\tilde{E}_j}},x_{_{\chi_\beta^0}};
x_{_{\tilde{\nu}_i}},x_{_{\chi_\alpha^\pm}})
\nonumber\\&&
+\Bigg\{{1\over8}{\partial^3\over\partial x_{_{\tilde{E}_j}}
\partial^2x_{_{\chi_\alpha^\pm}}}\Big[\Psi_{_{2b}}-\Psi_{_{2d}}\Big]
+{1\over8}{\partial^2\over\partial^2x_{_{\chi_\alpha^\pm}}}
\Big[\Psi_{_{1a}}-\Psi_{_{1b}}\Big]
\nonumber\\&&
+{1\over24}{\partial^3\over\partial^3x_{_{\chi_\alpha^\pm}}}
\Big[\Psi_{_{2a}}-\Psi_{_{2b}}\Big]
\Bigg\}(0;x_{_{\tilde{\nu}_i}},x_{_{\chi_\alpha^\pm}};
x_{_{\tilde{E}_j}},x_{_{\chi_\beta^0}})\;.
%%%%%%%%%%%%%%%%%%%%%%%%%%%%%%%%%%%%%%%%%%%%%%%%%%%%%%%%%%%%%%%%%%%%%%%%%%%%%%%%%%%
\label{fnc3}
\end{eqnarray}

\begin{eqnarray}
%%%%%%%%%%%%%%%%%%%%%%%%%%%%%%%%%%%%%%%%%%%%%%%%%%%%%%%%%%%%%%%%%%%%%%%%%%%%%%%%%%%
&&{\rm F}_{_{M,4}}(0;x_{_{\tilde{\nu}_i}},x_{_{\chi_\alpha^\pm}};
x_{_{\tilde{E}_j}},x_{_{\chi_\beta^0}})
\nonumber\\
&&\hspace{-0.4cm}=
\Bigg\{-{1\over24}{\partial^3\over\partial^3x_{_{\tilde{E}_j}}}
\Big[\Psi_{_{2a}}-\Psi_{_{2b}}\Big]
-{1\over8}{\partial^3\over\partial^2x_{_{\tilde{E}_j}}
\partial x_{_{\chi_\alpha^\pm}}}\Big[\Psi_{_{2b}}-\Psi_{_{2d}}\Big]
\nonumber\\&&
+{1\over8}{\partial^2\over\partial^2x_{_{\tilde{E}_j}}}
\Big[\Psi_{_{1a}}-\Psi_{_{1b}}\Big]
\Bigg\}(0;x_{_{\tilde{E}_j}},x_{_{\chi_\beta^0}};
x_{_{\tilde{\nu}_i}},x_{_{\chi_\alpha^\pm}})
\nonumber\\&&
+\Bigg\{{1\over24}{\partial^3\over\partial x_{_{\tilde{E}_j}}
\partial^2x_{_{\chi_\alpha^\pm}}}\Big[3\Psi_{_{2b}}-4\Psi_{_{2c}}
+\Psi_{_{2d}}\Big]
-{1\over4}{\partial^2\over\partial x_{_{\tilde{E}_j}}
\partial x_{_{\chi_\alpha^\pm}}}\Big[\Psi_{_{1a}}-\Psi_{_{1b}}\Big]
\nonumber\\&&
-{1\over8}{\partial^2\over\partial^2x_{_{\chi_\alpha^\pm}}}
\Big[\Psi_{_{1a}}-\Psi_{_{1b}}\Big]
+{1\over24}{\partial^3\over\partial^3x_{_{\chi_\alpha^\pm}}}
\Big[\Psi_{_{2a}}-\Psi_{_{2b}}\Big]
\Bigg\}(0;x_{_{\tilde{\nu}_i}},x_{_{\chi_\alpha^\pm}};
x_{_{\tilde{E}_j}},x_{_{\chi_\beta^0}})\;.
%%%%%%%%%%%%%%%%%%%%%%%%%%%%%%%%%%%%%%%%%%%%%%%%%%%%%%%%%%%%%%%%%%%%%%%%%%%%%%%%%%%
\label{fnc4}
\end{eqnarray}

\begin{eqnarray}
%%%%%%%%%%%%%%%%%%%%%%%%%%%%%%%%%%%%%%%%%%%%%%%%%%%%%%%%%%%%%%%%%%%%%%%%%%%%%%%%%%%
&&{\rm F}_{_{M,5}}(0;x_{_{\tilde{\nu}_i}},x_{_{\chi_\alpha^\pm}};
x_{_{\tilde{E}_j}},x_{_{\chi_\beta^0}})
\nonumber\\
&&\hspace{-0.4cm}=
\Bigg\{-{1\over4}{\partial^2\over\partial^2x_{_{\tilde{E}_j}}}
\Big[\Psi_{_{2a}}-\Psi_{_{2b}}\Big]
-{1\over6}{\partial^2\over\partial x_{_{\tilde{E}_j}}
\partial x_{_{\chi_\alpha^\pm}}}\Big[3\Psi_{_{2b}}-4\Psi_{_{2c}}
+\Psi_{_{2d}}\Big]
\nonumber\\&&
+{1\over2}{\partial\over\partial x_{_{\tilde{E}_j}}}
\Big[\Psi_{_{1a}}-\Psi_{_{1b}}\Big]
\Bigg\}(0;x_{_{\tilde{E}_j}},x_{_{\chi_\beta^0}};
x_{_{\tilde{\nu}_i}},x_{_{\chi_\alpha^\pm}})
\nonumber\\&&
+{1\over4}{\partial^2\over\partial^2x_{_{\chi_\alpha^\pm}}}
\Big[\Psi_{_{2b}}-\Psi_{_{2d}}\Big]
(0;x_{_{\tilde{\nu}_i}},x_{_{\chi_\alpha^\pm}};
x_{_{\tilde{E}_j}},x_{_{\chi_\beta^0}})\;.
%%%%%%%%%%%%%%%%%%%%%%%%%%%%%%%%%%%%%%%%%%%%%%%%%%%%%%%%%%%%%%%%%%%%%%%%%%%%%%%%%%%
\label{fnc5}
\end{eqnarray}

\begin{eqnarray}
%%%%%%%%%%%%%%%%%%%%%%%%%%%%%%%%%%%%%%%%%%%%%%%%%%%%%%%%%%%%%%%%%%%%%%%%%%%%%%%%%%%
&&{\rm F}_{_{M,6}}(0;x_{_{\tilde{\nu}_i}},x_{_{\chi_\alpha^\pm}};
x_{_{\tilde{E}_j}},x_{_{\chi_\beta^0}})
\nonumber\\
&&\hspace{-0.4cm}=
{1\over4}{\partial^2\over\partial^2x_{_{\tilde{E}_j}}}
\Big[\Psi_{_{2b}}-\Psi_{_{2d}}\Big]
(0;x_{_{\tilde{E}_j}},x_{_{\chi_\beta^0}}
;x_{_{\tilde{\nu}_i}},x_{_{\chi_\alpha^\pm}})
\nonumber\\&&
+\Bigg\{-{1\over4}{\partial^2\over\partial^2x_{_{\chi_\alpha^\pm}}}
\Big[\Psi_{_{2a}}-\Psi_{_{2b}}\Big]
-{1\over6}{\partial^2\over\partial x_{_{\tilde{E}_j}}
\partial x_{_{\chi_\alpha^\pm}}}\Big[3\Psi_{_{2b}}-4\Psi_{_{2c}}
+\Psi_{_{2d}}\Big]
\nonumber\\&&
+{1\over2}\Big[{\partial\over\partial x_{_{\chi_\alpha^\pm}}}
-{\partial\over\partial x_{_{\tilde{E}_j}}}\Big]
\Big[\Psi_{_{1a}}-\Psi_{_{1b}}\Big]
\Bigg\}(0;x_{_{\tilde{\nu}_i}},x_{_{\chi_\alpha^\pm}};
x_{_{\tilde{E}_j}},x_{_{\chi_\beta^0}})\;.
%%%%%%%%%%%%%%%%%%%%%%%%%%%%%%%%%%%%%%%%%%%%%%%%%%%%%%%%%%%%%%%%%%%%%%%%%%%%%%%%%%%
\label{fnc6}
\end{eqnarray}

\begin{eqnarray}
%%%%%%%%%%%%%%%%%%%%%%%%%%%%%%%%%%%%%%%%%%%%%%%%%%%%%%%%%%%%%%%%%%%%%%%%%%%%%%%%%%%
&&{\rm F}_{_{C,1}}(x_{_{l^J}};x_{_{\tilde{\nu}_i}},x_{_{\chi_\alpha^\pm}};
x_{_{\tilde{\nu}_j}},x_{_{\chi_\beta^\pm}})
\nonumber\\
&&\hspace{-0.4cm}=
\Bigg\{-{1\over4}{\partial^3\over\partial^3x_{_{\tilde{\nu}_j}}}\Psi_{_{3a}}
+{1\over4}{\partial^2\over\partial^2x_{_{\tilde{\nu}_j}}}
\Big[\Psi_{_{2b}}-\Psi_{_{2d}}\Big]
-{\partial^3\over\partial^2x_{_{\tilde{\nu}_j}}\partial x_{_{\chi_\alpha^\pm}}}
\Big[{1\over4}\Psi_{_{3b}}+{1\over3}\Psi_{_{3c}}\Big]
\nonumber\\&&
+{1\over4}{\partial^2\over\partial x_{_{\tilde{\nu}_j}}
\partial x_{_{\chi_\alpha^\pm}}}\Psi_{_{2b}}
-{1\over4}{\partial\over\partial x_{_{\tilde{\nu}_j}}}
\Big[\Psi_{_{1a}}-\Psi_{_{1b}}\Big]
\Bigg\}(x_{_{l^J}};x_{_{\tilde{\nu}_j}},x_{_{\chi_\beta^\pm}}
;x_{_{\tilde{\nu}_i}},x_{_{\chi_\alpha^\pm}})
\nonumber\\&&
+\Bigg\{{1\over4}{\partial^3\over\partial x_{_{\tilde{\nu}_j}}
\partial^2x_{_{\chi_\alpha^\pm}}}\Psi_{_{3b}}
-{1\over8}{\partial^2\over\partial x_{_{\tilde{\nu}_j}}
\partial x_{_{\chi_\alpha^\pm}}}\Big[2\Psi_{_{2b}}+\Psi_{_{2d}}\Big]
-{1\over8}{\partial^2\over\partial x_{_{\tilde{\nu}_i}}
\partial x_{_{\tilde{\nu}_j}}}\Big[2\Psi_{_{2b}}
\nonumber\\&&
-\Psi_{_{2d}}\Big]
+{1\over4}{\partial\over\partial x_{_{\tilde{\nu}_j}}}\Psi_{_{1a}}
+{1\over4}{\partial^3\over\partial^3x_{_{\chi_\alpha^\pm}}}\Psi_{_{3a}}
-{1\over8}{\partial^2\over\partial^2x_{_{\chi_\alpha^\pm}}}
\Big[\Psi_{_{2a}}+\Psi_{_{2b}}-\Psi_{_{2d}}\Big]
\nonumber\\&&
-{1\over8}{\partial^2\over\partial x_{_{\chi_\alpha^\pm}}
\partial x_{_{\tilde{\nu}_i}}}\Psi_{_{2a}}
+{3\over8}{\partial\over\partial x_{_{\chi_\alpha^\pm}}}
\Psi_{_{1a}}\Bigg\}(x_{_{l^J}};x_{_{\tilde{\nu}_i}},x_{_{\chi_\alpha^\pm}};
x_{_{\tilde{\nu}_j}},x_{_{\chi_\beta^\pm}})\;.
%%%%%%%%%%%%%%%%%%%%%%%%%%%%%%%%%%%%%%%%%%%%%%%%%%%%%%%%%%%%%%%%%%%%%%%%%%%%%%%%%%%
\label{fc1}
\end{eqnarray}

\begin{eqnarray}
%%%%%%%%%%%%%%%%%%%%%%%%%%%%%%%%%%%%%%%%%%%%%%%%%%%%%%%%%%%%%%%%%%%%%%%%%%%%%%%%%%%
&&{\rm F}_{_{C,2}}(x_{_{l^J}};x_{_{\tilde{\nu}_i}},x_{_{\chi_\alpha^\pm}};
x_{_{\tilde{\nu}_j}},x_{_{\chi_\beta^\pm}})
\nonumber\\
&&\hspace{-0.4cm}=
\Bigg\{{1\over24}{\partial^3\over\partial^3x_{_{\tilde{\nu}_j}}}
\Big[\Psi_{_{2a}}-\Psi_{_{2b}}\Big]
+{1\over24}{\partial^3\over\partial^2x_{_{\tilde{\nu}_j}}
\partial x_{_{\chi_\alpha^\pm}}}\Big[3\Psi_{_{2b}}
-4\Psi_{_{2c}}+\Psi_{_{2d}}\Big]
\nonumber\\&&
-{1\over4}{\partial^2\over\partial^2x_{_{\tilde{\nu}_j}}}
\Big[\Psi_{_{1a}}-\Psi_{_{1b}}\Big]
-{1\over4}{\partial\over\partial x_{_{\chi_\alpha^\pm}}}\Psi_{_{0}}
\Bigg\}(x_{_{l^J}};x_{_{\tilde{\nu}_j}},x_{_{\chi_\beta^\pm}}
;x_{_{\tilde{\nu}_i}},x_{_{\chi_\alpha^\pm}})
\nonumber\\&&
+\Bigg\{{1\over24}{\partial^3\over\partial x_{_{\tilde{\nu}_j}}
\partial^2x_{_{\chi_\alpha^\pm}}}\Big[\Psi_{_{2b}}
+2\Psi_{_{2c}}-3\Psi_{_{2d}}\Big]
+{1\over8}\Big[{\partial^2\over\partial x_{_{\chi_\alpha^\pm}}
\partial x_{_{\tilde{\nu}_j}}}
\nonumber\\&&
+{\partial^2\over\partial x_{_{\tilde{\nu}_i}}
\partial x_{_{\tilde{\nu}_j}}}\Big]\Psi_{_{1a}}
-{1\over24}{\partial^3\over\partial^3x_{_{\chi_\alpha^\pm}}}
\Big[\Psi_{_{2a}}-\Psi_{_{2b}}\Big]
+{1\over8}{\partial^2\over\partial^2x_{_{\chi_\alpha^\pm}}}\Psi_{_{1b}}
\nonumber\\&&
+{1\over8}{\partial^2\over\partial x_{_{\chi_\alpha^\pm}}
\partial x_{_{\tilde{\nu}_i}}}\Psi_{_{1a}}
\Bigg\}(x_{_{l^J}};x_{_{\tilde{\nu}_i}},x_{_{\chi_\alpha^\pm}};
x_{_{\tilde{\nu}_j}},x_{_{\chi_\beta^\pm}})\;.
%%%%%%%%%%%%%%%%%%%%%%%%%%%%%%%%%%%%%%%%%%%%%%%%%%%%%%%%%%%%%%%%%%%%%%%%%%%%%%%%%%%
\label{fc2}
\end{eqnarray}

\begin{eqnarray}
%%%%%%%%%%%%%%%%%%%%%%%%%%%%%%%%%%%%%%%%%%%%%%%%%%%%%%%%%%%%%%%%%%%%%%%%%%%%%%%%%%%
&&{\rm F}_{_{C,3}}(x_{_{l^J}};x_{_{\tilde{\nu}_i}},x_{_{\chi_\alpha^\pm}};
x_{_{\tilde{\nu}_j}},x_{_{\chi_\beta^\pm}})
\nonumber\\
&&\hspace{-0.4cm}=
\Bigg\{{1\over4}{\partial^2\over\partial^2x_{_{\tilde{\nu}_j}}}
\Big[\Psi_{_{2a}}-\Psi_{_{2b}}\Big]
+{1\over6}{\partial^2\over\partial x_{_{\tilde{\nu}_j}}
\partial x_{_{\chi_\alpha^\pm}}}\Big[3\Psi_{_{2b}}-4\Psi_{_{2c}}
+\Psi_{_{2d}}\Big]
\nonumber\\&&
-{\partial^2\over\partial^2x_{_{\tilde{\nu}_j}}}
\Big[\Psi_{_{1a}}-\Psi_{_{1b}}\Big]
\Bigg\}(x_{_{l^J}};x_{_{\tilde{\nu}_j}},x_{_{\chi_\beta^\pm}}
;x_{_{\tilde{\nu}_i}},x_{_{\chi_\alpha^\pm}})
\nonumber\\&&
-{1\over4}\Bigg\{{\partial^2\over\partial^2x_{_{\chi_\alpha^\pm}}}
\Big[\Psi_{_{2b}}-\Psi_{_{2d}}\Big]
-\Big[{\partial\over\partial x_{_{\chi_\alpha^\pm}}}
+{\partial\over\partial x_{_{\tilde{\nu}_i}}}\Big]\Psi_{_{1a}}
\Bigg\}(x_{_{l^J}};x_{_{\tilde{\nu}_i}},x_{_{\chi_\alpha^\pm}};
x_{_{\tilde{\nu}_j}},x_{_{\chi_\beta^\pm}})\;.
%%%%%%%%%%%%%%%%%%%%%%%%%%%%%%%%%%%%%%%%%%%%%%%%%%%%%%%%%%%%%%%%%%%%%%%%%%%%%%%%%%%
\label{fc3}
\end{eqnarray}

\begin{eqnarray}
%%%%%%%%%%%%%%%%%%%%%%%%%%%%%%%%%%%%%%%%%%%%%%%%%%%%%%%%%%%%%%%%%%%%%%%%%%%%%%%%%%%
&&{\rm F}_{_{C,4}}(x_{_{l^J}};x_{_{\tilde{\nu}_i}},x_{_{\chi_\alpha^\pm}};
x_{_{\tilde{\nu}_j}},x_{_{\chi_\beta^\pm}})
\nonumber\\
&&\hspace{-0.4cm}=
-{1\over4}{\partial^2\over\partial^2x_{_{\tilde{\nu}_j}}}
\Big[\Psi_{_{2b}}-\Psi_{_{2d}}\Big](x_{_{l^J}};x_{_{\tilde{\nu}_j}},x_{_{\chi_\beta^\pm}}
;x_{_{\tilde{\nu}_i}},x_{_{\chi_\alpha^\pm}})
\nonumber\\&&
+\Bigg\{{1\over6}{\partial^2\over\partial x_{_{\tilde{\nu}_j}}
\partial x_{_{\chi_\alpha^\pm}}}\Big[3\Psi_{_{2b}}-4\Psi_{_{2c}}
+\Psi_{_{2d}}\Big]
+{1\over4}{\partial^2\over\partial^2x_{_{\chi_\alpha^\pm}}}
\Big[\Psi_{_{2a}}-\Psi_{_{2b}}\Big]
\nonumber\\&&
+{1\over2}\Big[{\partial\over\partial x_{_{\chi_\beta^\pm}}}
-{\partial\over\partial x_{_{\tilde{\nu}_j}}}\Big]
\Big[\Psi_{_{1a}}-\Psi_{_{1b}}\Big]
+{1\over4}{\partial\over\partial x_{_{l^J}}}\Big[\Psi_{_{1a}}
-\Psi_{_{1b}}\Big]
\nonumber\\&&
-{1\over2}{\partial\over\partial x_{_{\chi_\alpha^\pm}}}\Big[\Psi_{_{1a}}
-\Psi_{_{1b}}\Big]
\Bigg\}(x_{_{l^J}};x_{_{\tilde{\nu}_i}},x_{_{\chi_\alpha^\pm}};
x_{_{\tilde{\nu}_j}},x_{_{\chi_\beta^\pm}})\;.
%%%%%%%%%%%%%%%%%%%%%%%%%%%%%%%%%%%%%%%%%%%%%%%%%%%%%%%%%%%%%%%%%%%%%%%%%%%%%%%%%%%
\label{fc4}
\end{eqnarray}

\section{The results in the on-shell scheme\label{ap4}}
\indent\indent

\begin{eqnarray}
%%%%%%%%%%%%%%%%%%%%%%%%%%%%%%%%%%%%%%%%%%%%%%%%%%%%%%%%%%%%%%%%%%%%%%%%%%%%%%%%%%%
&&\delta\Big(\Delta a_{_{l^I}}^{2L,\;\chi^0\chi^0}\Big)
=\Delta a_{_{l^I}}^{2L,\;\chi^0\chi^0}({\rm on-shell})
-\Delta a_{_{l^I}}^{2L,\;\chi^0\chi^0}({\overline{MS}})
\nonumber\\
&&\hspace{2.5cm}=
{e^4\over(4\pi)^4s_{_{\rm w}}^4c_{_{\rm w}}^4}
\Bigg\{{x_{_{l^I}}\over24}\Bigg[\Big(1+\ln x_{_{\rm RE}}
+\vartheta_{_2}(x_{_{l^J}},x_{_{\chi_\alpha^0}},x_{_{\tilde{E}_i}})\Big)
\rho_1(x_{_{\chi_\beta^0}},x_{_{\tilde{E}_j}})
\nonumber\\
&&\hspace{3.0cm}
+\Big(1+\ln x_{_{\rm RE}}
+\vartheta_{_2}(x_{_{l^J}},x_{_{\chi_\beta^0}},x_{_{\tilde{E}_j}})\Big)
\rho_1(x_{_{\chi_\alpha^0}},x_{_{\tilde{E}_i}})\Bigg]
\nonumber\\
&&\hspace{3.0cm}\times
\Bigg[{\bf Re}\Big((\xi_{_N}^I)_{_{j\beta}}(\eta_{_N}^J)_{_{i\beta}}
(\eta_{_N}^J)_{_{\alpha j}}^\dagger(\xi_{_N}^I)_{_{\alpha i}}^\dagger\Big)
\nonumber\\
&&\hspace{3.0cm}
+{\bf Re}\Big((\eta_{_N}^I)_{_{j\beta}}(\xi_{_N}^J)_{_{i\beta}}
(\xi_{_N}^J)_{_{\alpha j}}^\dagger(\eta_{_N}^I)_{_{\alpha i}}^\dagger\Big)\Bigg]
\nonumber\\
&&\hspace{3.0cm}
-{x_{_{l^I}}\over48}(x_{_{\chi_\alpha^0}}x_{_{\chi_\beta^0}})^{1/2}
\Bigg[\vartheta_{_1}(x_{_{l^J}},x_{_{\chi_\alpha^0}},x_{_{\tilde{E}_i}})
\rho_1(x_{_{\chi_\beta^0}},x_{_{\tilde{E}_j}})
\nonumber\\
&&\hspace{3.0cm}
+\vartheta_{_1}(x_{_{l^J}},x_{_{\chi_\beta^0}},x_{_{\tilde{E}_j}})
\rho_1(x_{_{\chi_\alpha^0}},x_{_{\tilde{E}_i}})\Bigg]
\nonumber\\
&&\hspace{3.0cm}\times
\Bigg[{\bf Re}\Big((\xi_{_N}^I)_{_{j\beta}}(\xi_{_N}^J)_{_{i\beta}}
(\xi_{_N}^J)_{_{\alpha j}}^\dagger(\xi_{_N}^I)_{_{\alpha i}}^\dagger\Big)
\nonumber\\
&&\hspace{3.0cm}
+{\bf Re}\Big((\eta_{_N}^I)_{_{j\beta}}(\eta_{_N}^J)_{_{i\beta}}
(\eta_{_N}^J)_{_{\alpha j}}^\dagger(\eta_{_N}^I)_{_{\alpha i}}^\dagger\Big)\Bigg]
\nonumber\\
&&\hspace{3.0cm}
+\Bigg[{(x_{_{l^I}}x_{_{\chi_\beta^0}})^{1/2}\over4}\Big(1+\ln x_{_{\rm RE}}
+\vartheta_{_2}(x_{_{l^J}},x_{_{\chi_\alpha^0}},x_{_{\tilde{E}_i}})\Big)
\rho_2(x_{_{\tilde{E}_j}},x_{_{\chi_\beta^0}})
\nonumber\\
&&\hspace{3.0cm}
-{(x_{_{l^I}}x_{_{\chi_\alpha^0}}^2x_{_{\chi_\beta^0}})^{1/2}\over8}
\rho_2(x_{_{\tilde{E}_i}},x_{_{\chi_\alpha^0}})
\vartheta_{_1}(x_{_{l^J}},x_{_{\chi_\beta^0}},x_{_{\tilde{E}_j}})\Bigg]
\nonumber\\
&&\hspace{3.0cm}\times
{\bf Re}\Big((\eta_{_N}^I)_{_{j\beta}}(\eta_{_N}^J)_{_{i\beta}}
(\eta_{_N}^J)_{_{\alpha j}}^\dagger(\xi_{_N}^I)_{_{\alpha i}}^\dagger\Big)
\nonumber\\
&&\hspace{3.0cm}
+\Bigg[{(x_{_{l^I}}x_{_{\chi_\alpha^0}})^{1/2}\over4}
\rho_2(x_{_{\tilde{E}_i}},x_{_{\chi_\alpha^0}})\Big(1+\ln x_{_{\rm RE}}
+\vartheta_{_2}(x_{_{l^J}},x_{_{\chi_\beta^0}},x_{_{\tilde{E}_j}})\Big)
\nonumber\\
&&\hspace{3.0cm}
-{(x_{_{l^I}}x_{_{\chi_\alpha^0}}x_{_{\chi_\beta^0}}^2)^{1/2}\over8}
\vartheta_{_1}(x_{_{l^J}},x_{_{\chi_\alpha^0}},x_{_{\tilde{E}_i}})
\rho_2(x_{_{\tilde{E}_j}},x_{_{\chi_\beta^0}})\Bigg]
\nonumber\\
&&\hspace{3.0cm}\times
{\bf Re}\Big((\eta_{_N}^I)_{_{j\beta}}(\xi_{_N}^J)_{_{i\beta}}
(\xi_{_N}^J)_{_{\alpha j}}^\dagger(\xi_{_N}^I)_{_{\alpha i}}^\dagger\Big)\Bigg\}\;.
%%%%%%%%%%%%%%%%%%%%%%%%%%%%%%%%%%%%%%%%%%%%%%%%%%%%%%%%%%%%%%%%%%%%%%%%%%%%%%%%%%%
\label{dmdm-nn}
\end{eqnarray}

\begin{eqnarray}
%%%%%%%%%%%%%%%%%%%%%%%%%%%%%%%%%%%%%%%%%%%%%%%%%%%%%%%%%%%%%%%%%%%%%%%%%%%%%%%%%%%
&&\delta\Big(\Delta d_{_{l^I}}^{2L,\;\chi^0\chi^0}\Big)
=\Delta d_{_{l^I}}^{2L,\;\chi^0\chi^0}({\rm on-shell})
-\Delta d_{_{l^I}}^{2L,\;\chi^0\chi^0}({\overline{MS}})
\nonumber\\
&&\hspace{2.5cm}=
{e^5\over2(4\pi)^4s_{_{\rm w}}^4c_{_{\rm w}}^4\Lambda_{_{\rm NP}}}
\Bigg\{{x_{_{l^I}}^{1/2}\over24}\Bigg[\Big(1+\ln x_{_{\rm RE}}
+\vartheta_{_2}(x_{_{l^J}},x_{_{\chi_\alpha^0}},x_{_{\tilde{E}_i}})\Big)
\rho_1(x_{_{\chi_\beta^0}},x_{_{\tilde{E}_j}})
\nonumber\\
&&\hspace{3.0cm}
+\Big(1+\ln x_{_{\rm RE}}
+\vartheta_{_2}(x_{_{l^J}},x_{_{\chi_\beta^0}},x_{_{\tilde{E}_j}})\Big)
\rho_1(x_{_{\chi_\alpha^0}},x_{_{\tilde{E}_i}})\Bigg]
\nonumber\\
&&\hspace{3.0cm}\times
\Bigg[{\bf Im}\Big((\xi_{_N}^I)_{_{j\beta}}(\eta_{_N}^J)_{_{i\beta}}
(\eta_{_N}^J)_{_{\alpha j}}^\dagger(\xi_{_N}^I)_{_{\alpha i}}^\dagger\Big)
\nonumber\\
&&\hspace{3.0cm}
-{\bf Im}\Big((\eta_{_N}^I)_{_{j\beta}}(\xi_{_N}^J)_{_{i\beta}}
(\xi_{_N}^J)_{_{\alpha j}}^\dagger(\eta_{_N}^I)_{_{\alpha i}}^\dagger\Big)\Bigg]
\nonumber\\
&&\hspace{3.0cm}
-{(x_{_{l^I}}x_{_{\chi_\alpha^0}}x_{_{\chi_\beta^0}})^{1/2}\over48}
\Bigg[\vartheta_{_1}(x_{_{l^J}},x_{_{\chi_\alpha^0}},x_{_{\tilde{E}_i}})
\rho_1(x_{_{\chi_\beta^0}},x_{_{\tilde{E}_j}})
\nonumber\\
&&\hspace{3.0cm}
+\vartheta_{_1}(x_{_{l^J}},x_{_{\chi_\beta^0}},x_{_{\tilde{E}_j}})
\rho_1(x_{_{\chi_\alpha^0}},x_{_{\tilde{E}_i}})\Bigg]
\nonumber\\
&&\hspace{3.0cm}\times
\Bigg[{\bf Im}\Big((\xi_{_N}^I)_{_{j\beta}}(\xi_{_N}^J)_{_{i\beta}}
(\xi_{_N}^J)_{_{\alpha j}}^\dagger(\xi_{_N}^I)_{_{\alpha i}}^\dagger\Big)
\nonumber\\
&&\hspace{3.0cm}
-{\bf Im}\Big((\eta_{_N}^I)_{_{j\beta}}(\eta_{_N}^J)_{_{i\beta}}
(\eta_{_N}^J)_{_{\alpha j}}^\dagger(\eta_{_N}^I)_{_{\alpha i}}^\dagger\Big)\Bigg]
\nonumber\\
&&\hspace{3.0cm}
+\Bigg[{(x_{_{\chi_\beta^0}})^{1/2}\over4}\Big(1+\ln x_{_{\rm RE}}
+\vartheta_{_2}(x_{_{l^J}},x_{_{\chi_\alpha^0}},x_{_{\tilde{E}_i}})\Big)
\rho_2(x_{_{\tilde{E}_j}},x_{_{\chi_\beta^0}})
\nonumber\\
&&\hspace{3.0cm}
-{x_{_{\chi_\alpha^0}}x_{_{\chi_\beta^0}}^{1/2}\over8}
\rho_2(x_{_{\tilde{E}_i}},x_{_{\chi_\alpha^0}})
\vartheta_{_1}(x_{_{l^J}},x_{_{\chi_\beta^0}},x_{_{\tilde{E}_j}})\Bigg]
\nonumber\\
&&\hspace{3.0cm}\times
{\bf Im}\Big((\eta_{_N}^I)_{_{j\beta}}(\eta_{_N}^J)_{_{i\beta}}
(\eta_{_N}^J)_{_{\alpha j}}^\dagger(\xi_{_N}^I)_{_{\alpha i}}^\dagger\Big)
\nonumber\\
&&\hspace{3.0cm}
+\Bigg[{(x_{_{\chi_\alpha^0}})^{1/2}\over4}
\rho_2(x_{_{\tilde{E}_i}},x_{_{\chi_\alpha^0}})\Big(1+\ln x_{_{\rm RE}}
+\vartheta_{_2}(x_{_{l^J}},x_{_{\chi_\beta^0}},x_{_{\tilde{E}_j}})\Big)
\nonumber\\
&&\hspace{3.0cm}
-{x_{_{\chi_\alpha^0}}^{1/2}x_{_{\chi_\beta^0}}\over8}
\vartheta_{_1}(x_{_{l^J}},x_{_{\chi_\alpha^0}},x_{_{\tilde{E}_i}})
\rho_2(x_{_{\tilde{E}_j}},x_{_{\chi_\beta^0}})\Bigg]
\nonumber\\
&&\hspace{3.0cm}\times
{\bf Im}\Big((\eta_{_N}^I)_{_{j\beta}}(\xi_{_N}^J)_{_{i\beta}}
(\xi_{_N}^J)_{_{\alpha j}}^\dagger(\xi_{_N}^I)_{_{\alpha i}}^\dagger\Big)\Bigg\}\;.
%%%%%%%%%%%%%%%%%%%%%%%%%%%%%%%%%%%%%%%%%%%%%%%%%%%%%%%%%%%%%%%%%%%%%%%%%%%%%%%%%%%
\label{dedm-nn}
\end{eqnarray}

\begin{eqnarray}
%%%%%%%%%%%%%%%%%%%%%%%%%%%%%%%%%%%%%%%%%%%%%%%%%%%%%%%%%%%%%%%%%%%%%%%%%%%%%%%%%%%
&&\delta\Big(\Delta a_{_{l^I}}^{2L,\;\chi^0\chi^\pm}\Big)
=\Delta a_{_{l^I}}^{2L,\;\chi^0\chi^\pm}({\rm on-shell})
-\Delta a_{_{l^I}}^{2L,\;\chi^0\chi^\pm}({\overline{MS}})
\nonumber\\
&&\hspace{2.5cm}=
{e^4\over4(4\pi)^2s_{_{\rm w}}^4c_{_{\rm w}}^2}
\Bigg\{{x_{_{l^I}}(x_{_{\chi_\alpha^\pm}}x_{_{\chi_\beta^0}})^{1/2}\over3}
\Bigg[\rho_1(x_{_{\tilde{\nu}_i}},x_{_{\chi_\alpha^\pm}})
\varrho_{_{0,1}}(x_{_{\chi_\beta^0}},x_{_{\tilde{E}_j}})
\nonumber\\
&&\hspace{3.0cm}
-\varrho_{_{0,1}}(x_{_{\chi_\alpha^\pm}},x_{_{\tilde{\nu}_i}})
\rho_1(x_{_{\chi_\beta^0}},x_{_{\tilde{E}_j}})\Bigg]
{\bf Re}\Big((\xi_{_N}^I)_{_{j\beta}}(\lambda_{_N}^J)_{_{\beta i}}^\dagger
(\zeta_{_C}^J)_{_{\alpha j}}^\dagger(\xi_{_C}^I)_{_{\alpha i}}^\dagger\Big)
%%%%%%%%%%%%%%%%%%%%%%%%%%%%%%%%%%%%%%%%%%%%%%%%%%%%%%%%%%%%%%%%%%%%%%%%%%%%%%%%%%%
\nonumber\\
&&\hspace{3.0cm}
+{\sqrt{2}\over3}\Bigg[\Big(
1+\ln x_{_{\rm RE}}-\varrho_{_{1,1}}(x_{_{\chi_\alpha^\pm}}
,x_{_{\tilde{\nu}_i}})\Big)\rho_1(x_{_{\chi_\beta^0}},x_{_{\tilde{E}_j}})
\nonumber\\
&&\hspace{3.0cm}
-\rho_1(x_{_{\tilde{\nu}_i}},x_{_{\chi_\alpha^\pm}})\Big(1+\ln x_{_{\rm RE}}
-\varrho_{_{1,1}}(x_{_{\chi_\beta^0}},x_{_{\tilde{E}_j}})\Big)\Bigg]
\nonumber\\
&&\hspace{3.0cm}\times
{x_{_{l^I}}m_{_{l^I}}\over m_{_{\rm w}}c_{_\beta}}{\bf Re}\Big((\eta_{_N}^I)_{_{j\beta}}
(\lambda_{_N}^J)_{_{\beta i}}^\dagger(\zeta_{_C}^J)_{_{\alpha j}}^\dagger
(\eta_{_C}^I)_{_{\alpha i}}^\dagger\Big)
%%%%%%%%%%%%%%%%%%%%%%%%%%%%%%%%%%%%%%%%%%%%%%%%%%%%%%%%%%%%%%%%%%%%%%%%%%%%%%%%%%%
\nonumber\\
&&\hspace{3.0cm}
+2(x_{_{l^I}}x_{_{\chi_\alpha^\pm}})^{1/2}\Bigg[
\varphi_3(x_{_{\chi_\alpha^\pm}},x_{_{\tilde{\nu}_i}})
\Big(1+\ln x_{_{\rm RE}}-\varrho_{_{1,1}}(x_{_{\chi_\beta^0}}
,x_{_{\tilde{E}_j}})\Big)
\nonumber\\
&&\hspace{3.0cm}
-{1\over4}x_{_{\chi_\beta^0}}
\varrho_{_{0,1}}(x_{_{\chi_\alpha^\pm}},x_{_{\tilde{\nu}_i}})
\rho_2(x_{_{\tilde{E}_j}},x_{_{\chi_\beta^0}})\Bigg]
\nonumber\\
&&\hspace{3.0cm}\times
{\bf Re}\Big((\eta_{_N}^I)_{_{j\beta}}
(\lambda_{_N}^J)_{_{\beta i}}^\dagger(\zeta_{_C}^J)_{_{\alpha j}}^\dagger
(\xi_{_C}^I)_{_{\alpha i}}^\dagger\Big)
%%%%%%%%%%%%%%%%%%%%%%%%%%%%%%%%%%%%%%%%%%%%%%%%%%%%%%%%%%%%%%%%%%%%%%%%%%%%%%%%%%%
\nonumber\\
&&\hspace{3.0cm}
+{m_{_{l^I}}(x_{_{l^I}}x_{_{\chi_\beta^0}})^{1/2}\over\sqrt{2}m_{_{\rm w}}c_{_\beta}}
\Bigg[\rho_2(x_{_{\tilde{E}_j}},x_{_{\chi_\beta^0}})
\Big(1+\ln x_{_{\rm RE}}
\nonumber\\
&&\hspace{3.0cm}
-\varrho_{_{1,1}}(x_{_{\chi_\alpha^\pm}},x_{_{\tilde{\nu}_i}})\Big)
-x_{_{\chi_\alpha^\pm}}
\varphi_3(x_{_{\chi_\alpha^\pm}},x_{_{\tilde{\nu}_i}})
\varrho_{_{0,1}}(x_{_{\chi_\beta^0}},x_{_{\tilde{E}_j}})\Bigg]
\nonumber\\
&&\hspace{3.0cm}\times
{\bf Re}\Big((\xi_{_N}^I)_{_{j\beta}}(\lambda_{_N}^J)_{_{\beta i}}^\dagger
(\zeta_{_C}^J)_{_{\alpha j}}^\dagger(\eta_{_C}^I)_{_{\alpha i}}^\dagger
\Big)\Bigg\}\;.
%%%%%%%%%%%%%%%%%%%%%%%%%%%%%%%%%%%%%%%%%%%%%%%%%%%%%%%%%%%%%%%%%%%%%%%%%%%%%%%%%%%
\label{dmdm-nc}
\end{eqnarray}

\begin{eqnarray}
%%%%%%%%%%%%%%%%%%%%%%%%%%%%%%%%%%%%%%%%%%%%%%%%%%%%%%%%%%%%%%%%%%%%%%%%%%%%%%%%%%%
&&\delta\Big(\Delta d_{_{l^I}}^{2L,\;\chi^0\chi^\pm}\Big)
=\Delta d_{_{l^I}}^{2L,\;\chi^0\chi^\pm}({\rm on-shell})
-\Delta d_{_{l^I}}^{2L,\;\chi^0\chi^\pm}({\overline{MS}})
\nonumber\\
&&\hspace{2.5cm}=
{e^5\over4(4\pi)^2\Lambda_{_{\rm NP}}s_{_{\rm w}}^4c_{_{\rm w}}^2}
\Bigg\{x_{_{\chi_\alpha^\pm}}^{1/2}\Bigg[\varphi_3(x_{_{\chi_\alpha^\pm}}
,x_{_{\tilde{\nu}_i}})\Big(1+\ln x_{_{\rm RE}}
\nonumber\\
&&\hspace{3.0cm}
-\varrho_{_{1,1}}(x_{_{\chi_\beta^0}},x_{_{\tilde{E}_j}})\Big)
-{1\over4}x_{_{\chi_\beta^0}}
\varrho_{_{0,1}}(x_{_{\chi_\alpha^\pm}},x_{_{\tilde{\nu}_i}})
\rho_2(x_{_{\tilde{E}_j}},x_{_{\chi_\beta^0}})\Bigg]
\nonumber\\
&&\hspace{3.0cm}\times
{\bf Re}\Big((\eta_{_N}^I)_{_{j\beta}}
(\lambda_{_N}^J)_{_{\beta i}}^\dagger(\zeta_{_C}^J)_{_{\alpha j}}^\dagger
(\xi_{_C}^I)_{_{\alpha i}}^\dagger\Big)
%%%%%%%%%%%%%%%%%%%%%%%%%%%%%%%%%%%%%%%%%%%%%%%%%%%%%%%%%%%%%%%%%%%%%%%%%%%%%%%%%%%
\nonumber\\
&&\hspace{3.0cm}
-{m_{_{l^I}}x_{_{\chi_\beta^0}}^{1/2}\over2\sqrt{2}m_{_{\rm w}}c_{_\beta}}\Bigg[
\rho_2(x_{_{\tilde{E}_j}},x_{_{\chi_\beta^0}})
\Big(1+\ln x_{_{\rm RE}}-\varrho_{_{1,1}}(x_{_{\chi_\alpha^\pm}}
,x_{_{\tilde{\nu}_i}})\Big)
\nonumber\\
&&\hspace{3.0cm}
-x_{_{\chi_\alpha^\pm}}
\varphi_3(x_{_{\chi_\alpha^\pm}},x_{_{\tilde{\nu}_i}})
\varrho_{_{0,1}}(x_{_{\chi_\beta^0}},x_{_{\tilde{E}_j}})\Bigg]
\nonumber\\
&&\hspace{3.0cm}\times
{\bf Re}\Big((\xi_{_N}^I)_{_{j\beta}}(\lambda_{_N}^J)_{_{\beta i}}^\dagger
(\zeta_{_C}^J)_{_{\alpha j}}^\dagger(\eta_{_C}^I)_{_{\alpha i}}^\dagger
\Big)\Bigg\}\;.
%%%%%%%%%%%%%%%%%%%%%%%%%%%%%%%%%%%%%%%%%%%%%%%%%%%%%%%%%%%%%%%%%%%%%%%%%%%%%%%%%%%
\label{dedm-nc}
\end{eqnarray}

\begin{eqnarray}
%%%%%%%%%%%%%%%%%%%%%%%%%%%%%%%%%%%%%%%%%%%%%%%%%%%%%%%%%%%%%%%%%%%%%%%%%%%%%%%%%%%
&&\delta\Big(\Delta a_{_{l^I}}^{2L,\;\chi^\pm\chi^\pm}\Big)
=\Delta a_{_{l^I}}^{2L,\;\chi^\pm\chi^\pm}({\rm on-shell})
-\Delta a_{_{l^I}}^{2L,\;\chi^\pm\chi^\pm}({\overline{MS}})
\nonumber\\
&&\hspace{2.6cm}=
-{e^4x_{_{l^I}}\over(4\pi)^2s_{_{\rm w}}^4}
\Bigg\{{1\over12}\Bigg[\Big(1+\ln x_{_{\rm RE}}
-\vartheta_{_2}(x_{_{l^J}},x_{_{\chi_\alpha^\pm}},x_{_{\tilde{\nu}_i}})\Big)
\rho_1(x_{_{\tilde{\nu}_j}},x_{_{\chi_\beta^\pm}})
\nonumber\\
&&\hspace{3.1cm}
+\Big(1+\ln x_{_{\rm RE}}
-\vartheta_{_2}(x_{_{l^J}},x_{_{\chi_\beta^\pm}},x_{_{\tilde{\nu}_j}})\Big)
\rho_1(x_{_{\tilde{\nu}_i}},x_{_{\chi_\alpha^\pm}})\Bigg]
\nonumber\\
&&\hspace{3.1cm}\times
\Bigg[{m_{_{l^J}}^2\over m_{_{\rm w}}^2c_{_\beta}^2}
{\bf Re}\Big((\xi_{_C}^I)_{_{j\beta}}(\eta_{_C}^J)_{_{i\beta}}
(\eta_{_C}^J)_{_{\alpha j}}^\dagger(\xi_{_C}^I)_{_{\alpha i}}^\dagger\Big)
\nonumber\\
&&\hspace{3.1cm}
+{m_{_{l^I}}^2\over m_{_{\rm w}}^2c_{_\beta}^2}
{\bf Re}\Big((\eta_{_C}^I)_{_{j\beta}}(\xi_{_C}^J)_{_{i\beta}}
(\xi_{_C}^J)_{_{\alpha j}}^\dagger(\eta_{_C}^I)_{_{\alpha i}}^\dagger\Big)\Bigg]
%%%%%%%%%%%%%%%%%%%%%%%%%%%%%%%%%%%%%%%%%%%%%%%%%%%%%%%%%%%%%%%%%%%%%%%%%%%%%%%%%%%
\nonumber\\
&&\hspace{3.1cm}
-{1\over12}(x_{_{\chi_\alpha^\pm}}x_{_{\chi_\beta^\pm}})^{1/2}
\Bigg[\vartheta_{_1}(x_{_{l^J}},x_{_{\chi_\alpha^\pm}},x_{_{\tilde{\nu}_i}})
\rho_1(x_{_{\tilde{\nu}_j}},x_{_{\chi_\beta^\pm}})
\nonumber\\
&&\hspace{3.1cm}
+\vartheta_{_1}(x_{_{l^J}},x_{_{\chi_\beta^\pm}},x_{_{\tilde{\nu}_j}})
\rho_1(x_{_{\tilde{\nu}_i}},x_{_{\chi_\alpha^\pm}})\Bigg]
\nonumber\\
&&\hspace{3.1cm}\times
\Bigg[{\bf Re}\Big((\xi_{_C}^I)_{_{j\beta}}(\xi_{_C}^J)_{_{i\beta}}
(\xi_{_C}^J)_{_{\alpha j}}^\dagger(\xi_{_C}^I)_{_{\alpha i}}^\dagger\Big)
\nonumber\\
&&\hspace{3.1cm}
+{m_{_{l^I}}^2m_{_{l^J}}^2\over4m_{_{\rm w}}^4c_{_\beta}^4}
{\bf Re}\Big((\eta_{_C}^I)_{_{j\beta}}(\eta_{_C}^J)_{_{i\beta}}
(\eta_{_C}^J)_{_{\alpha j}}^\dagger(\eta_{_C}^I)_{_{\alpha i}}^\dagger\Big)\Bigg]
%%%%%%%%%%%%%%%%%%%%%%%%%%%%%%%%%%%%%%%%%%%%%%%%%%%%%%%%%%%%%%%%%%%%%%%%%%%%%%%%%%%
\nonumber\\
&&\hspace{3.1cm}
-{1\over2\sqrt{2}}\Bigg[{m_{_{\chi_\beta^\pm}}m_{_{l^J}}^2\over m_{_{\rm w}}^3c_{_\beta}^3}
\Big(1+\ln x_{_{\rm RE}}-\vartheta_{_2}(x_{_{l^J}},x_{_{\chi_\alpha^\pm}}
,x_{_{\tilde{\nu}_i}})\Big)
\varphi_3(x_{_{\chi_\beta^\pm}},x_{_{\tilde{\nu}_j}})
\nonumber\\
&&\hspace{3.1cm}
-{1\over2}{m_{_{\chi_\alpha^\pm}}m_{_{l^J}}^2\over m_{_{\rm w}}^3c_{_\beta}^3}
(x_{_{\chi_\alpha^\pm}}x_{_{\chi_\beta^\pm}})^{1/2}
\vartheta_{_1}(x_{_{l^J}},x_{_{\chi_\beta^\pm}},x_{_{\tilde{\nu}_j}})
\varphi_3(x_{_{\chi_\alpha^\pm}},x_{_{\tilde{\nu}_i}})\Bigg]
\nonumber\\
&&\hspace{3.1cm}\times
{\bf Re}\Big((\eta_{_C}^I)_{_{j\beta}}(\eta_{_C}^J)_{_{i\beta}}
(\eta_{_C}^J)_{_{\alpha j}}^\dagger(\xi_{_C}^I)_{_{\alpha i}}^\dagger\Big)
%%%%%%%%%%%%%%%%%%%%%%%%%%%%%%%%%%%%%%%%%%%%%%%%%%%%%%%%%%%%%%%%%%%%%%%%%%%%%%%%%%%
\nonumber\\
&&\hspace{3.1cm}
-{1\over\sqrt{2}}\Bigg[{m_{_{\chi_\alpha^\pm}}\over m_{_{\rm w}}c_{_\beta}}
\Big(1+\ln x_{_{\rm RE}}-\vartheta_{_2}(x_{_{l^J}}
,x_{_{\chi_\beta^\pm}},x_{_{\tilde{\nu}_j}})\Big)
\varphi_3(x_{_{\chi_\alpha^\pm}},x_{_{\tilde{\nu}_i}})
\nonumber\\
&&\hspace{3.1cm}
-{1\over2}{m_{_{\chi_\beta^\pm}}\over m_{_{\rm w}}c_{_\beta}}
(x_{_{\chi_\alpha^\pm}}x_{_{\chi_\beta^\pm}})^{1/2}
\vartheta_{_1}(x_{_{l^J}},x_{_{\chi_\alpha^\pm}},x_{_{\tilde{\nu}_i}})
\varphi_3(x_{_{\chi_\beta^\pm}},x_{_{\tilde{\nu}_j}})\Bigg]
\nonumber\\
&&\hspace{3.1cm}\times
{\bf Re}\Big((\eta_{_C}^I)_{_{j\beta}}(\xi_{_C}^J)_{_{i\beta}}
(\xi_{_C}^J)_{_{\alpha j}}^\dagger(\xi_{_C}^I)_{_{\alpha i}}^\dagger\Big)
\Bigg\}\;.
%%%%%%%%%%%%%%%%%%%%%%%%%%%%%%%%%%%%%%%%%%%%%%%%%%%%%%%%%%%%%%%%%%%%%%%%%%%%%%%%%%%
\label{dmdm-cc}
\end{eqnarray}

\begin{eqnarray}
%%%%%%%%%%%%%%%%%%%%%%%%%%%%%%%%%%%%%%%%%%%%%%%%%%%%%%%%%%%%%%%%%%%%%%%%%%%%%%%%%%%
&&\delta\Big(\Delta d_{_{l^I}}^{2L,\;\chi^\pm\chi^\pm}\Big)
=\Delta d_{_{l^I}}^{2L,\;\chi^\pm\chi^\pm}({\rm on-shell})
-\Delta d_{_{l^I}}^{2L,\;\chi^\pm\chi^\pm}({\overline{MS}})
\nonumber\\
&&\hspace{2.6cm}=
-{e^5x_{_{l^I}}^{1/2}\over2(4\pi)^2s_{_{\rm w}}^4\Lambda_{_{\rm NP}}}
\Bigg\{{1\over12}\Bigg[\Big(1+\ln x_{_{\rm RE}}
-\vartheta_{_2}(x_{_{l^J}},x_{_{\chi_\alpha^\pm}},x_{_{\tilde{\nu}_i}})\Big)
\rho_1(x_{_{\tilde{\nu}_j}},x_{_{\chi_\beta^\pm}})
\nonumber\\
&&\hspace{3.1cm}
+\Big(1+\ln x_{_{\rm RE}}
-\vartheta_{_2}(x_{_{l^J}},x_{_{\chi_\beta^\pm}},x_{_{\tilde{\nu}_j}})\Big)
\rho_1(x_{_{\tilde{\nu}_i}},x_{_{\chi_\alpha^\pm}})\Bigg]
\nonumber\\
&&\hspace{3.1cm}\times
\Bigg[{m_{_{l^J}}^2\over m_{_{\rm w}}^2c_{_\beta}^2}
{\bf Im}\Big((\xi_{_C}^I)_{_{j\beta}}(\eta_{_C}^J)_{_{i\beta}}
(\eta_{_C}^J)_{_{\alpha j}}^\dagger(\xi_{_C}^I)_{_{\alpha i}}^\dagger\Big)
\nonumber\\
&&\hspace{3.1cm}
-{m_{_{l^I}}^2\over m_{_{\rm w}}^2c_{_\beta}^2}
{\bf Im}\Big((\eta_{_C}^I)_{_{j\beta}}(\xi_{_C}^J)_{_{i\beta}}
(\xi_{_C}^J)_{_{\alpha j}}^\dagger(\eta_{_C}^I)_{_{\alpha i}}^\dagger\Big)\Bigg]
%%%%%%%%%%%%%%%%%%%%%%%%%%%%%%%%%%%%%%%%%%%%%%%%%%%%%%%%%%%%%%%%%%%%%%%%%%%%%%%%%%%
\nonumber\\
&&\hspace{3.1cm}
-{1\over12}(x_{_{\chi_\alpha^\pm}}x_{_{\chi_\beta^\pm}})^{1/2}
\Bigg[\vartheta_{_1}(x_{_{l^J}},x_{_{\chi_\alpha^\pm}},x_{_{\tilde{\nu}_i}})
\rho_1(x_{_{\tilde{\nu}_j}},x_{_{\chi_\beta^\pm}})
\nonumber\\
&&\hspace{3.1cm}
+\vartheta_{_1}(x_{_{l^J}},x_{_{\chi_\beta^\pm}},x_{_{\tilde{\nu}_j}})
\rho_1(x_{_{\tilde{\nu}_i}},x_{_{\chi_\alpha^\pm}})\Bigg]
\nonumber\\
&&\hspace{3.1cm}\times
\Bigg[{\bf Im}\Big((\xi_{_C}^I)_{_{j\beta}}(\xi_{_C}^J)_{_{i\beta}}
(\xi_{_C}^J)_{_{\alpha j}}^\dagger(\xi_{_C}^I)_{_{\alpha i}}^\dagger\Big)
\nonumber\\
&&\hspace{3.1cm}
-{m_{_{l^I}}^2m_{_{l^J}}^2\over4m_{_{\rm w}}^4c_{_\beta}^4}
{\bf Im}\Big((\eta_{_C}^I)_{_{j\beta}}(\eta_{_C}^J)_{_{i\beta}}
(\eta_{_C}^J)_{_{\alpha j}}^\dagger(\eta_{_C}^I)_{_{\alpha i}}^\dagger\Big)\Bigg]
%%%%%%%%%%%%%%%%%%%%%%%%%%%%%%%%%%%%%%%%%%%%%%%%%%%%%%%%%%%%%%%%%%%%%%%%%%%%%%%%%%%
\nonumber\\
&&\hspace{3.1cm}
-{1\over2\sqrt{2}}\Bigg[{m_{_{\chi_\beta^\pm}}m_{_{l^J}}^2\over m_{_{\rm w}}^3c_{_\beta}^3}
\Big(1+\ln x_{_{\rm RE}}-\vartheta_{_2}(x_{_{l^J}},x_{_{\chi_\alpha^\pm}}
,x_{_{\tilde{\nu}_i}})\Big)
\varphi_3(x_{_{\chi_\beta^\pm}},x_{_{\tilde{\nu}_j}})
\nonumber\\
&&\hspace{3.1cm}
-{1\over2}{m_{_{\chi_\alpha^\pm}}m_{_{l^J}}^2\over m_{_{\rm w}}^3c_{_\beta}^3}
(x_{_{\chi_\alpha^\pm}}x_{_{\chi_\beta^\pm}})^{1/2}
\vartheta_{_1}(x_{_{l^J}},x_{_{\chi_\beta^\pm}},x_{_{\tilde{\nu}_j}})
\varphi_3(x_{_{\chi_\alpha^\pm}},x_{_{\tilde{\nu}_i}})\Bigg]
\nonumber\\
&&\hspace{3.1cm}\times
{\bf Im}\Big((\eta_{_C}^I)_{_{j\beta}}(\eta_{_C}^J)_{_{i\beta}}
(\eta_{_C}^J)_{_{\alpha j}}^\dagger(\xi_{_C}^I)_{_{\alpha i}}^\dagger\Big)
%%%%%%%%%%%%%%%%%%%%%%%%%%%%%%%%%%%%%%%%%%%%%%%%%%%%%%%%%%%%%%%%%%%%%%%%%%%%%%%%%%%
\nonumber\\
&&\hspace{3.1cm}
-{1\over\sqrt{2}}\Bigg[{m_{_{\chi_\alpha^\pm}}\over m_{_{\rm w}}c_{_\beta}}
\Big(1+\ln x_{_{\rm RE}}-\vartheta_{_2}(x_{_{l^J}}
,x_{_{\chi_\beta^\pm}},x_{_{\tilde{\nu}_j}})\Big)
\varphi_3(x_{_{\chi_\alpha^\pm}},x_{_{\tilde{\nu}_i}})
\nonumber\\
&&\hspace{3.1cm}
-{1\over2}{m_{_{\chi_\beta^\pm}}\over m_{_{\rm w}}c_{_\beta}}
(x_{_{\chi_\alpha^\pm}}x_{_{\chi_\beta^\pm}})^{1/2}
\vartheta_{_1}(x_{_{l^J}},x_{_{\chi_\alpha^\pm}},x_{_{\tilde{\nu}_i}})
\varphi_3(x_{_{\chi_\beta^\pm}},x_{_{\tilde{\nu}_j}})\Bigg]
\nonumber\\
&&\hspace{3.1cm}\times
{\bf Im}\Big((\eta_{_C}^I)_{_{j\beta}}(\xi_{_C}^J)_{_{i\beta}}
(\xi_{_C}^J)_{_{\alpha j}}^\dagger(\xi_{_C}^I)_{_{\alpha i}}^\dagger\Big)
\Bigg\}\;.
%%%%%%%%%%%%%%%%%%%%%%%%%%%%%%%%%%%%%%%%%%%%%%%%%%%%%%%%%%%%%%%%%%%%%%%%%%%%%%%%%%%
\label{dedm-cc}
\end{eqnarray}

\end{document}